\pdfoutput=1
\newcommand*{\ATLASLATEXPATH}{}
\documentclass[cernpreprint,coverpage=false, LANGEDIT=true, atlasdraft=true, texlive=2016, UKenglish]{\ATLASLATEXPATH atlasdoc}

\usepackage[biblatex=false]{\ATLASLATEXPATH atlaspackage}
\usepackage{\ATLASLATEXPATH atlasbiblatex}

\usepackage{\ATLASLATEXPATH atlasphysics}

\graphicspath{{logos/}{figures/}}

\AtlasTitle{Fluctuations of anisotropic flow in Pb+Pb collisions at $\sqrt{s_{\mathrm{NN}}}=5.02$~$\TeV$ with the ATLAS detector}

\AtlasAbstract{%
Multi-particle azimuthal cumulants are measured as a function of centrality and transverse momentum using 470~$\mu$b$^{-1}$ of Pb+Pb collisions at $\sqrt{s_{\mathrm{NN}}}=5.02$~$\TeV$ with the ATLAS detector at the LHC. These cumulants provide information on the event-by-event fluctuations of harmonic flow coefficients $v_n$ and correlated fluctuations between two harmonics $v_n$ and $v_m$. For the first time, a non-zero four-particle cumulant is observed for dipolar flow, $v_1$. The four-particle cumulants for elliptic flow, $v_2$, and triangular flow, $v_3$, exhibit a strong centrality dependence and change sign in ultra-central collisions. This sign change is consistent with significant non-Gaussian fluctuations in $v_2$ and $v_3$. The four-particle cumulant for quadrangular flow, $v_4$, is found to change sign in mid-central collisions. Correlations between two harmonics are studied with three- and four-particle mixed-harmonic cumulants, which indicate an anti-correlation between $v_2$ and $v_3$, and a positive correlation between $v_2$ and $v_4$. These correlations decrease in strength towards central collisions and either approach zero or change sign in ultra-central collisions. To investigate the possible flow fluctuations arising from intrinsic centrality or volume fluctuations, the results are compared between two different event classes used for centrality definitions. In peripheral and mid-central collisions where the cumulant signals are large, only small differences are observed. In ultra-central collisions, the differences are much larger and transverse momentum dependent. These results provide new information to disentangle flow fluctuations from the initial and final states, as well as new insights on the influence of centrality fluctuations. 
}

\AtlasRefCode{HION-2017-09}

\PreprintIdNumber{CERN-EP-2019-023}

\AtlasJournalRef{JHEP 01 (2020) 51}
\AtlasDOI{10.1007/JHEP01(2020)051}

\usepackage{float} 
\usepackage{url}
\usepackage{harpoon}
\usepackage{amsmath}
\usepackage{mathrsfs}
\usepackage{MnSymbol}
\usepackage{multirow}
\usepackage{multicol}
\usepackage{bm}
\usepackage{cite}

\newcommand{\lr}[1]{\left\langle #1\right\rangle}
\newcommand{\llrr}[1]{\left\llangle #1\right\rrangle}

\newcommand{\pA}{\mbox{$p$+A}}

\newcommand{\etfcal}{\mbox{$\Sigma E_{\mathrm{T}}$}}

\newcommand{\nchrec}{\mbox{$N_{\mathrm{ch}}^{\mathrm{rec}}$}}

\newcommand{\npart}{\mbox{$N_{\mathrm{part}}$}}

\newcommand{\sqrtsnn}{\mbox{$\sqrt{s_{\mathrm{NN}}}$}}

\newcommand{\scn}{\mathrm{sc}_{n,m}\{4\}}
\newcommand{\sca}{\mathrm{sc}_{2,3}\{4\}}
\newcommand{\scb}{\mathrm{sc}_{2,4}\{4\}}
\newcommand{\nscn}{\mathrm{nsc}_{n,m}\{4\}}
\newcommand{\nsca}{\mathrm{nsc}_{2,3}\{4\}}
\newcommand{\nscb}{\mathrm{nsc}_{2,4}\{4\}}
\newcommand{\acn}{\mathrm{ac}_n\{3\}}
\newcommand{\aca}{\mathrm{ac}_2\{3\}}
\newcommand{\nacn}{\mathrm{nac}_n\{3\}}
\newcommand{\naca}{\mathrm{nac}_2\{3\}}

\hypersetup{pdftitle={ATLAS document},pdfauthor={The ATLAS Collaboration}}

\begin{document}

\maketitle

\tableofcontents
\section{Introduction}
\label{intro}
Heavy-ion collisions at RHIC and the LHC create hot, dense matter whose space-time evolution is well described by relativistic viscous hydrodynamics~\cite{Gale:2013da,Heinz:2013th,Busza:2018rrf}. Owing to strong event-by-event energy density fluctuations in the initial state, the distributions of the final-state particles also fluctuate event by event. These fluctuations produce an effect in the azimuthal angle $\phi$ distribution of the final-state particles, characterized by a Fourier expansion  d$N/{\textrm d}\phi\propto 1+2\sum_{n=1}^{\infty}v_{n}\cos n(\phi-\Phi_{n})$, where $v_n$ and $\Phi_n$ represent the magnitude and event-plane angle of the $n^{\mathrm{th}}$-order harmonic flow. These quantities also are conveniently represented by the `flow vector' ${\bm V}_n=v_n{\mathrm e}^{{\textrm i}n\Phi_n}$ in each event. The ${\bm V}_n$ value reflects the hydrodynamic response of the produced medium to the $n^{\textrm{th}}$-order initial-state eccentricity vector~\cite{Gardim:2011xv,Gale:2012rq}, denoted by ${\mathcal{E}}_n=\epsilon_n {\mathrm e}^{{\textrm i}n\Psi_n}$. Model calculations show that ${\bm V}_n$ is approximately proportional to $\mathcal{E}_n$ in general for $n=2$ and 3, and for $n=4$ in the case of central collisions~\cite{Teaney:2010vd,Gardim:2011xv,Niemi:2012aj}. The measurements of $v_n$ and $\Phi_n$~\cite{Adare:2011tg,ALICE:2011ab,Aad:2012bu,Chatrchyan:2013kba,Aad:2013xma,Aad:2014fla,Aad:2015lwa,ALICE:2016kpq}  place important constraints on the properties of the medium and on the density fluctuations in the initial state~\cite{Luzum:2012wu,Teaney:2010vd,Gale:2012rq,Niemi:2012aj,Qiu:2012uy,Teaney:2013dta}.

In order to disentangle the initial- and final-state effects, one needs detailed knowledge of the probability density distribution (or the event-by-event fluctuation) for single harmonics, $p(v_n)$, and two harmonics, $p(v_n,v_m)$. These distributions are often studied through multi-particle azimuthal correlations within the cumulant framework~\cite{Borghini:2000sa,Borghini:2001vi,Bilandzic:2010jr,Bilandzic:2013kga,Jia:2014jca}.  In this framework, the moments of the $p(v_n)$ distributions are measured by the $2k$-particle cumulants, $c_n\{2k\}$, for instance, $c_n\{2\}=\lr{v_n^2}$ and $c_n\{4\}=\lr{v_n^4}-2\lr{v_n^2}^2$ which are then used to define flow harmonics $v_n\{2k\}$ such as $v_n\{2\}=\left(c_n\{2\}\right)^{1/2}$ and $v_n\{4\}=\left(-c_n\{4\}\right)^{1/4}$. The four-particle cumulants $c_2\{4\}$ and $c_3\{4\}$ have been measured at RHIC and the LHC~\cite{Aad:2014vba,Abelev:2014mda,Chatrchyan:2013nka,Adamczyk:2015obl,Sirunyan:2017pan,Aaboud:2017blb,Sirunyan:2017fts,Acharya:2019vdf}. Most models of the initial state of A+A collisions predict a $p(v_n)$ with shape that  is close to Gaussian, and these models predict zero or negative values for $c_n\{4\}$~\cite{Yan:2013laa,Yan:2014nsa}. The values of $c_2\{4\}$ and $c_3\{4\}$ are found to be negative, except that $c_2\{4\}$ in very central Au+Au collisions at RHIC is positive~\cite{Adamczyk:2015obl}.  Six- and eight-particle cumulants for $v_2$ have also been measured~\cite{Aad:2014vba,Sirunyan:2017pan,Acharya:2018lmh}. 

In the cumulant framework, the $p(v_n,v_m)$ distribution is studied using the four-particle `symmetric cumulants', $\scn=\lr{v_n^2v_m^2}-\lr{v_n^2}\lr{v_m^2}$~\cite{Bilandzic:2013kga}, or the three-particle `asymmetric cumulants', $\acn=\lr{{\bm V}_n^2{\bm V}_{2n}^{*}}=\lr{v_n^2v_{2n}\cos2n(\Phi_n-\Phi_{2n})}$~\cite{Jia:2017hbm}. The asymmetric cumulants involve both the magnitude and phase of the flow vectors, and are often referred to as the `event-plane correlators'~\cite{Aad:2014fla}. The $\sca$, $\scb$ and $\aca$ values have been measured in A+A collisions~\cite{Aad:2014fla,ALICE:2016kpq,Aad:2015lwa,Acharya:2017gsw,STAR:2018fpo}. The values of $\sca$ are found to be negative, reflecting an anti-correlation between $v_2$ and $v_3$, while the positive values of $\scb$ and $\aca$ suggest a positive correlation between $v_2$ and $v_4$.

Assuming that the scaling between ${\bm V}_n$ and ${\mathcal{E}}_n$ is exactly linear, then $p(v_n)$ and $p(v_n,v_m)$ should be the same as $p(\epsilon_n)$ and $p(\epsilon_n,\epsilon_m)$ up to a global rescaling factor. In order to isolate the initial eccentricity fluctuations, it was proposed in Ref.~\cite{Giacalone:2017uqx} to measure the ratios of two cumulants of different order, for instance nc$_n\{4\}\equiv c_n\{4\}/\left(c_n\{2\}\right)^2=-\left(v_n\{4\}/v_n\{2\}\right)^4$. Similar cumulant ratios can be constructed for symmetric and asymmetric cumulants such as $\nscn\equiv\scn/(\lr{v_n^2}\lr{v_m^2})$ and $\nacn=\acn/(\lr{v_n^4}\lr{v_{2n}^2})^{1/2}$. In addition, hydrodynamic model calculations suggest strong $\pT$-dependent fluctuations of $v_n$ and $\Phi_n$ even in a single event~\cite{Gardim:2012im,Heinz:2013bua}. Such final-state intra-event flow fluctuations may change the shape of $p(v_n)$ or $p(v_n,v_m)$ in a $\pT$-dependent way and can be quantified by comparing cumulant ratios using particles from different $\pT$ ranges.

In heavy-ion collisions, $v_n$ coefficients are calculated for events with similar centrality, defined by the particle multiplicity in a fixed pseudorapidity range, which is also referred to as the reference multiplicity. The event ensemble, selected using a given reference multiplicity, is referred to as a reference event class. Due to fluctuations in the particle production process, the true centrality for events with the same reference multiplicity still fluctuates from event to event. Since the $v_n$ values vary with centrality, the fluctuations of centrality can lead to additional fluctuations of $v_n$ and change the underlying $p(v_n)$ and $p(v_n,v_m)$ distributions~\cite{Zhou:2018fxx}. Consequently, the cumulants $c_n\{2k\}$, $\scn$, and $\acn$ could be affected by the centrality resolution effects that are associated with the definition of the reference event class. Such centrality fluctuations, also known as volume fluctuations, have been shown to contribute significantly to event-by-event fluctuations of conserved quantities, especially in ultra-central collisions~\cite{Skokov:2012ds,Luo:2013bmi,Xu:2016qzd}. Recently, the centrality fluctuations were found to affect flow fluctuations as indicated  by the sign change of $c_2\{4\}$ measured in ultra-central collisions~\cite{Zhou:2018fxx}. A detailed study of $c_n\{2k\}$, $\scn$ and $\acn$ for different choices of the reference event class helps  clarify the meaning of centrality and provides insight into the sources of particle production in heavy-ion collisions. In this paper, two reference event-class definitions are used to study the influence of centrality fluctuations on flow cumulants. The total transverse energy in the forward pseudorapidity range $3.2<|\eta|<4.9$ is taken as the default definition and a second definition uses the number of reconstructed charged particles in the mid-rapidity range $|\eta|<2.5$.

This paper presents a measurement of $c_n\{2k\}$ for $n=2,3,4$ and $k=1,2,3$, $c_1\{4\}$, $\sca$, $\scb$ and $\aca$ in Pb+Pb collisions at $\sqrtsnn=5.02$~$\TeV$ with the ATLAS detector at the LHC. The corresponding normalized cumulants nc$_n\{2k\}$, cumulant ratios $v_n\{4\}/v_n\{2\}$ and $v_n\{6\}/v_n\{4\}$, as well as normalized mixed-harmonic cumulants $\nscn$ and $\naca$, are calculated in order to shed light on the nature of $p(v_n)$ and $p(v_n, v_m)$. Results are obtained with the standard cumulant method as well as with the recently proposed three-subevent cumulant method~\cite{Jia:2017hbm,Aaboud:2017blb} in order to quantify the influence of non-flow correlations such as resonance decays and jets. Results using the two reference event-class definitions are compared in order to understand the role of centrality fluctuations and to probe the particle production mechanism which directly influences the size of centrality fluctuations.

The paper is organized as follows. Sections~\ref{sec:det} and \ref{sec:sel} describe  the detector, trigger and datasets, as well as event and track selections. The mathematical framework for the multi-particle cumulants and the list of cumulant observables are provided in Section~\ref{sec:obs}.  The correlation analysis and systematic uncertainties are described in Sections~\ref{sec:ana} and~\ref{sec:syscheck}, respectively. Section~\ref{sec:res} first presents the results for various cumulant observables and then  investigates the role of centrality fluctuations by making a detailed comparison of the cumulants calculated using two reference event classes. A summary is given in Section~\ref{sec:sum}.

\section{ATLAS detector and trigger}
\label{sec:det}
The ATLAS detector~\cite{Aad:2008zzm} provides nearly full solid-angle coverage with tracking detectors, calorimeters, and muon chambers, and is well suited for measurements of multi-particle azimuthal correlations over a large pseudorapidity range.\footnote{ATLAS uses a right-handed coordinate system with its origin at the nominal interaction point (IP) in the centre of the detector and the $z$-axis along the beam pipe. The $x$-axis points from the IP to the centre of the LHC ring, and the $y$-axis points upward. Cylindrical coordinates $(r,\phi)$ are used in the transverse plane, $\phi$ being the azimuthal angle around the beam pipe. The pseudorapidity is defined in terms of the polar angle $\theta$ as $\eta=-\ln\tan(\theta/2)$.} The measurements are performed using the inner detector~(ID), the forward calorimeters~(FCal), and the zero-degree calorimeters (ZDC). The ID detects charged particles within $|\eta| < 2.5$ using a combination of silicon pixel detectors, silicon microstrip detectors~(SCT), and a straw-tube transition-radiation tracker, all immersed in a 2 T axial magnetic field~\cite{Aad:ID}. An additional pixel layer, the `insertable B-layer'~\cite{ATLAS-TDR-19,Abbott:2018ikt}, was installed during the 2013--2015 shutdown between Run 1 and Run 2, and is used in the present analysis. The FCal consists of three sampling layers, longitudinal in shower depth, and covers $3.2<|\eta|< 4.9$. The ZDC, positioned at $\pm$140~m from the IP, detects neutrons and photons with $|\eta|>8.3$. 

The ATLAS trigger system~\cite{Aaboud:2016leb} consists of a level-1 (L1) trigger implemented using a combination of dedicated electronics and programmable logic, and a high-level trigger (HLT), which uses software algorithms similar to those applied in the offline event reconstruction. Events for this analysis were selected by two types of trigger. The minimum-bias trigger required either a scalar sum, over the whole calorimeter system, of transverse energy $\Sigma \eT^{\mathrm{tot}}$ greater than 0.05 $\TeV$  or the presence of at least one neutron on both sides of the ZDC in coincidence with a track identified by the HLT. This trigger selected 22~$\mu\textrm{b}^{-1}$ of Pb+Pb data. The number of recorded events from very central Pb+Pb collisions was increased by using a dedicated trigger selecting on the $\Sigma \eT^{\mathrm{tot}}$ at L1 and $\Sigma \eT$, the total transverse energy in the FCal, at HLT. The combined trigger selects events with $\Sigma \eT$ larger than one of the three threshold values: 4.21 $\TeV$, 4.37 $\TeV$ and 4.54 $\TeV$. This ultra-central trigger has a very sharp turn-on as a function of $\etfcal$ and for these thresholds was fully efficient for the 1.3\%, 0.5\% and 0.1\% of events with the highest transverse energy in the FCal. The trigger collected 52~$\mu$b$^{-1}$, 140~$\mu$b$^{-1}$ and 470~$\mu$b$^{-1}$ of Pb+Pb collisions for the three thresholds, respectively.

In the offline data analysis, events from the minimum-bias and ultra-central triggers are combined as a function of $\etfcal$ by applying an event-by-event weight calculated as the ratio of the number of minimum-bias events to the total number of events. This procedure ensures that the weighted distribution as a function of $\etfcal$ for the combined dataset follows the distribution of the minimum-bias events, and the results measured as a function of $\etfcal$ or centrality (see Section~\ref{sec:sel}) are not biased in their $\etfcal$ or centrality values.
 
\section{Event and track selection}
\label{sec:sel}
The analysis uses approximately 470~$\mu\mathrm{b}^{-1}$ of $\sqrtsnn = 5.02$~$\TeV$ Pb+Pb data collected in 2015. The offline event selection requires a reconstructed primary vertex with a $z$ position satisfying $|\zvtx|< 100$~mm. A coincidence between the ZDC signals at forward and backward pseudorapidity rejects a variety of background processes such as elastic collisions and non-collision backgrounds, while maintaining high efficiency for inelastic processes. The contribution from events containing more than one inelastic interaction (pile-up) is studied by exploiting the correlation between the transverse energy, $\Sigma \eT$, measured in the FCal or the estimated number of neutrons $N_{\mathrm{n}}$ in the ZDC and the number of tracks associated with a primary vertex $\nchrec$. Since the distribution of $\Sigma \eT$ or $N_{\mathrm{n}}$ in events with pile-up is broader than that for the events without pile-up, pile-up events are suppressed by rejecting events with an abnormally large $\Sigma \eT$ or $N_{\mathrm{n}}$ as a function of $\nchrec$. The remaining pile-up contribution after this procedure is estimated to be less than 0.1\% in the most central collisions.

The Pb+Pb event centrality~\cite{Aad:2011yr} is characterized by the $\Sigma \eT$ deposited in the FCal over the pseudorapidity range $3.2 < |\eta| < 4.9$. The FCal $\Sigma \eT$ distribution is divided into a set of centrality intervals. A centrality interval refers to a percentile range, starting at 0\% relative to the most central collisions at the largest $\Sigma \eT$ value. Thus the 0--5\% centrality interval, for example, corresponds to the most central 5\% of the events. The ultra-central trigger mentioned in Section~\ref{sec:det} enhances the number of events in the 0--1.3\%, 0--0.5\% and 0--0.1\% centrality intervals with full efficiency for the three L1 $\etfcal$ thresholds, respectively. Centrality percentiles are set by using a Monte Carlo Glauber analysis~\cite{Miller:2007ri,Aad:2011yr} to provide a correspondence between the $\Sigma \eT$ distribution and the sampling fraction of the total inelastic Pb+Pb cross section.

Charged-particle tracks~\cite{Aad:2015wga} are reconstructed from hits in the ID and are then used to construct the primary vertices. Tracks are required to have $\pT>0.5$~$\GeV$ and $|\eta|<2.5$. They are required to have at least one pixel hit, with the additional requirement of a hit in the first pixel layer when one is expected, and at least six SCT hits. In order to reduce contribution from resonance decays, each track must have transverse and longitudinal impact parameters relative to the primary vertex which satisfy $|d_0|<1.5$~mm and $|z_0\sin\theta|<1.5$~mm, respectively~\cite{Aad:2011yk}.

The efficiency $\epsilon(\pT,\eta)$ of the track reconstruction and track selection criteria is evaluated using Pb+Pb Monte Carlo events produced with the HIJING event generator~\cite{Gyulassy:1994ew}. The generated particles in each event are rotated in azimuthal angle according to the procedure described in Ref.~\cite{Masera:2009zz} in order to produce a harmonic flow that is consistent with the previous ATLAS measurements~\cite{Aad:2011yk,Aad:2012bu}. The response of the detector is simulated using $\GEANT$4~\cite{Agostinelli:2002hh,Aad:2010ah} and the resulting events are reconstructed with the same algorithms as are applied to the data. For peripheral collisions, the efficiency ranges from 75\% at $\eta\approx 0$ to about 50\% for $|\eta| > 2$ for charged particles with $\pT > 0.8$~$\GeV$. The efficiency falls by about 5\% for a $\pT$ of 0.5~$\GeV$.  The efficiency in central collisions ranges from 71\% at $\eta\approx 0$ to about 40\% for $|\eta| > 2$ for charged particles with $\pT > 0.8$~$\GeV$, falling by about 8\% for a $\pT$ of 0.5~$\GeV$. The rate of falsely reconstructed tracks (`fake' tracks) is also estimated and found to be significant only at $\pT<1$~$\GeV$ in central collisions where it ranges from 2\% for $|\eta|<1$ to 8\% at larger $|\eta|$. The fake-track rate drops rapidly for higher $\pT$ and for more peripheral collisions. The fake-track rate is accounted for in the tracking efficiency correction following the procedure in Ref.~\cite{Aad:2014vba}.

\section{Observables}
\label{sec:obs}
Both the standard cumulant method~\cite{Borghini:2001vi} and the three-subevent cumulant method~\cite{Jia:2017hbm,Aaboud:2017blb,Aaboud:2018syf,DiFrancesco:2016srj} are used to calculate the cumulants $c_n\{4\}$, $\scn$ and $\acn$. However, only the standard method is used to calculate the six-particle cumulants $c_n\{6\}$.

\subsection{Cumulants in the standard method}
The standard cumulant method calculates the $2k$-particle ($k=1$,2...) cumulants $c_n\{2k\}$ from the $2m$-particle ($m=1$,2...$k$) azimuthal correlations $\lr{\{2m\}_n}$, which are calculated for each event as~\cite{Bilandzic:2010jr,Bilandzic:2013kga}
\begin{align}\label{eq:1}
\lr{\{2\}_n} &= \lr{{\mathrm{e}}^{{\textrm i}n(\phi_1-\phi_2)}}, & \lr{\{4\}_n} &= \lr{{\mathrm{e}}^{{\textrm i}n(\phi_1+\phi_2-\phi_3-\phi_4)}}, &  \lr{\{6\}_n} &= \lr{{\mathrm{e}}^{{\textrm i}n(\phi_1+\phi_2+\phi_3-\phi_4-\phi_5-\phi_6)}}\;,
\end{align}
where `$\lr{}$' denotes a single-event average over all pairs, quadruplets or sextuplets, respectively. The averages from Eq.~\eqref{eq:1} can be expressed in terms of per-particle normalized flow vectors ${\bm q}_{n;l}$ with $l=1,2...$ in each event~\cite{Bilandzic:2010jr}
\begin{equation}\label{eq:2}
{\bm q}_{n;l} \equiv {\sum\limits_j  \left(w_j\right)^l{\mathrm{e}}^{{\textrm i}n\phi_j}}{\left.\vphantom{\sum\limits_j}\right/}{\sum\limits_j\left(w_j\right)^l}\;, 
\end{equation}
where the sum runs over all particles in the event and $w_j$ is a weight assigned to the $j^{\textrm{th}}$ particle. This weight is constructed to correct for both detector non-uniformity and tracking inefficiency as explained in Section~\ref{sec:ana}. 

The multi-particle cumulants are obtained from the azimuthal correlations using
\begin{align}\nonumber
c_n\{2\} &=\llrr{\{2\}_n} =\lr{v_n^2},\\\label{eq:3}
c_n\{4\} &=\llrr{\{4\}_n}-2\llrr{\{2\}_n}^2=\lr{v_n^4}-2\lr{v_n^2}^2\;,\\\nonumber
c_n\{6\} &=\llrr{\{6\}_n}-9\llrr{\{4\}_n}\llrr{\{2\}_n}+12\llrr{\{2\}_n}^3=\lr{v_n^6}-9\lr{v_n^4}\lr{v_n^2}+12\lr{v_n^2}^3\;,
\end{align}
where `$\llrr{}$' represents a weighted average of $\lr{\{2k\}_n}$ over an event ensemble with similar $\etfcal$ or $\nchrec$. In the absence of non-flow correlations, the $c_n\{2k\}$ values are related to  the moments of the $p(v_n)$ distribution by the expression given in the  last part of each equation chain. In particular, the higher moments of $p(v_n)$ can be obtained by combining the cumulants of different order, for example $\lr{v_n^4}=2c_n\{2\}^2+c_n\{4\}$. If the amplitude of the flow vector does not fluctuate event by event, then Eq.~\eqref{eq:3} gives a negative $c_n\{4\}=-v_n^4$ and a positive $c_n\{2\}=v_n^2$ and $c_n\{6\}=4v_n^6$, which directly measure the true $v_n$. Flow coefficients from multi-particle cumulants $v_n\{2k\}$ are defined in this analysis as

\noindent\begin{minipage}{.2\linewidth}
 \begin{eqnarray}\nonumber
\vspace*{-0.1cm} v_{n}\{2\} =\sqrt{c_n\{2\}}\;,
  \end{eqnarray}
\end{minipage}%
\begin{minipage}{.37\linewidth}
 \begin{eqnarray}\nonumber
v_{n}\{4\}  = \left\{\begin{array}{ll} \sqrt[4]{-c_n\{4\}} & c_n\{4\}\leq0\\\nonumber
-\sqrt[4]{c_n\{4\}} & c_n\{4\}>0\;\;,   \end{array}\right.
  \end{eqnarray}
\end{minipage}%
\begin{minipage}{.35\linewidth}
 \begin{eqnarray}
v_{n}\{6\}  = \left\{\begin{array}{ll} \sqrt[6]{\frac{1}{4}c_n\{6\}} & c_n\{6\}\geq0\\\label{eq:4}
-\sqrt[6]{-\frac{1}{4}c_n\{6\}} & c_n\{6\}<0\;\;,  \end{array}\right.
  \end{eqnarray}
\end{minipage}

\noindent 
which extends the standard definition~\cite{Borghini:2001vi} of $v_{n}\{2k\}$ to regions where $c_n\{4\}>0$ and $c_n\{6\}<0$. 

If the fluctuation of the event-by-event flow-vector ${\bm V}_n=v_n{\mathrm e}^{{\textrm i}n\Phi_n}$ is described in the plane transverse to the beam by a two-dimensional Gaussian function~\footnote{Also known as a Bessel-Gaussian function.} given by
\begin{equation}\label{eq:4b}
 p({\bm V}_n) =\frac{1}{\pi\delta^2_{n}}{\mathrm e}^{-\left|{\bm V}_n-v_n^{\;_0}\right|^2\big{/}\left(\delta^2_{n}\right)}\;,
\end{equation}
then $v_{n}\{2\}=\sqrt{(v_n^{\;_0})^2+\delta^2_{n}}$ and $v_{n}\{4\}=v_{n}\{6\}=v_n^{\;_0}$~\cite{Aad:2013xma,Voloshin:2008dg}. The parameter $\delta_{n}$ is the width of the Gaussian function and $v_n^{\;_0}$ is related to the average geometry of the overlap region. However, if the shape of $p(v_n)$ has significant non-Gaussian fluctuations at large $v_{n}$, both $c_n\{4\}$ and $c_n\{6\}$ may change sign, giving negative values for $v_{n}\{4\}$ and $v_{n}\{6\}$~\cite{Jia:2014pza}.

The four-particle symmetric cumulants $\scn$ and three-particle asymmetric cumulants $\acn$ are related to multi-particle azimuthal correlations for two flow harmonics of different order by~\cite{Bilandzic:2013kga,Aaboud:2018syf}
\begin{align*}
\lr{\{4\}_{n,m}} &= \lr{{\mathrm{e}}^{{\mathrm i}n(\phi_1-\phi_2)+{\mathrm i}m(\phi_3-\phi_4)}}\,, & \lr{\{3\}_{n}} &= \lr{{\mathrm{e}}^{{\mathrm i}(n\phi_1+n\phi_2-2n\phi_3)}}\;,  \\
\scn &= \llrr{\{4\}_{n,m}}-\llrr{\{2\}_{n}}\llrr{\{2\}_{m}}\,, & \acn = \llrr{\{3\}_{n}} &= \llrr{{\mathrm{e}}^{{\mathrm i}(n\phi_1+n\phi_2-2n\phi_3)}}\;. 
\end{align*}

The first average is over all distinct quadruplets, triplets or pairs in one event to obtain $\lr{\{4\}_{n,m}}$, $\lr{\{3\}_{n}}$, $\lr{\{2\}_{n}}$ and $\lr{\{2\}_{m}}$, and the second average is over an event ensemble with the same $\etfcal$ or $\nchrec$ to obtain $\scn$ and $\acn$. In the absence of non-flow correlations, $\scn$ and $\acn$ are related to the correlation between $v_n$ and $v_m$ or between $v_n$ and $v_{2n}$, respectively:
\begin{align}\label{eq:6}
\scn &=\lr{v_n^2v_m^2}-\lr{v_n^2}\lr{v_m^2}\,, &  \acn &=\lr{v_n^2v_{2n}\cos2n(\Phi_n-\Phi_{2n})}\;.
\end{align}
Note that $\acn$ is also related to the correlation between $\Phi_n$ and $\Phi_{2n}$. This analysis measures three types of cumulants defined by Eq.\eqref{eq:6}: $\sca$, $\scb$ and $\aca$.

All the observables discussed above can be similarly defined for eccentricities by replacing $v_n$ and $\Phi_n$ with $\epsilon_n$ and $\Psi_n$ respectively. Denoted by $c_{n}\{2k,\epsilon\}$, $v_{n}\{2k,\epsilon\}$, $\mathrm{sc}_{n,m}\{4,\epsilon\}$ and $\mathrm{ac}_{n}\{3,\epsilon\}$, they describe the properties of $p(\epsilon_n)$ and $p(\epsilon_n, \epsilon_m)$. For example, $c_{n}\{4,\epsilon\} \equiv \lr{\epsilon_{2}^4}-2\lr{\epsilon_{2}^2}^2$ and $\mathrm{ac}_n\{3,\epsilon\}=\lr{\epsilon_n^2\epsilon_{2n}\cos2n(\Psi_n-\Psi_{2n})}$.

\subsection{Cumulants in the subevent method}
In the `standard' cumulant method described so far, all the $k$-particle multiplets involved in $\lr{\{k\}_n}$ and $\lr{\{k\}_{n,m}}$ are selected using charged tracks that are in the entire ID acceptance of $|\eta|<2.5$. In order to further suppress the non-flow correlations that typically involve particles emitted within a localized region in $\eta$, the charged tracks are grouped into three subevents, labelled $a$, $b$ and $c$, that each cover a unique $\eta$ range~\cite{Jia:2017hbm}:
\begin{equation*}
  \begin{array}{lcr}
-2.5<\eta_a<-\frac{2.5}{3}\;, & |\eta_b|<\frac{2.5}{3}\;, & \frac{2.5}{3}<\eta_c<2.5\;.
  \end{array}
\end{equation*}
Various subevent cumulants are then constructed by correlating particles between different subevents:
\begin{eqnarray*}
c_n^{a|c}\{2\} &\equiv& \llrr{\{2\}_n}_{a|c}\;,\\
c_n^{2a|b,c}\{4\} &\equiv& \llrr{\{4\}_n}_{2a|b,c}- 2\llrr{\{2\}_n}_{a|b}\llrr{\{2\}_n}_{a|c}\;,\\
\mathrm{sc}_{n,m}^{2a|b,c}\{4\} &\equiv& \llrr{\{4\}_{n,m}}_{2a|b,c}- \llrr{\{2\}_n}_{a|b}\llrr{\{2\}_m}_{a|c}\;,\\ \label{eq:7}
\mathrm{ac}_n^{a,b|c}\{3\} &\equiv& \llrr{\{3\}_{n}}_{a,b|c}\;,
\end{eqnarray*}
where
\begin{align*}
\lr{\{2\}_n}_{a|b} &= \lr{{\mathrm{e}}^{{\textrm i}n(\phi_1^a-\phi_2^b)}}\;, & \lr{\{4\}_n}_{2a|b,c} &= \lr{{\mathrm{e}}^{{\textrm i}n(\phi_1^a+\phi_2^a-\phi_3^b-\phi_4^c)}}\;,\\
\lr{\{4\}_{n,m}}_{2a|b,c} &= \lr{{\mathrm{e}}^{{\textrm i}n(\phi_1^a-\phi_2^b)+{\textrm i}m(\phi_3^a-\phi_4^c)}}\;,& \lr{\{3\}_{n}}_{a,b|c} &= \lr{{\mathrm{e}}^{{\mathrm i}(n\phi_1^a+n\phi_2^b-2n\phi_3^c)}}\;.
\end{align*}
The statistical precision is enhanced by interchanging  the $\eta$ range for subevent $a$  with that for subevent $b$ or $c$  which results in three independent measurements for each of $c_n\{4\}$, $\scn$ and $\acn$. They are averaged to obtain the final result.

It is well known that the values of $c_n\{2\}$ and $v_n\{2\}$ calculated using the standard cumulant method have a significant contribution from non-flow effects~\cite{Voloshin:2008dg}. Therefore, in this analysis, they are measured using the two-subevent method following the expressions used in previous publications~\cite{Aaboud:2017acw}:
\begin{align}
c_n\{2\} & \equiv c_n^{a|c}\{2\}\;, & v_n\{2\} & \equiv \sqrt{c_n^{a|c}\{2\}}\;.
\label{eq:10}
\end{align}
This definition ensures that the non-flow correlations in $v_n\{2\}$ are greatly reduced by requiring a minimum pseudorapidity gap of 1.67 between subevents $a$ and $c$. For $k$-particle cumulants with $k>2$, the standard method is used as the default since they are less influenced by non-flow correlations, and this assumption is additionally verified with the three-subevent method~\cite{Jia:2017hbm,Huo:2017nms,Zhang:2018lls}. 

\subsection{Normalized cumulants and cumulant ratios}
Any quantity which is linearly proportional to $v_n$ has the same cumulants, up to a global factor. Therefore the shapes of $p(v_n)$ and $p(v_n,v_m)$ can be more directly probed using the ratio of the cumulants~\cite{Giacalone:2016afq,Das:2017ned}:
\begin{eqnarray}\label{eq:9a}
\mathrm{nc}_n\{4\}&=&\frac{c_n\{4\}}{c_n^{a|c}\{2\}^2}=\frac{\lr{v_n^4}}{\lr{v_n^2}^2}-2\;,\\\label{eq:9b}
\mathrm{nc}_n\{6\}&=&\frac{c_n\{6\}}{4c_n^{a|c}\{2\}^3}\;,\\\label{eq:9c}
\nscn&=&\frac{\scn}{c_n^{a|c}\{2\}c_m^{a|c}\{2\}}= \frac{\lr{v_n^2v_m^2}}{\lr{v_n^2}\lr{v_m^2}}-1\;,\\\label{eq:9d}
\mathrm{nac}_n\{3\}&=&\frac{\acn}{\sqrt{\left(2c_n^{a|c}\{2\}^2+c_n\{4\}\right)c_{2n}^{a|c}\{2\}}} = \frac{\lr{v_n^2v_{2n}\cos2n(\Phi_n-\Phi_{2n})}}{\sqrt{\lr{v_n^4}\lr{v_{2n}^2}}}\;,
\end{eqnarray}
where the two-particle cumulants $c_n\{2\}$ in the denominator of these equations are calculated from subevents $a$ and $c$ using Eq.~\eqref{eq:10}. If $v_n$ is exactly proportional to  $\epsilon_n$, the normalized cumulants defined above would be the same as the normalized cumulants calculated from eccentricities in the initial state, i.e. \mbox{$\mathrm{nc}_n\{2k\} = \mathrm{nc}_n\{2k,\epsilon\}$}, $\mathrm{nsc}_{n,m}\{4\}=\mathrm{nsc}_{n,m}\{4,\epsilon\}$ and $\mathrm{nac}_n\{3\}=\mathrm{nac}_n\{3,\epsilon\}$. In practice, final-state effects, such as $\pT$-dependent fluctuations of $v_n$ and $\Phi_n$~\cite{Gardim:2012im,Heinz:2013bua}, hydrodynamic noise~\cite{Akamatsu:2016llw} and non-linear mode-mixing between harmonics of different order~\cite{Gardim:2011xv,Teaney:2012ke} can break this equality. Therefore, studying the $\pT$ dependence of these normalized cumulants can help in understanding the influence of dynamical effects from the final state.

The $\mathrm{nc}_n\{4\}$ and $\mathrm{nc}_n\{6\}$ cumulants defined above contain the same information as the previously proposed ratios of $v_n\{4\}$ to $v_n\{2\}$ and $v_n\{6\}$ to $v_n\{2\}$~\cite{Giacalone:2017uqx} given by,

\noindent\begin{minipage}{.5\linewidth}
\hspace*{0.5cm} \begin{eqnarray}\nonumber
\frac{v_n\{4\}}{v_n\{2\}}\equiv \left\{\begin{array}{ll}  \sqrt[4]{-\mathrm{nc}_n\{4\}} & \mathrm{nc}_n\{4\}\leq0\\\nonumber
-\sqrt[4]{\mathrm{nc}_n\{4\}} & \mathrm{nc}_n\{4\}>0\;\;,   \end{array}\right.
  \end{eqnarray}
\end{minipage}%
\begin{minipage}{.5\linewidth}
 \begin{eqnarray}\label{eq:9e}
\frac{v_n\{6\}}{v_n\{2\}} = \left\{\begin{array}{ll} \sqrt[6]{\mathrm{nc}_n\{6\}} & \mathrm{nc}_n\{6\}\geq0\\
-\sqrt[6]{-\mathrm{nc}_n\{6\}}  & \mathrm{nc}_n\{6\}<0\;\;.  \end{array}\right.
  \end{eqnarray}
\end{minipage}

\noindent 
The $\mathrm{nc}_n\{4\}$ and $\mathrm{nc}_n\{6\}$ values still vary smoothly as a function of centrality even if the $c_n\{4\}$ or $c_n\{6\}$ values change sign as a function of centrality. However, due to the fractional power in Eq.~\eqref{eq:9e}, this is not true for $v_n\{4\}$ and $v_n\{6\}$ in the region where the sign changes. For this reason, the results in this paper are often presented using $\mathrm{nc}_n\{4\}$ and $\mathrm{nc}_n\{6\}$ instead of $v_n\{4\}$ and $v_n\{6\}$.

\section{Data analysis}
\label{sec:ana}
The cumulants are calculated in three steps following examples from Refs.~\cite{Aaboud:2017blb,Aaboud:2018syf} using the standard and subevent methods. Since these steps are the same for $c_n\{2k\}$, $\scn$ and $\acn$, they are explained using $c_n\{2k\}$ as an example. 

In the first step, the multi-particle correlators $\lr{\{2k\}_n}$ are calculated for each event from particles in one of four $\pT$ ranges: $0.5<\pT<5$~$\GeV$, $1.0<\pT<5$~$\GeV$, $1.5<\pT<5$~$\GeV$, and $2<\pT<5$~$\GeV$. The upper $\pT$ cutoff is required to reduce the contribution from jet fragmentation.  In the second step, the correlators $\lr{\{2k\}_n}$ are averaged over an event ensemble, defined as events in either a narrow interval of $\etfcal$ (0.002 $\TeV$) or a narrow interval of $\nchrec$ (track bin width is 1) taken as the number of reconstructed charged particles in the range $0.5<\pT<5$~$\GeV$. The $c_n\{2k\}$ values are then calculated separately for these two types of reference event classes, denoted by $c_n\{2k,\etfcal\}$ and $c_n\{2k,\nchrec\}$, respectively. In order to obtain statistically significant results, in the final step the $c_n\{2k\}$ values from several neighbouring $\etfcal$ or $\nchrec$ intervals are combined, weighted by the number of events in each interval. The $\pT$ dependence of the cumulants is studied by simultaneously varying the  $\pT$ range  for all particles in each $2k$-multiplet in the cumulant analysis. This approach is different from previous studies where the $\pT$ range of only one particle in the multiplet is varied~\cite{Borghini:2001vi,Bilandzic:2013kga,Adler:2002pu,Aad:2014vba,Sirunyan:2017pan}. 

\begin{figure}[h!]
\begin{center}
\includegraphics[width=1\linewidth]{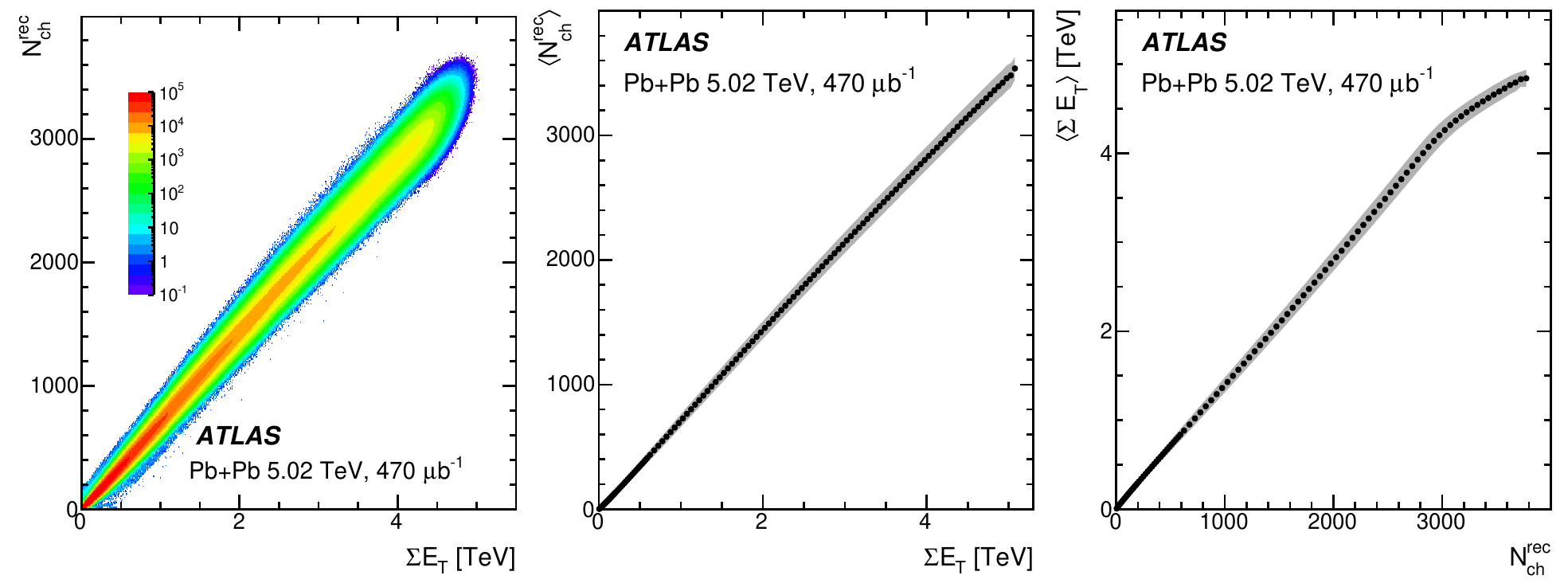}
\end{center}
\caption{\label{fig:0} The correlation between $\nchrec$ and $\etfcal$ (left panel), and the mean (solid points) and root-mean-square (shaded bands) of either the $\nchrec$ distributions for events in narrow slices of $\etfcal$ (middle panel) or the $\etfcal$ distributions for events in narrow slices of $\nchrec$ (right panel).}
\end{figure}

The left panel of Figure~\ref{fig:0} shows the correlation between $\etfcal$ and $\nchrec$.  The two quantities have an approximately linear correlation, but events with the same $\etfcal$ have significant fluctuations in $\nchrec$ and vice versa. Due to these relative fluctuations, the reference event class based on $\nchrec$ may have centrality fluctuations that differ from those of the reference event class based on $\etfcal$, even if both are matched to have the same $\lr{\etfcal}$ or the same $\lr{\nchrec}$. 

The correlation between $\etfcal$ and $\nchrec$ is studied using events  divided into narrow intervals in either $\etfcal$ or $\nchrec$. The mean and root-mean-square values of the $\nchrec$ ($\etfcal$) distributions are calculated for each $\etfcal$ ($\nchrec$) interval, and the results are shown in the middle and right panels of Figure~\ref{fig:0}, respectively. A linear relation is observed between $\lr{\nchrec}$ and $\etfcal$ over the full $\etfcal$ range, while a significant non-linear relation is observed between $\lr{\etfcal}$ and $\nchrec$ at large $\nchrec$. This latter behaviour suggests that, in ultra-central collisions, $\etfcal$ retains sensitivity to the $\lr{\nchrec}$ of the events, while $\nchrec$ has relatively poorer sensitivity to the $\lr{\etfcal}$ of the events. This implies that the true centrality is more smeared for events with the same $\nchrec$ than for events with the same $\etfcal$.  

Since $v_n$ changes with centrality, any centrality fluctuations could lead to additional fluctuation of $v_n$, and subsequently to a change in the flow cumulants. Indeed, previous ATLAS studies~\cite{Aaboud:2017acw,Aaboud:2017blb,Aaboud:2018syf} have shown that the $c_n\{2k\}$ values depend on the definition of the reference event class used for averaging. A comparison of the results based on these two reference event classes can shed light on the details of flow fluctuations and how they are affected by centrality fluctuations. 

\begin{figure}[h!]
\begin{center}
\includegraphics[width=0.9\linewidth]{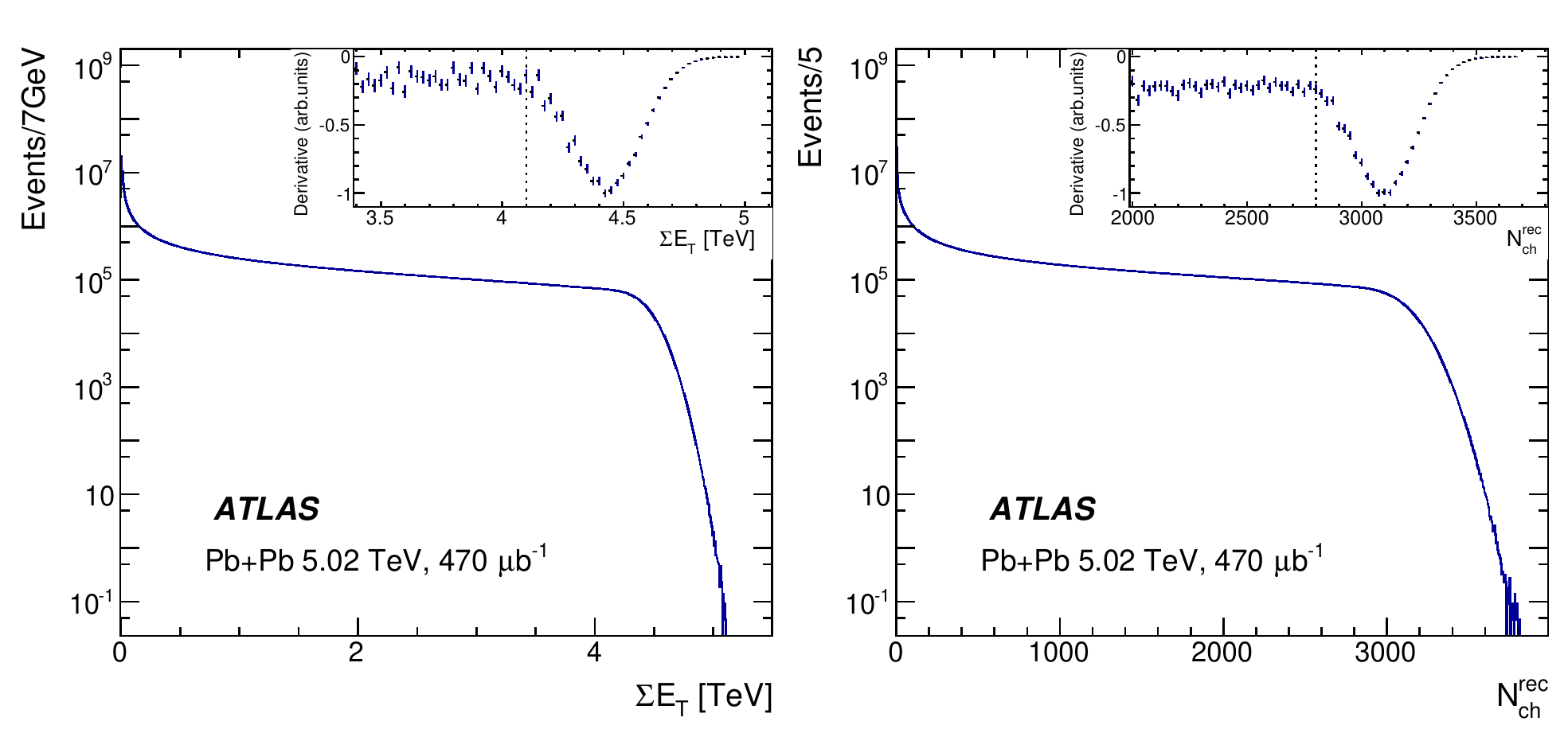}
\end{center}
\caption{\label{fig:1}
The distribution of $\etfcal$ (left panel) and the distribution of $\nchrec$ (right panel) for the Pb+Pb collisions. The insert panels show the first-order derivative of the corresponding one-dimensional distributions. The vertical dashed line indicates the location, $(\etfcal)_{\mathrm{knee}}=4.1$~$\TeV$ and $(\nchrec)_{\mathrm{knee}}=2800$ respectively, where the derivatives for $\etfcal$ and $\nchrec$ start to decrease. The values of the derivatives have been rescaled to a minimum value of $-1$.}
\end{figure}

Figure~\ref{fig:1} shows the distributions of $\nchrec$ and $\etfcal$ obtained from the projections of the two-dimensional correlation shown in the left panel of Figure~\ref{fig:0}. The inserted panels show the local first-order derivatives of the one-dimensional $\etfcal$ or $\nchrec$ distributions in the most central collisions. The derivative for the $\etfcal$ distribution is relatively independent of $\etfcal$ up to 4.1~$\TeV$ and then decreases and reaches a local minimum at around 4.4 $\TeV$. The derivative for the $\nchrec$ distribution is mostly flat up to 2800 and then decreases and reaches a local minimum at around 3100. The locations where the derivative starts to depart from a constant are defined as the knee of the $\etfcal$ or $\nchrec$ distribution and is given by $(\etfcal)_{\mathrm{knee}}=4.1$~$\TeV$ and $(\nchrec)_{\mathrm{knee}}=2800$. Events with $\etfcal>(\etfcal)_{\mathrm{knee}}$ correspond to the top 1.9\% centrality and events with $\nchrec>(\nchrec)_{\mathrm{knee}}$ correspond to top 2.7\% centrality when mapped to the equivalent $\lr{\etfcal}$. The  knees mark the locations where multiplicity distributions start to decrease sharply and the underlying centrality fluctuations are expected to deviate significantly from a Gaussian distribution~\cite{Xu:2016qzd,Zhou:2018fxx}. The knee values are important in discussing the trends of cumulants in ultra-central collisions in Section~\ref{sec:c}. 

The particle weights used in Eq.~\eqref{eq:2} that account for detector inefficiencies and non-uniformity  are defined as~\cite{Aaboud:2017acw}
\begin{eqnarray}\label{eq:11}
w_j(\phi,\eta,\pT) = d(\phi,\eta)/\epsilon(\eta,\pT)\;,
\end{eqnarray}
where $\epsilon(\eta,\pT)$ is the efficiency for reconstructing charged particles from Monte Carlo. The additional weight factor $d(\phi,\eta)$, determined from data, accounts for non-uniformities in the efficiency as a function of $\phi$ in each $\eta$ range. All reconstructed charged particles with $\pT>0.5$~$\GeV$ are entered into a two-dimensional histogram $N(\phi,\eta)$, and the weight factor is then obtained as $d(\phi,\eta) \equiv \lr{N(\eta)}/N(\phi,\eta)$, where $\lr{N(\eta)}$ is the track density averaged over $\phi$ in the given $\eta$ interval. This procedure corrects most of the $\phi$-dependent non-uniformity that results from track reconstruction~\cite{Aaboud:2017acw}.

\section{Systematic uncertainties}
\label{sec:syscheck}
The systematic uncertainties of the measurements presented in this paper are evaluated by varying different aspects of the analysis and comparing $c_n\{2k\}$, $\sca$, $\scb$ and $\aca$ with their baseline values. The main sources of systematic uncertainty are track selection, the track reconstruction efficiency, the pile-up contribution, and differences between data and Monte Carlo simulation. The uncertainties are generally small when the absolute values of the cumulants are large. The relative uncertainties are larger in central or very peripheral collisions where the signal is small. The uncertainties also decrease rapidly with increasing $\pT$, due to a larger flow signal at higher $\pT$ and are typically less than a few percent for $\pT>1$~$\GeV$. Therefore, the following discussion focuses mainly on the results obtained for charged particles in the $0.5<\pT<5$~$\GeV$ range. The systematic uncertainties are also found to be similar between the standard method and the three-subevent method.

The systematic uncertainty associated with track selection is evaluated by applying more restrictive requirements. The requirement on $|d_0|$ and $|z_0\sin\theta|$ is changed to be less than $1.0$~mm instead of the nominal value of $1.5$~mm. The numbers of pixel and SCT hits required are also increased, to two and eight respectively, to further reduce the fake-track rates. The uncertainties are less than 2\% for $c_n\{2\}$, less than 3\% for $c_2\{4\}$, $c_2\{6\}$ and $c_3\{4\}$, less than 5\% for $c_1\{4\}$ and $c_4\{4\}$, and are in the range of 1--5\% for $\sca$, $\scb$ and $\aca$.

Previous measurements~\cite{Aad:2012bu} show that the $v_n$ signal  has a strong dependence on $\pT$ but a relatively weak dependence on $\eta$. Therefore, a $\pT$-dependent uncertainty in the track reconstruction efficiency $\epsilon(\eta,\pT)$ could affect the measured cumulants through the particle weights in Eqs.~\eqref{eq:2} and \eqref{eq:11}. The uncertainty of $\epsilon(\eta,\pT)$ arises from differences in the detector conditions and known differences in the material between data and simulations. This uncertainty varies between 1\% and 4\%, depending on $\eta$ and $\pT$~\cite{Aad:2014vba}.  Its impact on cumulants is evaluated by repeating the analysis with the tracking efficiency varied up and down by its corresponding uncertainty. The impact on cumulants is in the range of 1--5\% for $c_n\{2\}$, 0.5--12\% for $c_n\{4\}$ and $c_n\{6\}$, and in the range of 2--8\% for $\scn$ and $\aca$.

Pile-up events are suppressed by exploiting the correlation, discussed in Section~\ref{sec:sel},  between  $\Sigma \eT$ measured in the FCal and the number of neutrons $N_{\mathrm{n}}$ in the ZDC. In the ultra-central collisions, where the pile-up fraction is the largest, the residual pile-up is estimated to be less than 0.1\%. The impact of the pile-up is evaluated by tightening and relaxing pile-up rejection criteria, and the resulting variation is included in the systematic uncertainty. The uncertainty is in the range of 0.1--1\% for all cumulants. 

The analysis procedure is also validated through Monte Carlo studies by comparing the observables calculated with generated particles with those obtained from reconstructed particles, using the same analysis chain and correction procedure as for data. In the low $\pT$ region, where tracking performance suffers from low efficiency and high fake-track rates, systematic differences are observed between the cumulants calculated at the generator level and at the reconstruction level. These differences are included as part of the systematic uncertainty. They amount to 0.1--3\% in mid-central and peripheral collisions and up to 10\% in the most central collisions. 

The systematic uncertainties from different sources are added in quadrature to determine the total systematic uncertainties.  These uncertainties for two-particle cumulants are in the range of 1--5\% for $c_2\{2\}$, 2--7\% for $c_3\{2\}$ and 4--9\% for $c_4\{2\}$. For multi-particle cumulants, the total uncertainties are in the range of 8--12\% for $c_1\{4\}$, 2--7 \% for $c_2\{4\}$, 1--9\% for $c_3\{4\}$, 4--15\% for $c_4\{4\}$ and  4--15 \% for $c_2\{6\}$. For symmetric and asymmetric cumulants, the total uncertainties are in the range of 2--7\% for $\sca$, 2--9\% for $\scb$ and 2--7\% for $\aca$. The total systematic uncertainties for the three-subevent cumulant method are comparable. The uncertainties in the flow coefficients $v_n\{2k\}$ are obtained from the total uncertainties of $c_n\{2k\}$ by using Eq.~\eqref{eq:3}.

The uncertainties for normalized cumulants, $\mathrm{nc}_n\{4\}$, $\mathrm{nc}_2\{6\}$, $\nsca$, $\nscb$ and $\naca$, are calculated separately for each source of systematic uncertainty discussed above, and are similar to the baseline results. Most of the systematic uncertainties cancel out in these ratios. In mid-central and peripheral collisions, the total uncertainties are in the range of 1--5\% depending on the observables. However, the total uncertainties are larger in ultra-central collisions, reaching as high as 10\% for $\mathrm{nc}_2\{6\}$ and $\naca$.

\section{Results}
\label{sec:res}
The results for various cumulant observables are presented in Sections~\ref{sec:a} and \ref{sec:b}. The cumulants are calculated using the reference event class based on $\etfcal$  and with the procedure discussed in Section~\ref{sec:ana}. The results are presented as a function of centrality calculated from $\etfcal$. Section~\ref{sec:a} discusses the cumulants related to single harmonics: $c_n\{2k,\etfcal\}$, $v_n\{2k,\etfcal\}$, and nc$_n\{2k,\etfcal\}$. Section~\ref{sec:b} presents correlations between two flow harmonics: $\mathrm{nsc}_{2,3}\{4,\etfcal\}$,  $\mathrm{nsc}_{2,4}\{4,\etfcal\}$ and  $\mathrm{nac}_{2}\{3,\etfcal\}$. The results are shown for four $\pT$ ranges: $0.5<\pT<5$~$\GeV$, $1.0<\pT<5$~$\GeV$, $1.5<\pT<5$~$\GeV$, and $2<\pT<5$~$\GeV$. The default results are obtained using the standard cumulant method and are compared with those obtained using the three-subevent cumulant method. The comparisons are shown only if significant differences are observed; otherwise, they are included in Appendix~\ref{sec:app2}.

Section~\ref{sec:c} discusses the influence of centrality fluctuations on flow cumulants. Each cumulant observable is calculated using both the $\etfcal$-based reference event class and the $\nchrec$-based reference event class. The results from the two reference event classes, for example $c_n\{2k,\etfcal\}$ and $c_n\{2k,\nchrec\}$, are compared as a function of $\lr{\etfcal}$ or $\lr{\nchrec}$. The differences are sensitive to the centrality fluctuations.

While most of the results are presented for $v_n\{2\}$, $\mathrm{nc}_n\{2k\}$, $\nscn$ and $\naca$, the results for $c_n\{4\}$, $c_n\{6\}$, $v_n\{4\}$ and $v_n\{6\}$, as well as $\sca$, $\scb$ and $\aca$, are not shown explicitly (although some are included in Appendix~\ref{sec:app1}). However, they can be obtained directly from $v_n\{2\}$, normalized cumulants and normalized mixed-harmonic cumulants according to Eqs.~\eqref{eq:9a}--\eqref{eq:9e}.

\subsection{Flow cumulants for $p(v_n)$}
\label{sec:a}
\begin{figure}[h!]
\begin{center}
\includegraphics[width=1\linewidth]{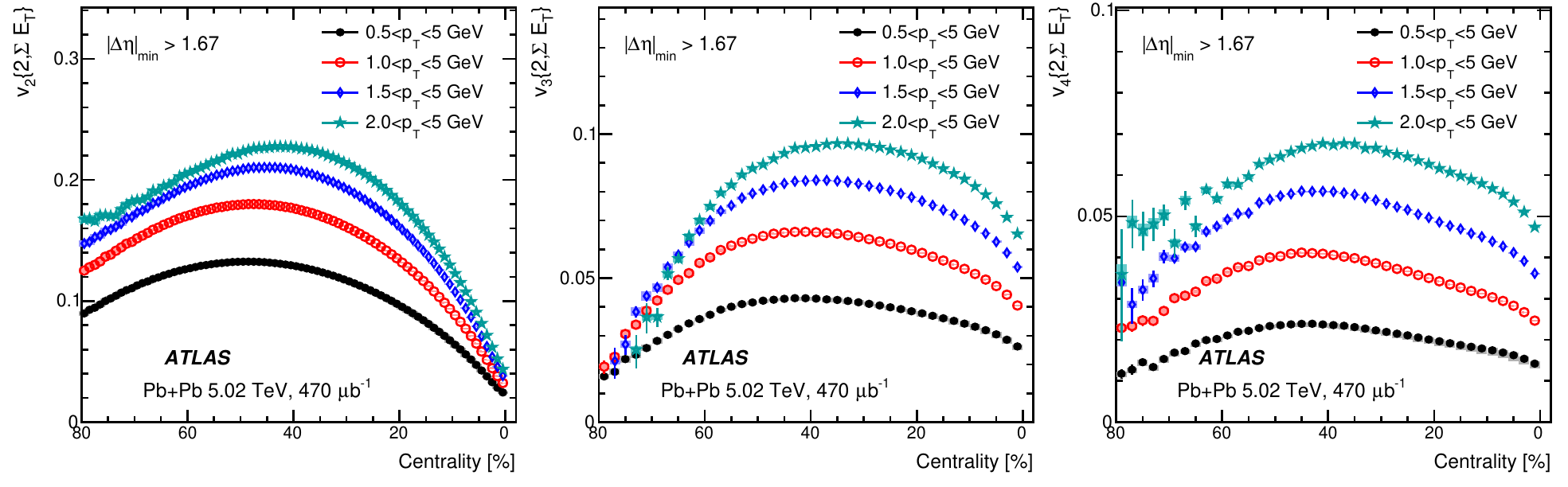}
\end{center}
\caption{\label{fig:a1} The centrality dependence of $v_2\{2, \etfcal\}$ (left panel), $v_3\{2, \etfcal\}$ (middle panel) and $v_4\{2, \etfcal\}$ (right panel) for four $\pT$ ranges. The error bars and shaded boxes represent the statistical and systematic uncertainties, respectively.}
\end{figure}

Figure~\ref{fig:a1} shows the $v_n\{2\}$ values for $n=2,3,4$ for charged particles in several $\pT$ ranges, calculated for the event class based on FCal $\etfcal$ and then plotted as a function of centrality. The $v_n\{2\}$ values are obtained from two-particle cumulants with a pseudorapidity gap according to Eq.~\eqref{eq:10}. For all $\pT$ ranges, $v_2\{2\}$ first increases and then decreases toward central collisions, reflecting the typical centrality dependence behaviour of the eccentricity $\epsilon_2$~\cite{Voloshin:2008dg}. The magnitude of $v_2\{2\}$ also increases strongly with $\pT$. The centrality and $\pT$ dependences of $v_3\{2\}$ and $v_4\{2\}$ are similar, but the tendency to decrease from mid-central toward central collisions is less pronounced. 
\begin{figure}[h!]
\begin{center}
\includegraphics[width=1\linewidth]{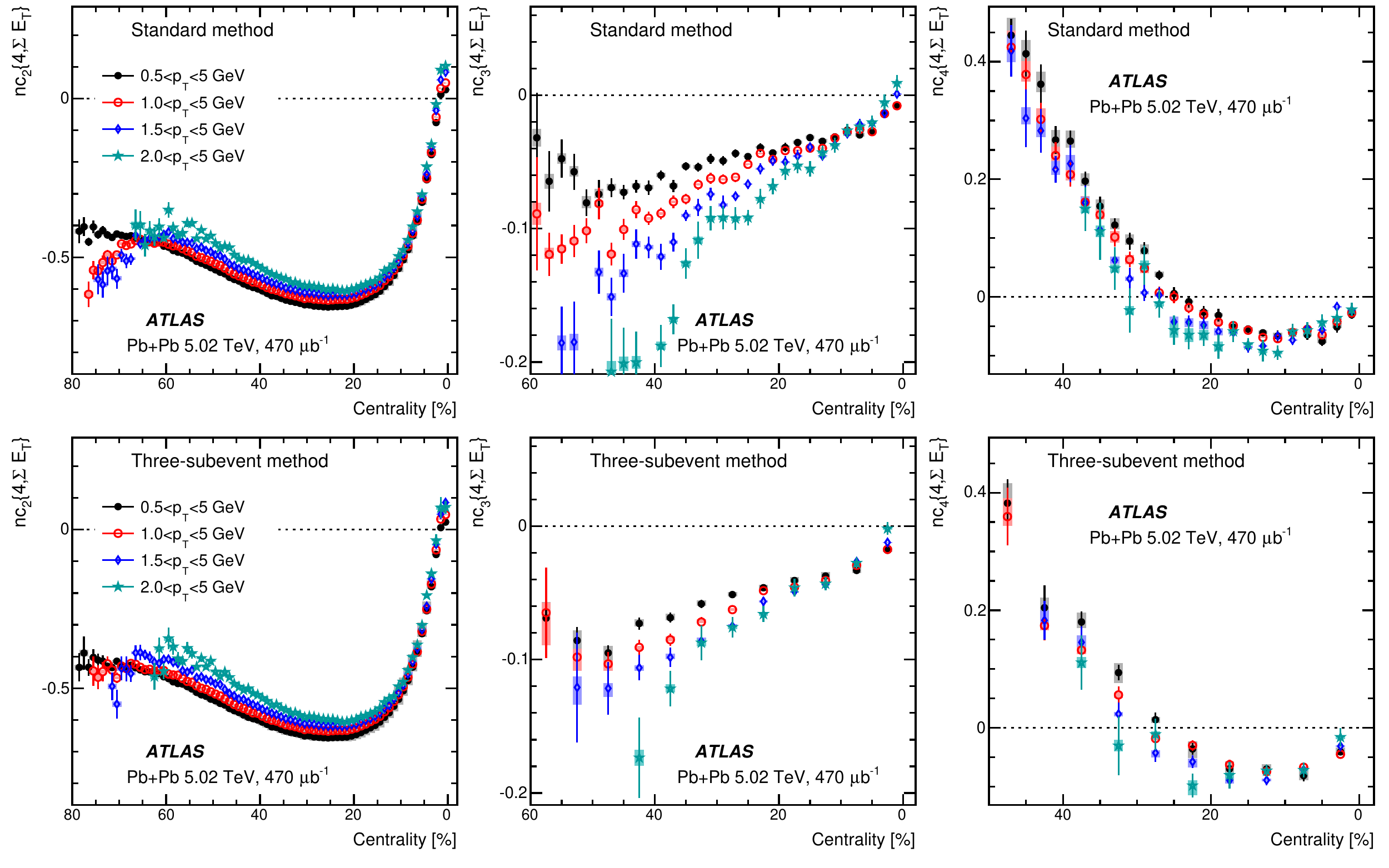}
\end{center}
\caption{\label{fig:a2} The centrality dependence of normalized four-particle cumulants nc$_2\{4, \etfcal\}$ (left panel), nc$_3\{4, \etfcal\}$ (middle panel), and nc$_4\{4, \etfcal\}$ (right panel) obtained with the standard method (top row) and the three-subevent method (bottom row) for four $\pT$ ranges. The error bars and shaded boxes represent the statistical and systematic uncertainties, respectively. Zero is indicated by a dotted line.}
\end{figure}

Figure~\ref{fig:a2} shows the centrality dependence of normalized four-particle cumulants nc$_2\{4\}$, nc$_3\{4\}$, and nc$_4\{4\}$ in four $\pT$ ranges using the standard method (top row) and the three-subevent method (bottom row). The advantage of using nc$_n\{4\}$ instead of $c_n\{4\}$ is that the $\pT$ dependence of $v_n$, seen in Figure~\ref{fig:a1}, is largely cancelled out and that nc$_n\{4\}$ directly reflects the shape of the $p(v_n)$ distributions~\cite{Aad:2013xma}. Overall, the results based on the three-subevent method behave similarly to those from the standard cumulant method, implying that the influence of non-flow correlations is small. Therefore, the remaining discussion is focused on the standard method in the top row.

Figure~\ref{fig:a2} shows that the values of nc$_2\{4\}$ and nc$_3\{4\}$ are negative in most of the centrality range. The values of $|\mathrm{nc}_2\{4\}|$ increase and then decrease toward central collisions, while the values of $|\mathrm{nc}_3\{4\}|$ decrease continuously toward central collisions. These centrality-dependent trends are shown in Refs.~\cite{Alver:2008zza,Aad:2014vba,Abelev:2014mda} to be driven by the centrality dependence of the four-particle cumulants for $\epsilon_2$ and $\epsilon_3$, respectively. The normalized cumulants still show some residual dependence on $\pT$. Namely, the $|\mathrm{nc}_2\{4\}|$ values are smaller for the higher-$\pT$ particles, while the values of $|\mathrm{nc}_3\{4\}|$ are larger for the higher $\pT$ range. Furthermore, the values of nc$_2\{4\}$ are also observed to change sign in ultra-central collisions and the pattern of these sign changes also has significant $\pT$ dependence. The observed behaviour of nc$_n\{4\}$ in ultra-central collisions is closely related to centrality fluctuations and is discussed further in Section~\ref{sec:c}.

The nc$_4\{4\}$ values, as shown in the right panels of Figure~\ref{fig:a2}, are negative in central collisions but change sign around a  centrality range of 25--30\%. The centrality value at which the sign change occurs shifts towards more peripheral collisions as the $\pT$ of the particles increases. It is well established that ${\bm V}_4$ in Pb+Pb collisions contains a linear contribution associated with the initial geometry and a mode-mixing contribution from lower-order harmonics due to the non-linear hydrodynamic response~\cite{Gardim:2011xv,Teaney:2012ke,Aad:2014fla,Aad:2015lwa,Qiu:2012uy},
\begin{eqnarray}
\label{eq:12}
{\bm V}_4 = {\bm V}_{4\textrm{L}} +\chi_{2} {\bm V}_2^2\;,
\end{eqnarray}
where the linear component ${\bm V}_{4\textrm{L}}$ is driven by the corresponding eccentricity $\epsilon_4$ in the initial geometry~\cite{Teaney:2010vd}, and $\chi_{2}$ is a constant. Previous measurements~\cite{Aad:2014fla,Aad:2015lwa} show that the ${\bm V}_{4\textrm{L}}$ term dominates in central collisions, while the ${\bm V}_2^2$ term dominates in more peripheral collisions. Therefore, the sign change of nc$_4\{4\}$ could reflect an interplay between these two contributions~\cite{Giacalone:2016mdr}. In central collisions, nc$_4\{4\}$ is dominated by a negative contribution from $p(v_{4\textrm{L}})$, while in peripheral collisions nc$_4\{4\}$ is dominated by a positive contribution from $p(v_2^2)$. The change of the crossing point with $\pT$ suggests that the relative contribution from these two sources is also a function of $\pT$. 

If the $v_n$ value is driven only by $\epsilon_n$, then $p(v_n)$ should have the same shape as $p(\epsilon_n)$. On the other hand, the significant $\pT$ dependence of nc$_n\{4\}$ in Figure~\ref{fig:a2} suggests that the shape of $p(v_n)$ also changes with $\pT$. Such $\pT$-dependent behaviour implies that the eccentricity fluctuations in the initial state are  not the only source for flow fluctuations. Dynamical fluctuations in the momentum space in the initial or final state may also change $p(v_n)$. 

\begin{figure}[h!]
\begin{center}
\includegraphics[width=1\linewidth]{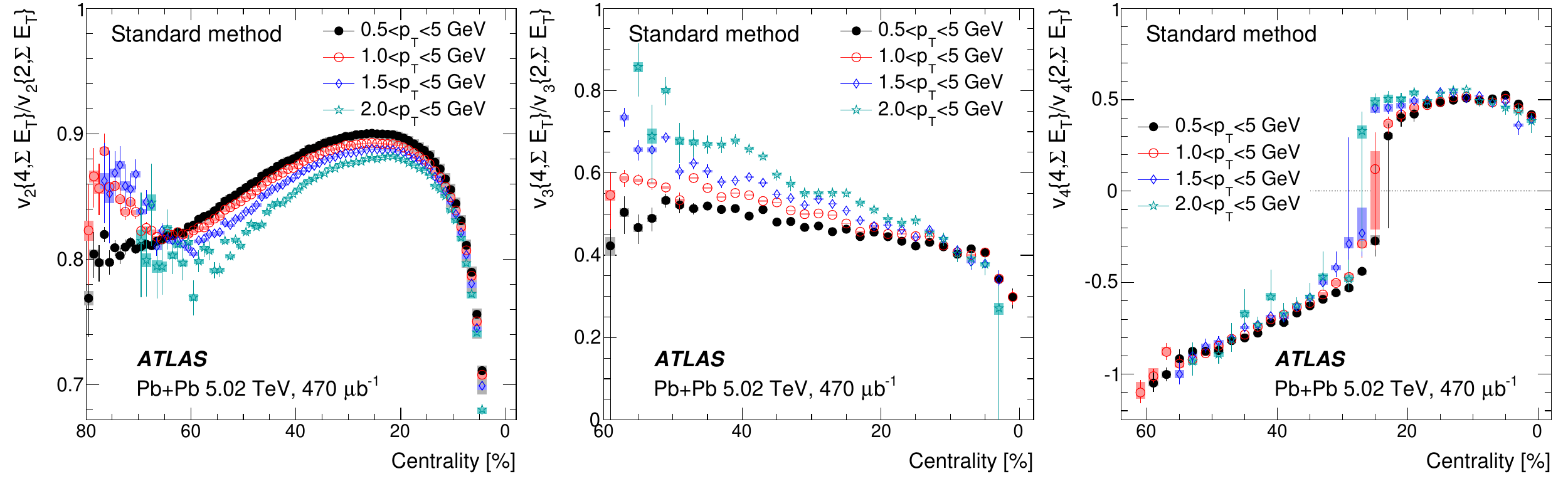}
\end{center}
\caption{\label{fig:a3} The centrality dependence of cumulant ratios $v_n\{4,\etfcal\}/v_n\{2,\etfcal\}$ for $n=2$ (left panel), $n=3$ (middle panel), and $n=4$ (right panel) for four $\pT$ ranges. The error bars and shaded boxes represent the statistical and systematic uncertainties, respectively. Zero is indicated by a dotted line.}
\end{figure}

Figure~\ref{fig:a3} shows the cumulant ratio, $v_n\{4\}/v_n\{2\}$, obtained from the nc$_n\{4\}$ data shown in Figure~\ref{fig:a2} using Eq.~\eqref{eq:9e}. This ratio is directly related to the magnitude of the relative fluctuation of the $p(v_n)$ distribution. For the Gaussian fluctuation model given in Eq.~\eqref{eq:4b}, it is \mbox{$v_n\{4\}/v_n\{2\}=v_n^{\;_0}/\sqrt{(v_n^{\;_0})^2+\delta^2_{n}}$}. A ratio close to one suggests a small flow fluctuation $\delta_{n}\ll v_n^{\;_0}$, while a ratio close to zero implies a large fluctuation $\delta_{n}\gg v_n^{\;_0}$. The results for $v_2\{4\}/v_2\{2\}$ imply that flow fluctuations are small relative to $v_2^{\;_0}$, but become larger in the most central collisions. The results for $v_3\{4\}/v_3\{2\}$ suggest that the relative fluctuation of $p(v_3)$ grows gradually from peripheral to central collisions. The values of $v_4\{4\}/v_4\{2\}$ are around 0.4--0.5 in the 0--20\% centrality range, comparable to slightly larger than the values of $v_3\{4\}/v_3\{2\}$. In peripheral collisions, $v_4\{4\}/v_4\{2\}$ is negative and its magnitude increases and reaches minus one in very peripheral collisions, suggesting a significant departure of $p(v_4)$ from a Gaussian shape. The large statistical uncertainties around the sign-change region is due to the divergence in the first derivative of the function $\sqrt[4]{|x|}$ around $x\equiv {\mathrm{nc}}_4\{4\}=0$.

\begin{figure}[h!]
\begin{center}
\includegraphics[width=1\linewidth]{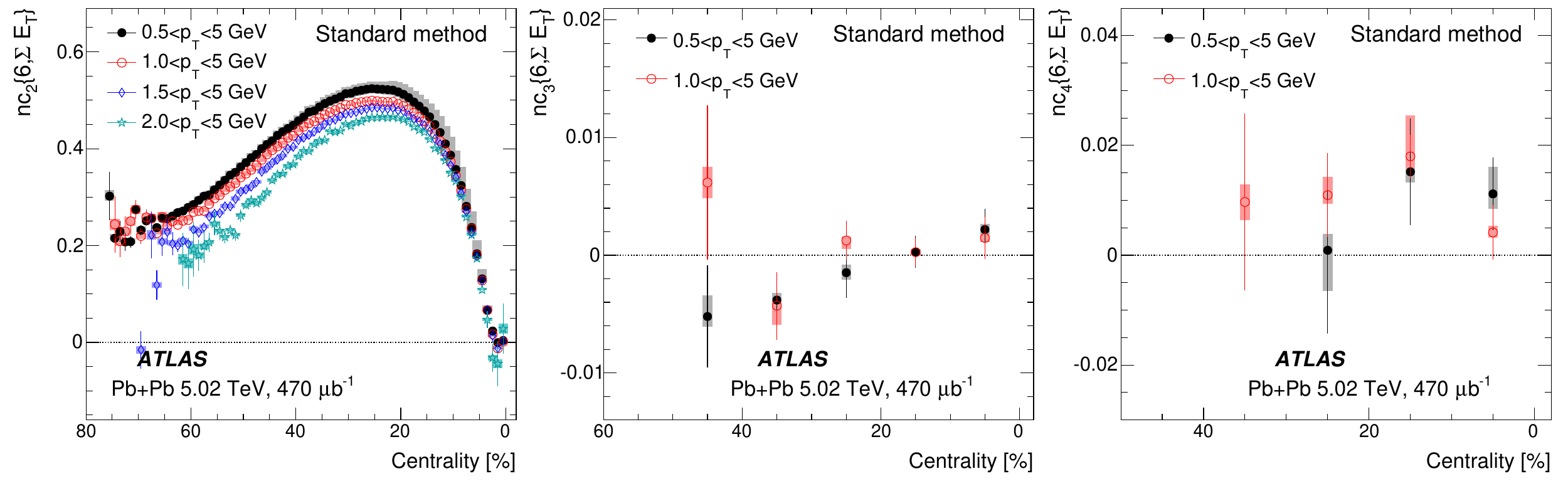}
\end{center}
\caption{\label{fig:a4} The centrality dependence of normalized six-particle cumulants nc$_2\{6, \etfcal\}$ (left panel), nc$_3\{6, \etfcal\}$ (middle panel), and nc$_4\{6, \etfcal\}$ (right panel) obtained with the standard method for several $\pT$ ranges. The error bars and shaded boxes represent the statistical and systematic uncertainties, respectively. Zero is indicated by a dotted line.}
\end{figure}

Figure~\ref{fig:a4} shows the centrality dependence of normalized six-particle cumulants nc$_2\{6\}$, nc$_3\{6\}$ and nc$_4\{6\}$. According to Eq.~\eqref{eq:9e}, these quantities are directly related to the cumulant ratios $v_n\{6\}/v_n\{2\}$. The values of nc$_2\{6\}$ are positive over most of the centrality range, but reach zero in ultra-central collisions. The centrality dependence of $|\mathrm{nc}_2\{6\}|$ is very similar to that of $|\mathrm{nc}_2\{4\}|$ in the left panel of Figure~\ref{fig:a2}.  The values of nc$_3\{6\}$ and nc$_4\{6\}$ are much smaller and have larger statistical uncertainties. Therefore, only the results from the two $\pT$ ranges with lower $\pT$ thresholds, which have the best statistical precision, are shown. The values are smaller than 0.005 and 0.01 for nc$_3\{6\}$ and nc$_4\{6\}$, which correspond to an upper limit of $|v_3\{6\}/v_3\{2\}|\lesssim\sqrt[6]{0.005}=0.38$ and $|v_4\{6\}/v_4\{2\}|\lesssim\sqrt[6]{0.01}=0.46$, respectively. 

\begin{figure}[h!]
\begin{center}
\includegraphics[width=1\linewidth]{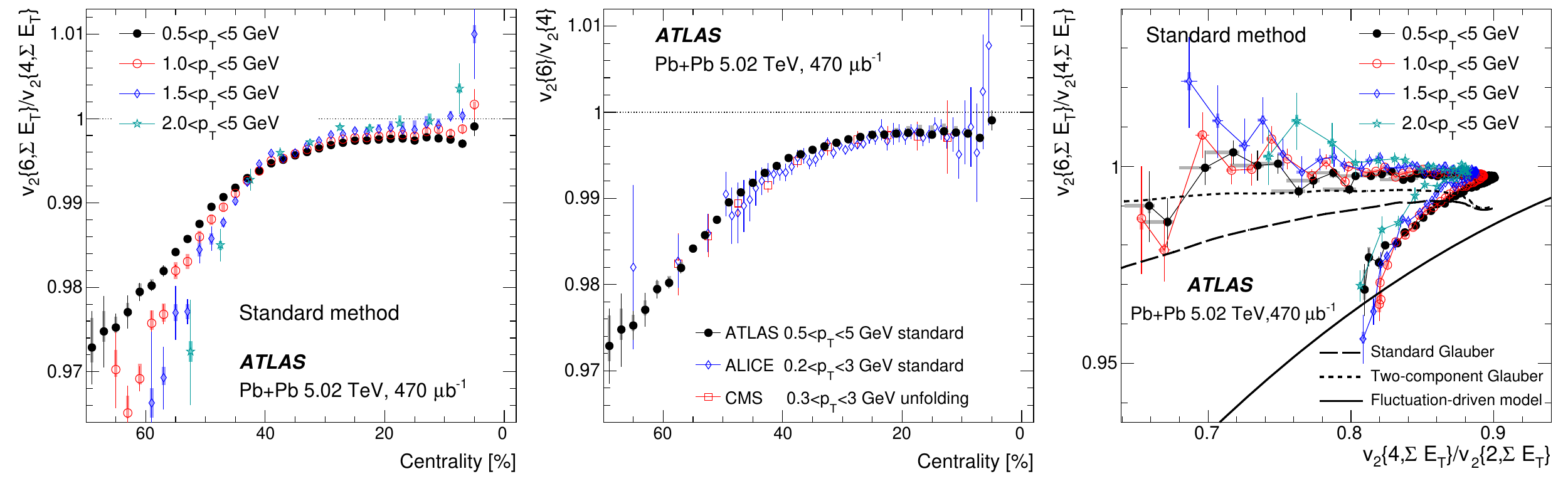}
\end{center}
\caption{\label{fig:a5} The centrality dependence of the cumulant ratio $v_2\{6,\etfcal\}/v_2\{4,\etfcal\}$ for four $\pT$ ranges (left panel) and comparison with results obtained with the standard method from ALICE Collaboration~\cite{Acharya:2018lmh} and unfolding technique from the CMS Collaboration~\cite{Sirunyan:2017fts} (middle panel), and correlation between $v_2\{6,\etfcal\}/v_2\{4,\etfcal\}$ and $v_2\{4,\etfcal\}/v_2\{2,\etfcal\}$ compared with models based on initial-state eccentricities (right panel). The error bars and shaded boxes represent the statistical and systematic uncertainties, respectively. One is indicated by a dotted line.}
\end{figure}

From the measured nc$_2\{6\}$ and nc$_2\{4\}$, the ratio of the six-particle cumulant to the fourth-particle cumulant, $v_2\{6\}/v_2\{4\}$, can be obtained. The results are shown in the left panel of Figure~\ref{fig:a5}. For the Gaussian fluctuation model in Eq.~\eqref{eq:4b}, this ratio is expected to be one. The apparent deviation of the ratio from one suggests non-Gaussianity of $p(v_2)$ over a broad centrality range. The results for different $\pT$ ranges are close to each other, but nevertheless show systematic- and  centrality-dependent differences. In general, the results from higher $\pT$ are larger in central collisions and smaller in peripheral collisions than those from lower $\pT$. The middle panel of Figure~\ref{fig:a5} compares the results for $0.5<\pT<5$ GeV with those obtained from ALICE and CMS Collaborations. Despite slight differences in the $\pT$ selections, good consistency is observed, although the ATLAS results have much smaller statistical and systematic uncertainties.

To further understand the nature of the $p(v_2)$ and its relation to $p(\epsilon_2)$, the right panel of Figure~\ref{fig:a5} shows directly the correlation between $v_2\{6\}/v_2\{4\}$ and $v_2\{4\}/v_2\{2\}$. Each data point is obtained by combining the information from the left panels of Figures~\ref{fig:a3} and \ref{fig:a5} from the same centrality range. The central region corresponds to the left-most points, while peripheral region corresponds to points near the bottom-middle of the panel. If $v_2$ values are driven by $\epsilon_2$, this correlation should be directly comparable to analogous correlation calculated directly from initial-state elliptic eccentricity: $v_2\{6,\epsilon\}/v_2\{4,\epsilon\}$ vs $v_2\{4,\epsilon\}/v_2\{2,\epsilon\}$. The data are compared to correlations from three initial state models: the standard Glauber model with $\epsilon_2$ calculated from the participating nucleons (long-dashed line)~\cite{Miller:2007ri,Zhou:2018fxx}, a two-component Glauber model with $\epsilon_2$ calculated from a combination of participating nucleons and binary nucleon-nucleon collisions (short-dashed line)~\cite{Miller:2007ri,Zhou:2018fxx}, or a fluctuation-driven model with $\epsilon_2$ calculated from random sources (solid line)~\cite{Yan:2013laa}. These models fail to describe quantitatively the overall correlation pattern, although the two-component Glauber model is closest to the data in central collisions, while the fluctuation-driven model is closest to the data in peripheral collisions.

\begin{figure}[h!]
\begin{center}
\includegraphics[width=1\linewidth]{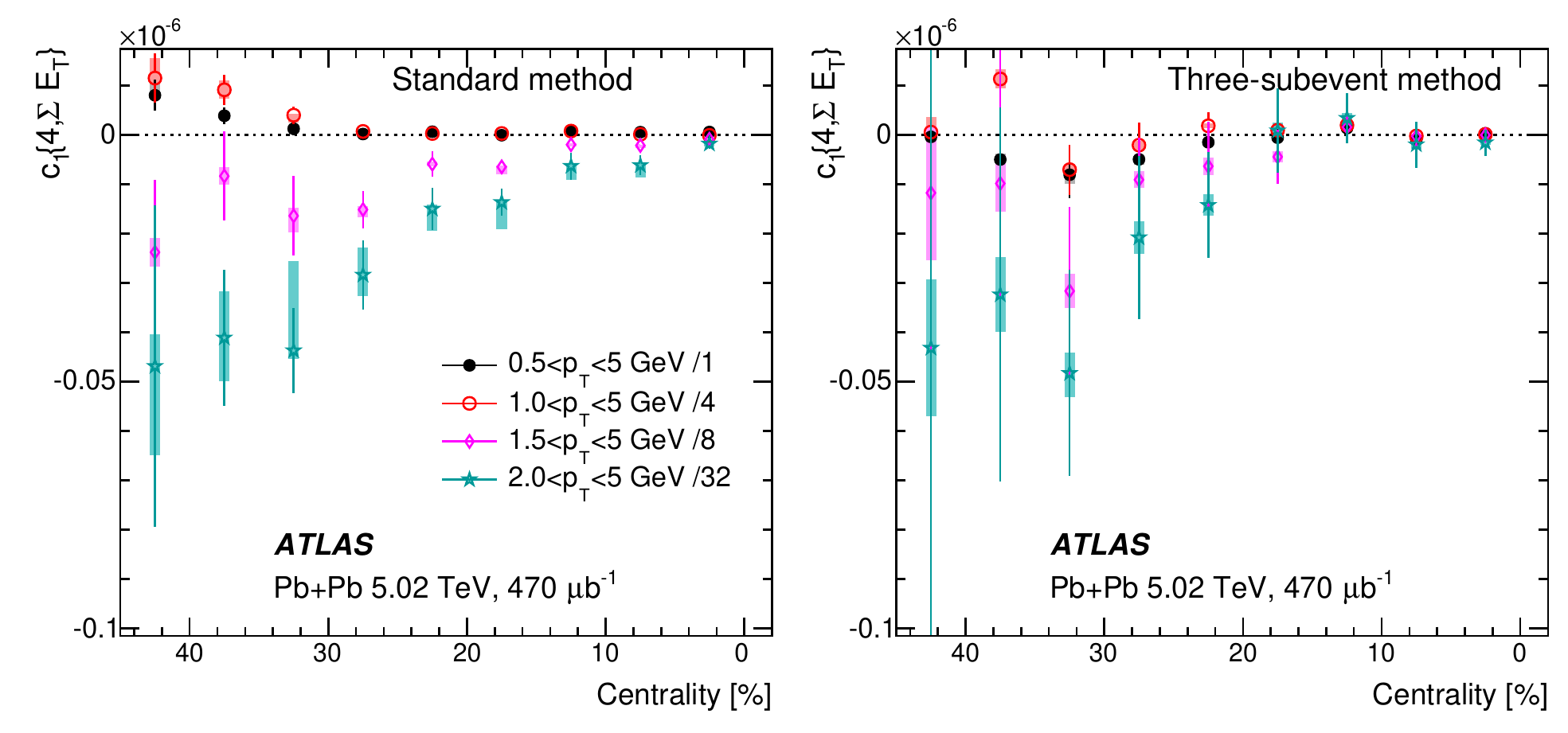}
\end{center}
\caption{\label{fig:a6} The centrality dependence of $c_1\{4\}$ calculated for charged particles in several $\pT$ ranges with the standard method (left panel) and three-subevent method (right panel). The error bars and shaded boxes represent the statistical and systematic uncertainties, respectively. The data for each $\pT$ range are scaled by a constant factor indicated in the legend for the purpose of presentation. Zero is indicated by a dotted line.}
\end{figure}

The multi-particle correlations are also calculated to obtain cumulants for the dipolar flow, $v_1$. Figure~\ref{fig:a6} shows the centrality dependence of $c_1\{4\}$ in several $\pT$ ranges, which is obtained from the reference event class based on $\etfcal$. In the hydrodynamic picture, $c_1\{4\}$ is sensitive to event-by-event fluctuations of the dipolar eccentricity $\epsilon_1$ associated with initial-state geometry~\cite{Teaney:2010vd}. This measurement has a large uncertainty because both $\llrr{\{4\}_1}$ and $\llrr{\{2\}_1}$ in Eq.~\eqref{eq:3} contain a significant contribution from global momentum-conservation effects~\cite{Borghini:2002mv,Aad:2012bu}. This contribution cancels out for $c_1\{4\}$ but leads to a large statistical uncertainty. A negative $c_1\{4\}$ for $\pT>1.5$~$\GeV$ is observed in both the standard and three-subevent cumulant methods, which reflects the event-by-event fluctuations of the dipolar eccentricity.

\begin{figure}[h!]
\begin{center}
\includegraphics[width=1\linewidth]{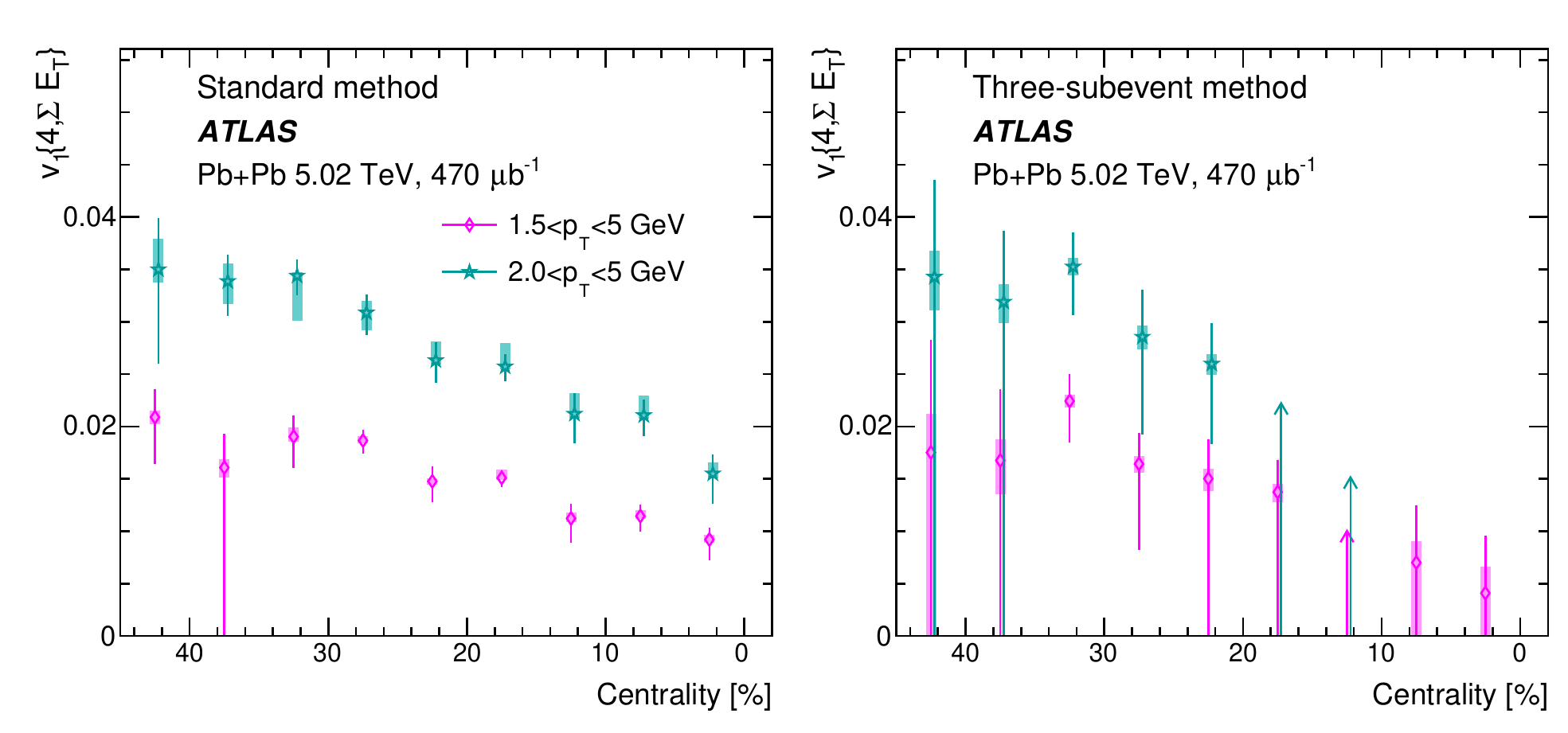}
\end{center}
\caption{\label{fig:a7} The centrality dependence of $v_1\{4\}$ calculated for charged particles in two $\pT$ ranges with the standard method (left panel) and three-subevent method (right panel). The error bars and shaded boxes represent the statistical and systematic uncertainties, respectively.}
\end{figure}

Previously, ATLAS measured $v_1$ using the two-particle correlation method in Pb+Pb collisions at $\sqrtsnn = 2.76$~$\TeV$ where an explicit procedure was employed to subtract the global momentum-conservation effects~\cite{Aad:2012bu}. The $v_1\{2\}$ values was observed to be negative at low $\pT$, change sign at $\pT\approx 1.2$~$\GeV$ and increase quickly for higher $\pT$. Therefore, a $c_1\{4\}$ signal is expected to be larger and easier to measure at higher $\pT$. Figure~\ref{fig:a7} shows the $v_1\{4\}$ values calculated from $c_1\{4\}$ for the two highest $\pT$ ranges: $1.5<\pT<5$~$\GeV$ and $2<\pT<5$~$\GeV$. The $v_1\{4\}$ values increase both with $\pT$ and in more peripheral collisions, and are in the range of 0.02--0.04 for $2<\pT<5$~$\GeV$. 

\subsection{Flow cumulants for $p(v_n,v_m)$ }
\label{sec:b}
\begin{figure}[h!]
\begin{center}
\includegraphics[width=0.8\linewidth]{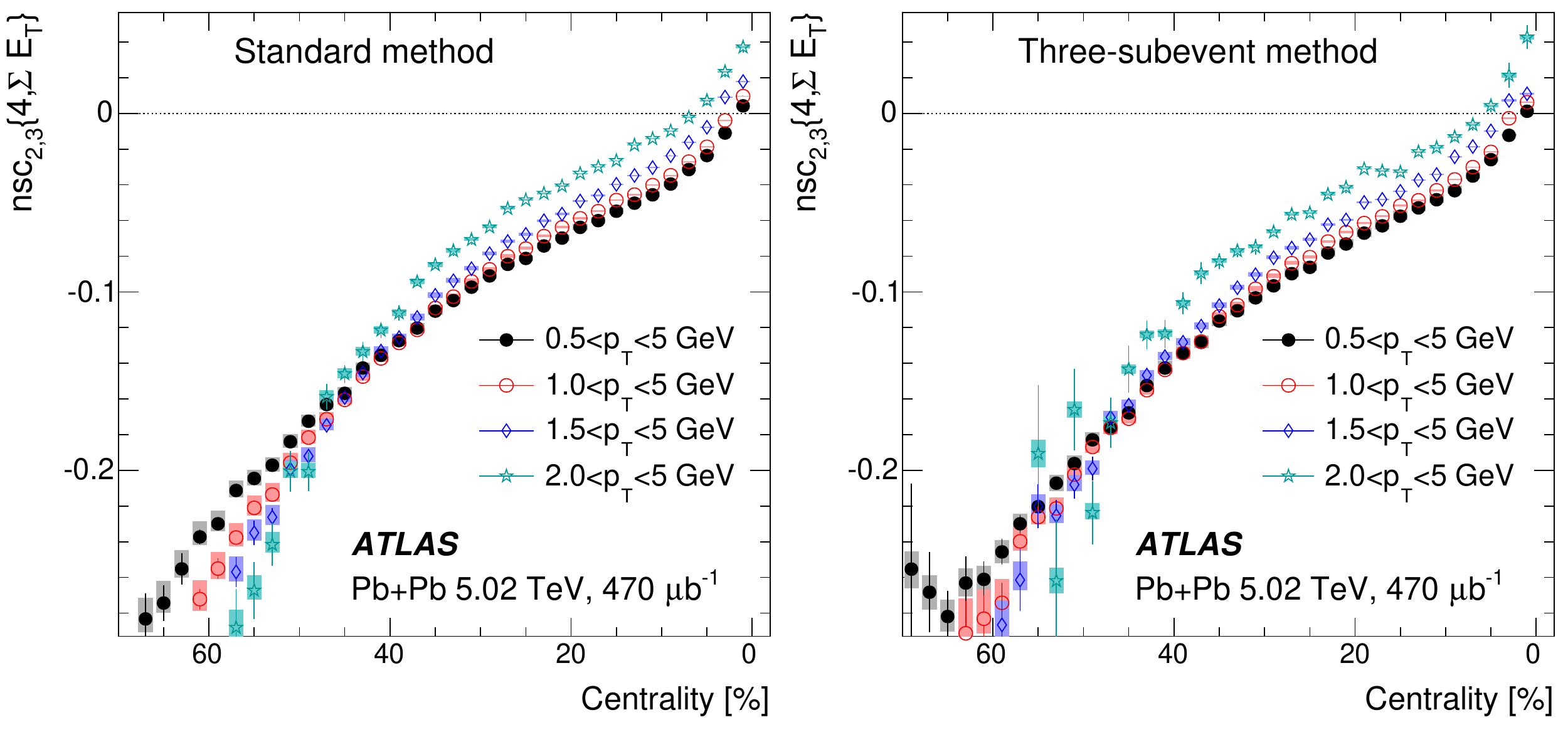}
\end{center}
\caption{\label{fig:b1} The centrality dependence of $\nsca$ calculated for charged particles in four $\pT$ ranges with the standard method (left panel) and three-subevent method (right panel). The error bars and shaded boxes represent the statistical and systematic uncertainties, respectively. Zero is indicated by a dotted line.}
\end{figure}

The correlation between flow harmonics of different order is studied using the four-particle normalized symmetric cumulant nsc$_{2,3}\{4\}$ and nsc$_{2,4}\{4\}$, and the three-particle normalized asymmetric cumulant nac$_{3}\{3\}$. Figure~\ref{fig:b1} shows the centrality dependence of $\mathrm{nsc}_{2,3}\{4\}$ in several $\pT$ ranges which probes the correlation between the $v_2$ and $v_3$. The $\mathrm{nsc}_{2,3}\{4\}$ is negative in most of the centrality range, indicating an anti-correlation between the $v_2$ and $v_3$. This anti-correlation has been observed in previous studies based on the same observable~\cite{ALICE:2016kpq} and using an event-shape engineering technique~\cite{Aad:2015lwa}. The strength of the anti-correlation has significant $\pT$ dependence. For higher-$\pT$ particles, the anti-correlation is stronger in peripheral collisions and weaker in central collisions. In the ultra-central collisions,  $\mathrm{nsc}_{2,3}\{4\}$ changes sign and becomes positive. This positive correlation is related to centrality fluctuations and is discussed further in Section~\ref{sec:c}. The behaviour of the overall centrality and $\pT$ dependence is also found to be similar between the standard cumulant method and the three-subevent cumulant method. This suggests that these features are not caused by non-flow correlations. 

\begin{figure}[h!]
\begin{center}
\includegraphics[width=0.8\linewidth]{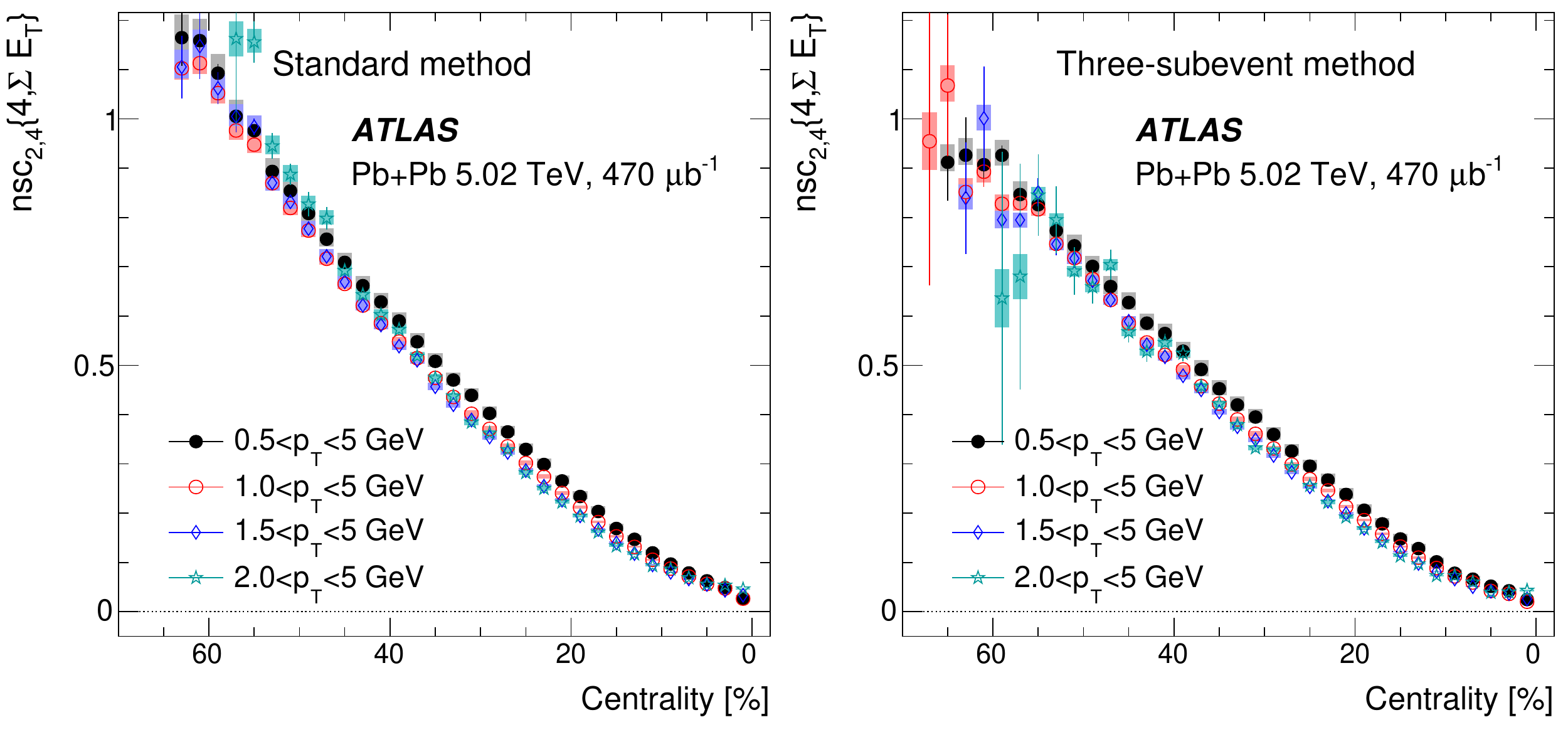}
\end{center}
\caption{\label{fig:b2} The centrality dependence of $\nscb$ calculated for charged particles in four $\pT$ ranges with the standard method (left panel) and three-subevent method (right panel). The error bars and shaded boxes represent the statistical and systematic uncertainties, respectively. Zero is indicated by a dotted line.}
\end{figure}

Figure~\ref{fig:b2} shows the centrality dependence of $\mathrm{nsc}_{2,4}\{4\}$ in several $\pT$ ranges which probes the correlation between $v_2$ and $v_4$. The $\mathrm{nsc}_{2,4}\{4\}$ value is positive over the entire centrality range, indicating a positive correlation between $v_2$ and $v_4$. The signal is very small in central collisions but increases rapidly towards peripheral collisions. The correlations are similar among different $\pT$ ranges in central collisions but are slightly weaker for higher-$\pT$ particles in mid-central collisions. This behaviour is also predicted by hydrodynamic models~\cite{Niemi:2012aj,Zhu:2016puf}. Compared with the three-subevent method, the $\mathrm{nsc}_{2,4}\{4\}$ values from the standard method have better statistical precision but slightly higher values in peripheral collisions, indicating that the non-flow effects may become significant for events beyond 60\% centrality. 

\begin{figure}[h!]
\begin{center}
\includegraphics[width=0.8\linewidth]{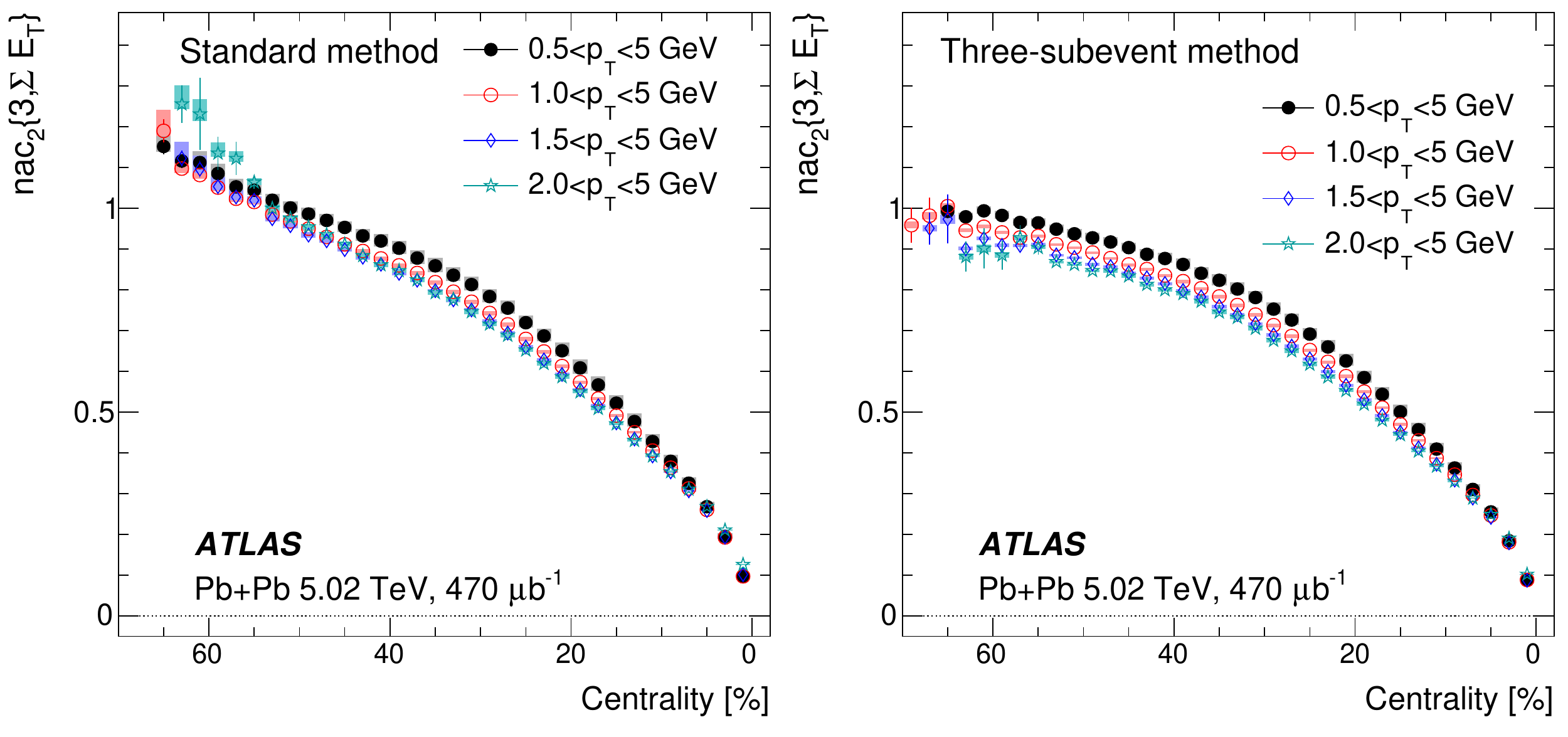}
\end{center}
\caption{\label{fig:b3} The centrality dependence of $\naca$ calculated for charged particles in four $\pT$ ranges with the standard method (left panel) and three-subevent method (right panel). The error bars and shaded boxes represent the statistical and systematic uncertainties, respectively. Zero is indicated by a dotted line.}
\end{figure}

Figure~\ref{fig:b3} shows the centrality dependence of $\mathrm{nac}_{2}\{3\}$ in several $\pT$ ranges which also probes the correlation between $v_2$ and $v_4$. The $\mathrm{nac}_{2}\{3\}$ value is positive over the entire centrality range. The correlation is weak in the central collisions, increases rapidly as the centrality approaches about 20--30\% and then increases slowly toward more peripheral collisions. The correlation patterns for different $\pT$ ranges are similar in central collisions but are slightly weaker for higher-$\pT$ particles in mid-central collisions. Compared with results obtained from the three-subevent method, the results from the standard method are slightly larger in peripheral collisions, indicating that non-flow fluctuations may contribute for events beyond 60\% centrality. The similar $\pT$ and centrality dependences for $\mathrm{nsc}_{2,4}\{4\}$ and $\mathrm{nac}_{2}\{3\}$ are related to the non-linear mode-mixing effects between $v_2$ and $v_4$ described by Eq.~\eqref{eq:12}~\cite{Giacalone:2016afq}.

\subsection{Dependence on reference event class and the role of centrality fluctuations}
\label{sec:c}

This section presents the $\lr{\etfcal}$ or $\lr{\nchrec}$ dependence of various cumulants for the two reference event classes. Section~\ref{sec:ana} describes how the role of centrality fluctuations associated with the reference event class used in the calculation of the cumulants can be understood by extracting the results for each observable in narrow ranges of $\etfcal$ and $\nchrec$. These results are  presented as a function of $\lr{\etfcal}/(\etfcal)_{\mathrm{knee}}$ and $\lr{\nchrec}/(\nchrec)_{\mathrm{knee}}$, where $(\etfcal)_{\mathrm{knee}}=4.1$~$\TeV$ and $(\nchrec)_{\mathrm{knee}}=2800$ are the knee values of the $\etfcal$ and $\nchrec$ distributions shown in Figure~\ref{fig:1}.  It should be noted that $c_n\{2k,\etfcal\}$ (and other observables as well) as a function of $\lr{\etfcal}/(\etfcal)_{\mathrm{knee}}$ contains the same information as the centrality dependence of $c_n\{2k,\etfcal\}$ shown in two previous sections. However, $x$-axes based on $\lr{\etfcal}/(\etfcal)_{\mathrm{knee}}$ and $\lr{\nchrec}/(\nchrec)_{\mathrm{knee}}$  more naturally characterize the size of the overlap region in Pb+Pb collisions and allow a more detailed visualization of the ultra-central region, where the impacts of centrality fluctuations is strongest.

\subsubsection{Two-particle cumulants}
\label{sec:c1}
\begin{figure}[h!]
\begin{center}
\includegraphics[width=1\linewidth]{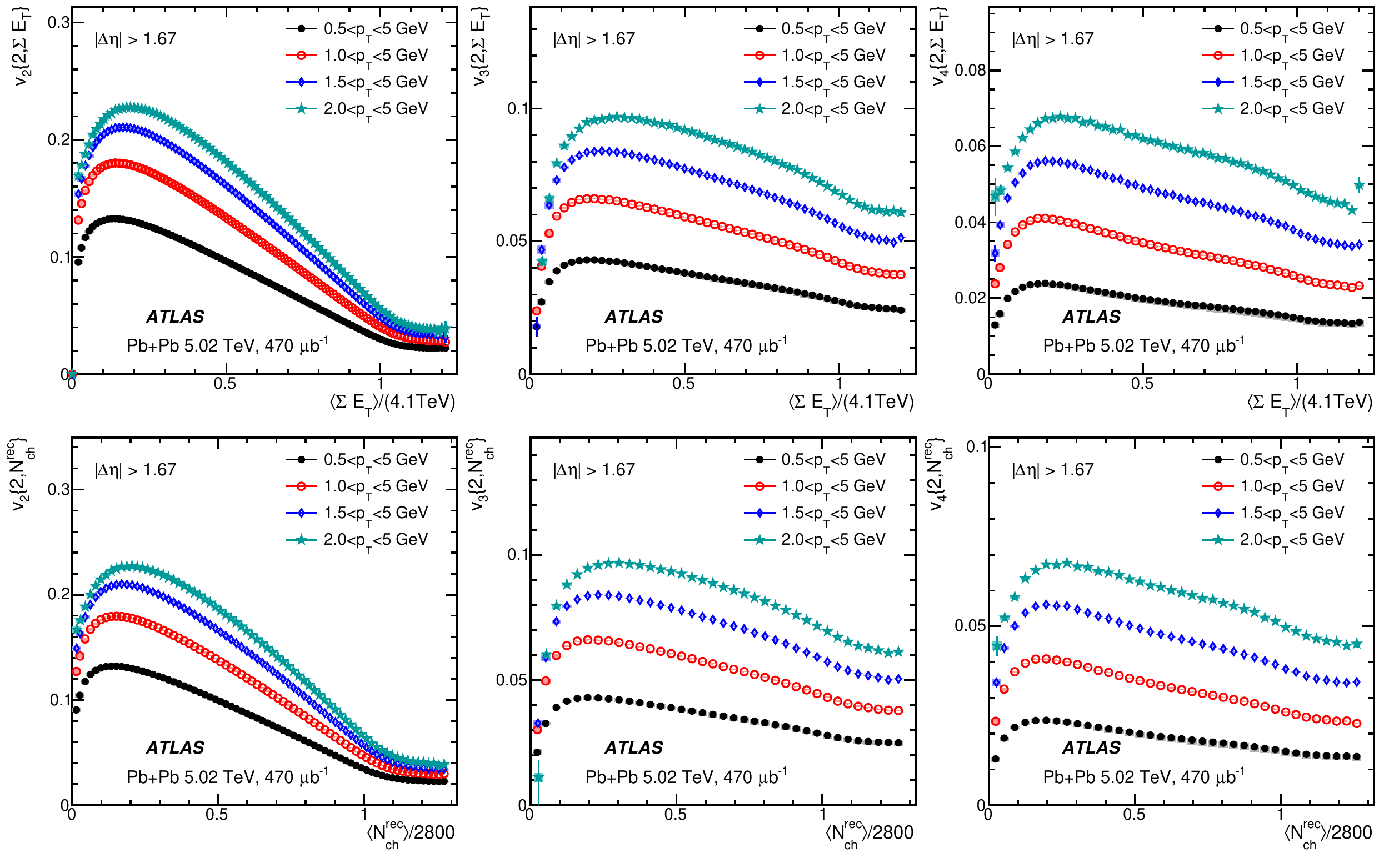}
\end{center}
\caption{\label{fig:c1} The $\lr{\etfcal}$ (top row) and $\lr{\nchrec}$ (bottom row) dependence of $v_2\{2, \etfcal\}$ (left panel), $v_3\{2, \etfcal\}$ (middle panel) and $v_4\{2, \etfcal\}$ (right panel) for four $\pT$ ranges. The error bars and shaded boxes represent the statistical and systematic uncertainties, respectively.}
\end{figure}

The top panels of Figure~\ref{fig:c1} show $v_n\{2, \etfcal\}$ as a function of $\lr{\etfcal}$. The $v_n\{2,\etfcal\}$ values are reflecting the same centrality and $\pT$ dependence behaviour already shown in Figure~\ref{fig:a1}. In ultra-central collisions, the $v_n\{2,\etfcal\}$ values are nearly constant. Similar trends are also observed for $v_n\{2, \nchrec\}$ which are shown in the bottom panels of Figure~\ref{fig:c1} as a function of $\lr{\nchrec}$. These results suggest that the underlying initial geometry, in terms of $\lr{\epsilon_n^2}$, is quite similar between the two reference event classes. 

\begin{figure}[h!]
\begin{center}
\includegraphics[width=1\linewidth]{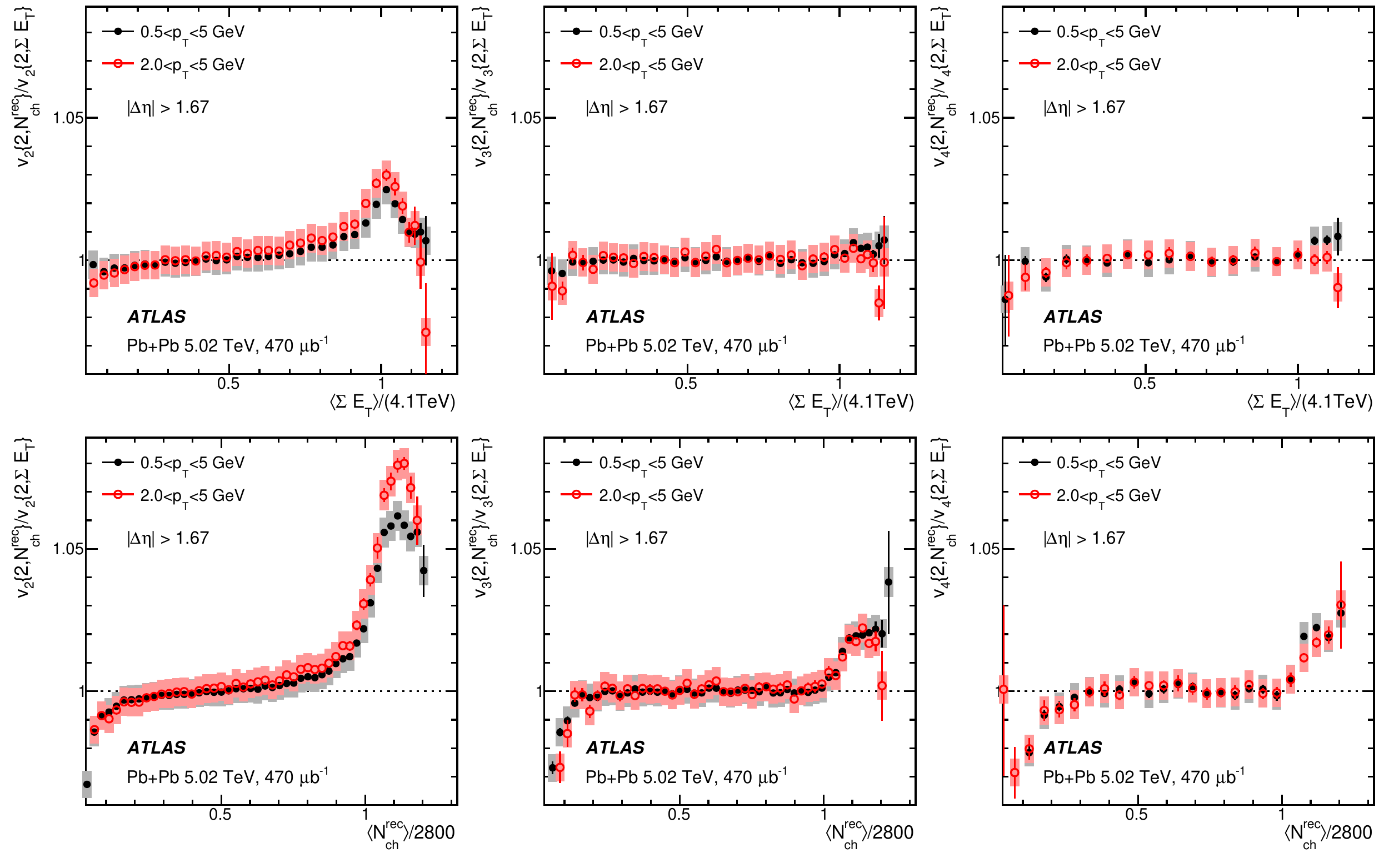}
\end{center}
\caption{\label{fig:c3} The ratios of flow harmonics between the two event-class definitions $v_n\{2, \nchrec\}/v_n\{2, \etfcal\}$ as a function of $\lr{\etfcal}$ (top row) and $\lr{\nchrec}$ (bottom row) for $n=2$ (left panel), $n=3$ (middle panel), and $n=4$ (right panel) for charged particles in two $\pT$ ranges. The error bars and shaded boxes represent the statistical and systematic uncertainties, respectively. Unity is indicated by a dotted line. See text for detailed description.}
\end{figure}

In order to quantify differences between the two reference event classes, $v_n\{2, \nchrec\}$ is mapped to a $\lr{\etfcal}$ dependence and $v_n\{2, \etfcal\}$ is mapped to a $\lr{\nchrec}$ dependence. The ratio $v_n\{2, \nchrec\}/v_n\{2, \etfcal\}$ is then calculated at a given $\lr{\etfcal}$ or at a given $\lr{\nchrec}$. The top row of Figure~\ref{fig:c3} shows $v_n\{2, \nchrec\}/v_n\{2, \etfcal\}$ as a function of $\lr{\etfcal}$. The ratios are very close to unity for $v_3$ and $v_4$ but show a few percent deviation in ultra-central collisions for $v_2$, i.e~$v_2\{2, \nchrec\}> v_2\{2, \etfcal\}$. This result implies that events in a narrow $\nchrec$ range have slightly larger $v_2$ than events in a narrow $\etfcal$, when the two ensembles have the same $\lr{\etfcal}$. This would be the case if the centrality resolution of $\nchrec$ was poorer than the centrality resolution of $\etfcal$. Consequently, $v_2\{2, \nchrec\}$ is expected to contain more events from less central regions, where $v_2$ is larger.

The bottom row of Figure~\ref{fig:c3} shows the same ratio, $v_n\{2, \nchrec\}/v_n\{2, \etfcal\}$, but instead as a function of $\lr{\nchrec}$. Compared with the upper row of Figure~\ref{fig:c3}, the ratio for $v_2$ shows a larger deviation from unity which reaches 7\% in ultra-central collisions. Smaller, but significant differences are also observed for $v_3$ and $v_4$ in ultra-central collisions. This is probably because $v_n\{2, \nchrec\}$ has even more contributions from less central events than $v_n\{2, \etfcal\}$ when both are matched to the same $\lr{\nchrec}$ instead of the same $\lr{\etfcal}$. This is consistent with the hypothesis in which $\nchrec$ has poorer centrality resolution and therefore larger centrality fluctuations than $\etfcal$, when mapped to the same average event activity in the final state. 

Due to the steep decrease of the  $\etfcal$ and $\nchrec$ distributions in the ultra-central region, the centrality fluctuations and the shapes of the  $p(\epsilon_n)$ and $p(v_n)$ distributions are expected to exhibit a significant departure from a Gaussian shape~\cite{Skokov:2012ds,Zhou:2018fxx}. The flow cumulants with four or more particles are more sensitive to a non-Gaussian shape of $p(v_n)$ than the two-particle cumulants. Therefore, they are expected to exhibit larger differences between the two reference event classes. This is the topic of the next section.

\subsubsection{Multi-particle cumulants}
\label{sec:c2}
\begin{figure}[h!]
\begin{center}
\includegraphics[width=1\linewidth]{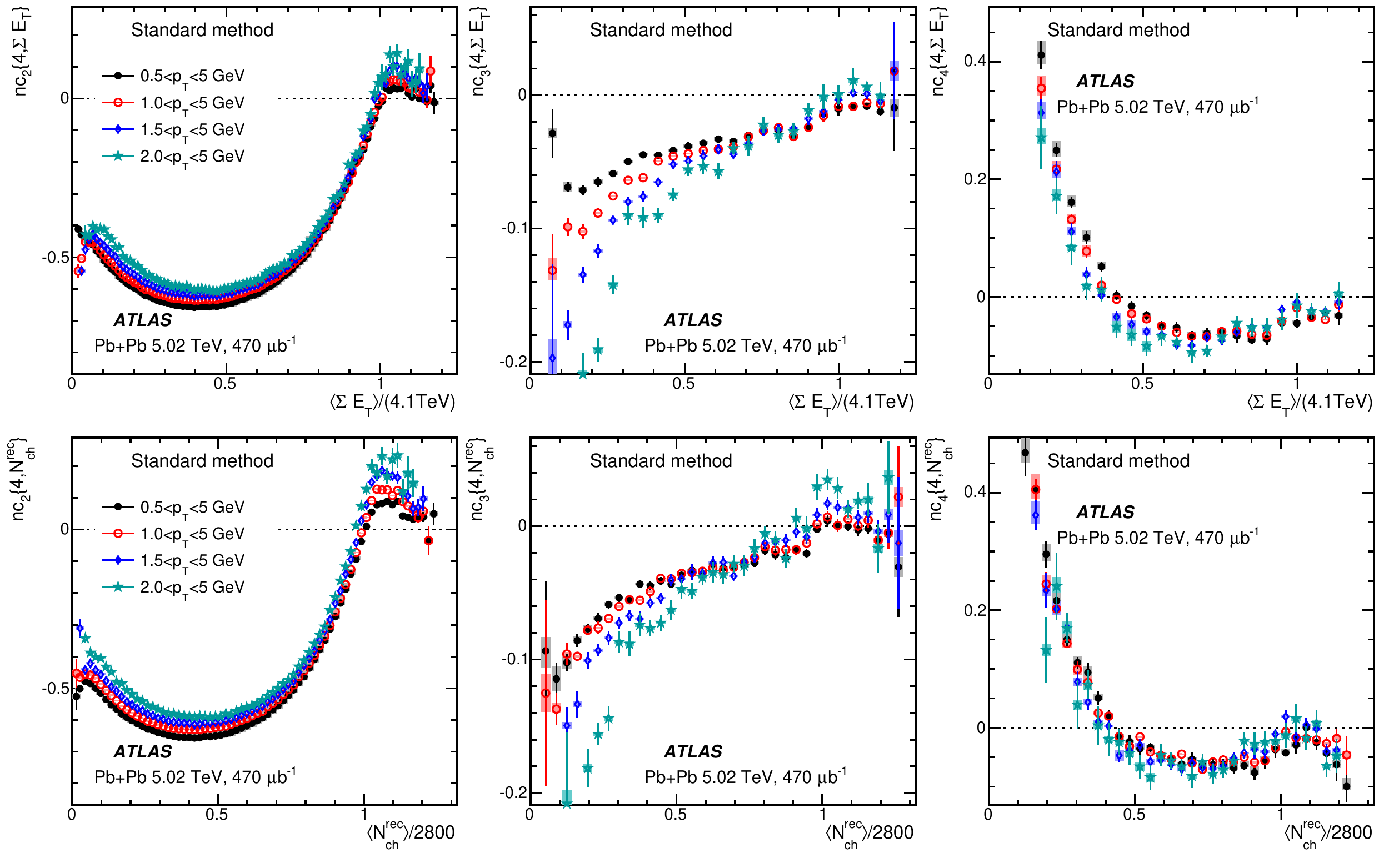}
\end{center}
\caption{\label{fig:c5} The normalized four-particle cumulants nc$_n\{4, \etfcal\}$ as a function of $\lr{\etfcal}$ (top row) and nc$_n\{4, \nchrec\}$ as a function of $\lr{\nchrec}$ (bottom row) for $n=2$ (left panel), $n=3$ (middle panel), and $n=4$ (right panel) for four $\pT$ ranges. The error bars and shaded boxes represent the statistical and systematic uncertainties, respectively. Zero is indicated by a dotted line.}
\end{figure}

The top panels of Figure~\ref{fig:c5} show  nc$_n\{4, \etfcal\}$ as a function of $\lr{\etfcal}$. This figure contains the same information as the results shown in Figure~\ref{fig:a2}, except for a change in the scale of the $x$-axis which shows the central region in more detail. The nc$_2\{4, \etfcal\}$ value changes sign for $\lr{\etfcal}\gtrsim (\etfcal)_{\mathrm{knee}}$, where it first increases, reaches a maximum and then decreases to close to zero. The value of the maximum also increases with the $\pT$ of the particles. The nc$_3\{4, \etfcal\}$ value is negative and approaches zero in ultra-central collisions and only changes sign for the highest $\pT$ range used in this analysis. The $\mathrm{nc}_4\{4, \etfcal\}$ value changes from positive in peripheral collisions to negative in mid-central collisions, reaches a minimum and then turns back and approaches zero in the ultra-central collisions.

\begin{figure}[h!]
\begin{center}
\includegraphics[width=1\linewidth]{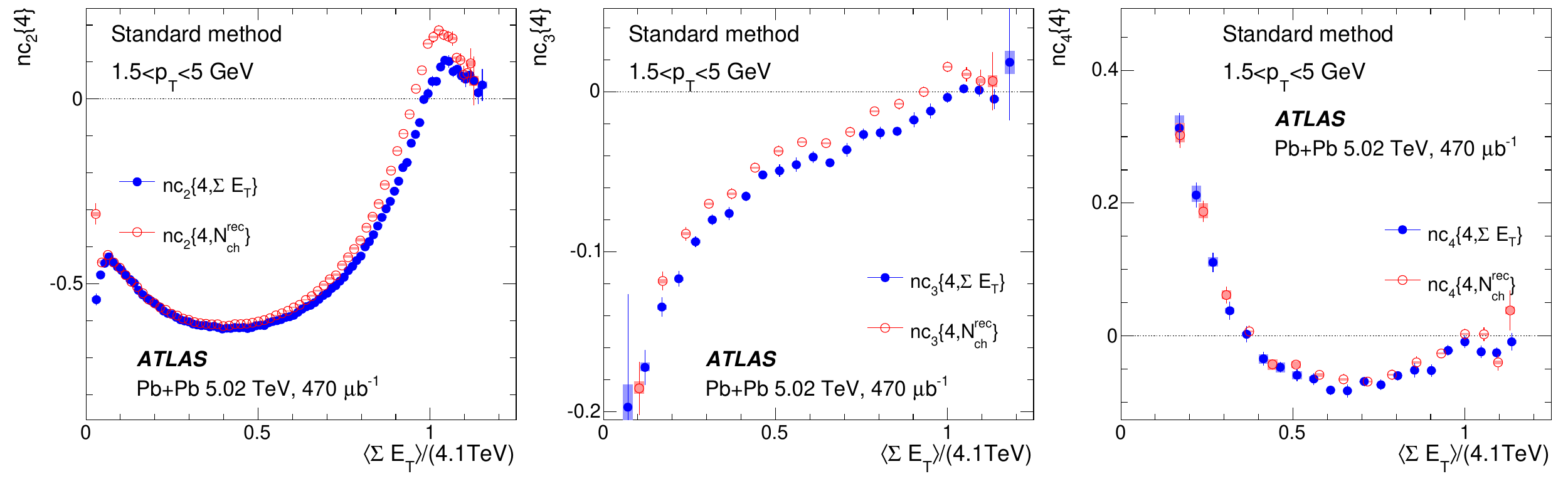}
\end{center}
\caption{\label{fig:c6} The comparison of normalized four-particle cumulants nc$_n\{4, \etfcal\}$ and nc$_n\{4, \nchrec\}$ as a function of $\lr{\etfcal}$ for $n=2$ (left panel), $n=3$ (middle panel), and $n=4$ (right panel) for charged particles $1.5<\pT<5$~$\GeV$. The error bars and shaded boxes represent the statistical and systematic uncertainties, respectively. Zero is indicated by a dotted line.}
\end{figure}

The bottom panels of Figure~\ref{fig:c5} show nc$_n\{4, \nchrec\}$ as a function of $\lr{\nchrec}$. The overall $\lr{\nchrec}$ and $\pT$-dependent trends are similar to those in the top panels. However, the maximum of nc$_2\{4, \nchrec\}$ is more than a factor of two larger, and nc$_3\{4, \nchrec\}$ shows a clear sign change for the two highest $\pT$ ranges used in this analysis. Furthermore, nc$_4\{4, \nchrec\}$ shows a local maximum in ultra-central collisions, a feature absent for nc$_4\{4, \etfcal\}$. 

If ${\bm V}_n\propto {\mathcal{E}}_n$ is valid, then the shape of $p(v_n)$ should be the same as the shape of $p(\epsilon_n)$ and \mbox{nc$_n\{4\}=\mathrm{nc}_{n}\{4,\epsilon\}$}~\cite{Giacalone:2017uqx,Zhou:2018fxx}. The $c_{n}\{4,\epsilon\}$ values can be estimated from a simple Glauber model framework using participating nucleons in the overlap region. The $c_{n}\{4,\epsilon\}$ value is found to be always negative when the reference event class is defined using the number of participating nucleons $\npart$ or the impact parameter of the collisions~\cite{Alver:2008zza}. However, a positive nc$_{n}\{4,\epsilon\}$ is observed in ultra-central collisions when the reference event class is defined using the final-state particle multiplicity~\cite{Agakishiev:2011eq,Zhou:2018fxx}. Due to multiplicity smearing, events with the same final-state multiplicity can have different $\npart$, and therefore different $\epsilon_n$. The positive nc$_{n}\{4,\epsilon\}$ reflects the non-Gaussian shape of $p(\epsilon_n)$ due to the smearing in $\npart$ for events with the same final-state multiplicity. The larger values of nc$_n\{4,\nchrec\}$ in comparison with nc$_n\{4,\etfcal\}$ in ultra-central collisions could be due to stronger multiplicity smearing for nc$_n\{4,\nchrec\}$. 
Figure~\ref{fig:c6} compares  nc$_n\{4, \etfcal\}$ and nc$_n\{4, \nchrec\}$ as a function of $\lr{\etfcal}$ obtained for $1.5<\pT<5$~$\GeV$. In both cases, the normalized cumulants for $v_2$ and $v_3$ show significant differences between the two reference event classes, while the difference is smaller for $v_4$. The values of nc$_n\{4, \nchrec\}$ for $n=2$ and 3 are significantly larger than those for nc$_n\{4, \etfcal\}$ over a broad centrality range, not only limited to the ultra-central collisions. This implies that the influence of centrality fluctuations on flow fluctuations is potentially important even in mid-central collisions. 

\begin{figure}[h!]
\begin{center}
\includegraphics[width=1\linewidth]{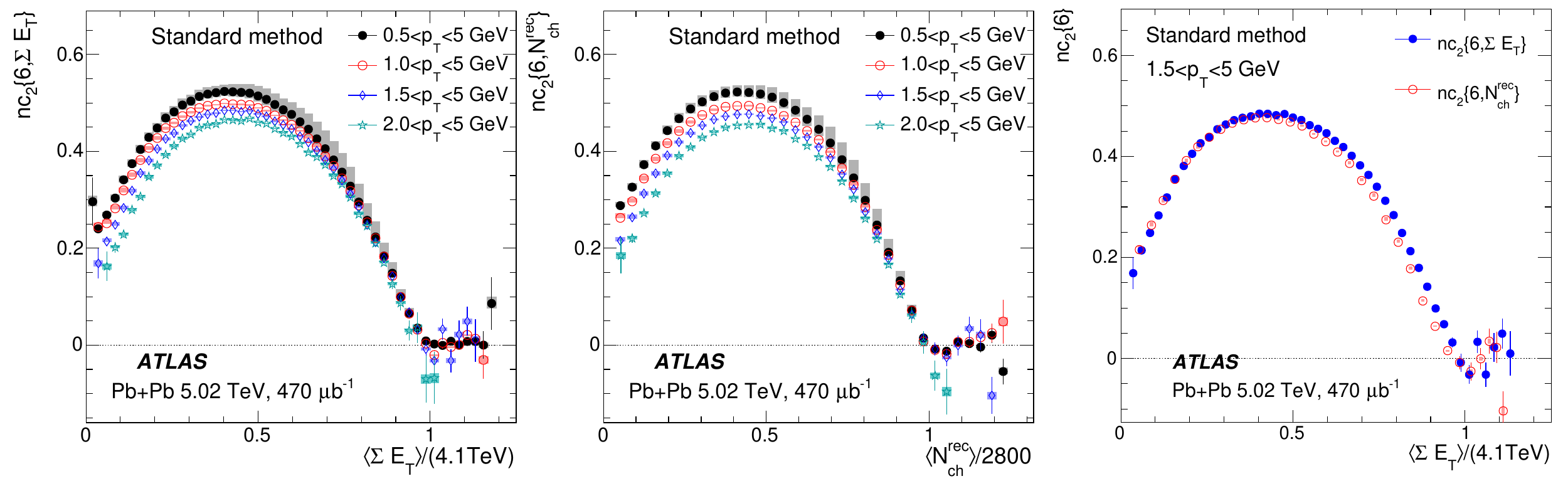}
\end{center}
\caption{\label{fig:c7} The normalized six-particle cumulants nc$_2\{6, \etfcal\}$ as a function of $\lr{\etfcal}$ (left panel) and nc$_2\{6, \nchrec\}$ as a function of $\lr{\nchrec}$ (middle panel) in four $\pT$ ranges. The nc$_2\{6, \etfcal\}$ and nc$_6\{4, \nchrec\}$ results for $1.5<\pT<5$~$\GeV$ are also compared directly as a function of $\lr{\etfcal}$ (right panel). The error bars and shaded boxes represent the statistical and systematic uncertainties, respectively. Zero is indicated by a dotted line.}
\end{figure}

The left two panels of Figure~\ref{fig:c7} show the six-particle normalized cumulants for $v_2$ obtained using the two reference event classes, nc$_2\{6, \etfcal\}$ and nc$_2\{6, \nchrec\}$, respectively. The nc$_2\{6\}$ values are positive in most of the centrality range but decrease to zero at around $\lr{\etfcal}=(\etfcal)_{\mathrm{knee}}$ or $\lr{\nchrec}=(\nchrec)_{\mathrm{knee}}$ and stay close to zero above that. The right panel of Figure~\ref{fig:c7} compares nc$_2\{6, \etfcal\}$ and nc$_2\{6, \nchrec\}$ as a function of $\lr{\etfcal}$. The values of nc$_2\{6, \nchrec\}$ are smaller than those for nc$_2\{6, \etfcal\}$ in central and mid-central collisions, suggesting that the centrality fluctuations influence the multi-particle cumulants of $p(v_2)$ over a broad centrality range. 

\begin{figure}[h!]
\begin{center}
\includegraphics[width=1\linewidth]{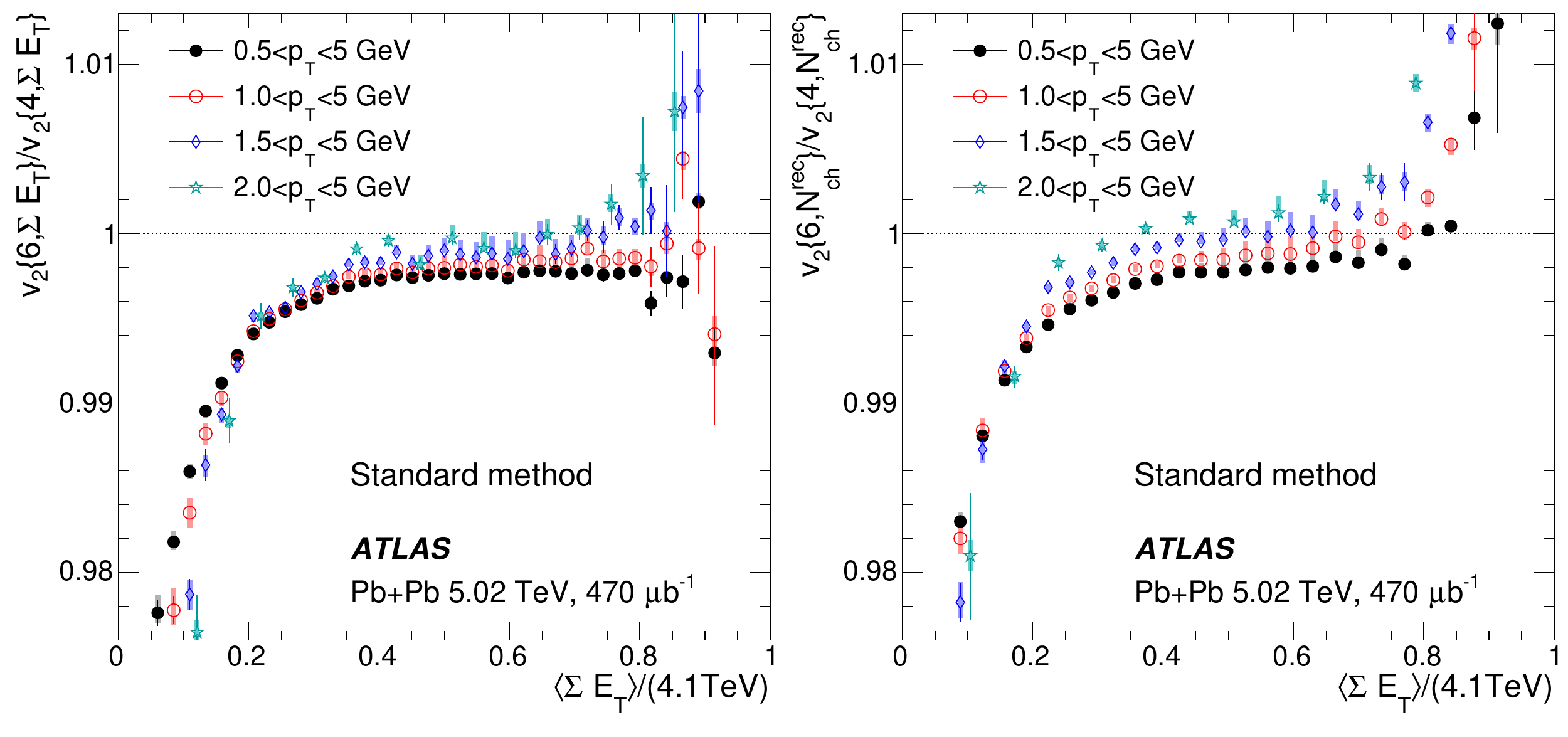}
\end{center}
\caption{\label{fig:c8} The $\lr{\etfcal}$ dependence of cumulant ratio $v_2\{6,\etfcal\}/v_2\{4,\etfcal\}$ (left panel) and $v_2\{6,\nchrec\}/v_2\{4,\nchrec\}$ (right panel) for charged particles in four $\pT$ ranges. The error bars and shaded boxes represent the statistical and systematic uncertainties, respectively. Unity is indicated by a dotted line.}
\end{figure}

The left panel in Figure~\ref{fig:c8} shows the cumulant ratio $v_2\{6\}/v_2\{4\}$ obtained using the event class based on $\etfcal$. This panel contains the same information as shown in Figure~\ref{fig:a5} except for a change in the scale of the $x$-axis made in order to show more detail in the central region. The data show significant differences between the four $\pT$ ranges. The value of $v_2\{6\}/v_2\{4\}$ is larger for higher $\pT$ and even exceeds one in ultra-central collisions. This behaviour is expected, as  $c_2\{4\}$ and therefore $v_2\{4\}$, changes sign in ultra-central collisions. The right panel of Figure~\ref{fig:c8} shows $v_2\{6\}/v_2\{4\}$ obtained using the event class based on $\nchrec$ but then mapped onto $\lr{\etfcal}$. The differences between the results for the various $\pT$ ranges are larger for most of the centrality range, which again implies that the centrality fluctuations influence the ratios between multi-particle cumulants over a broad centrality range. 

\subsubsection{Multi-particle mixed-harmonic cumulants}
\label{sec:c3}
\begin{figure}[h!]
\begin{center}
\includegraphics[width=1\linewidth]{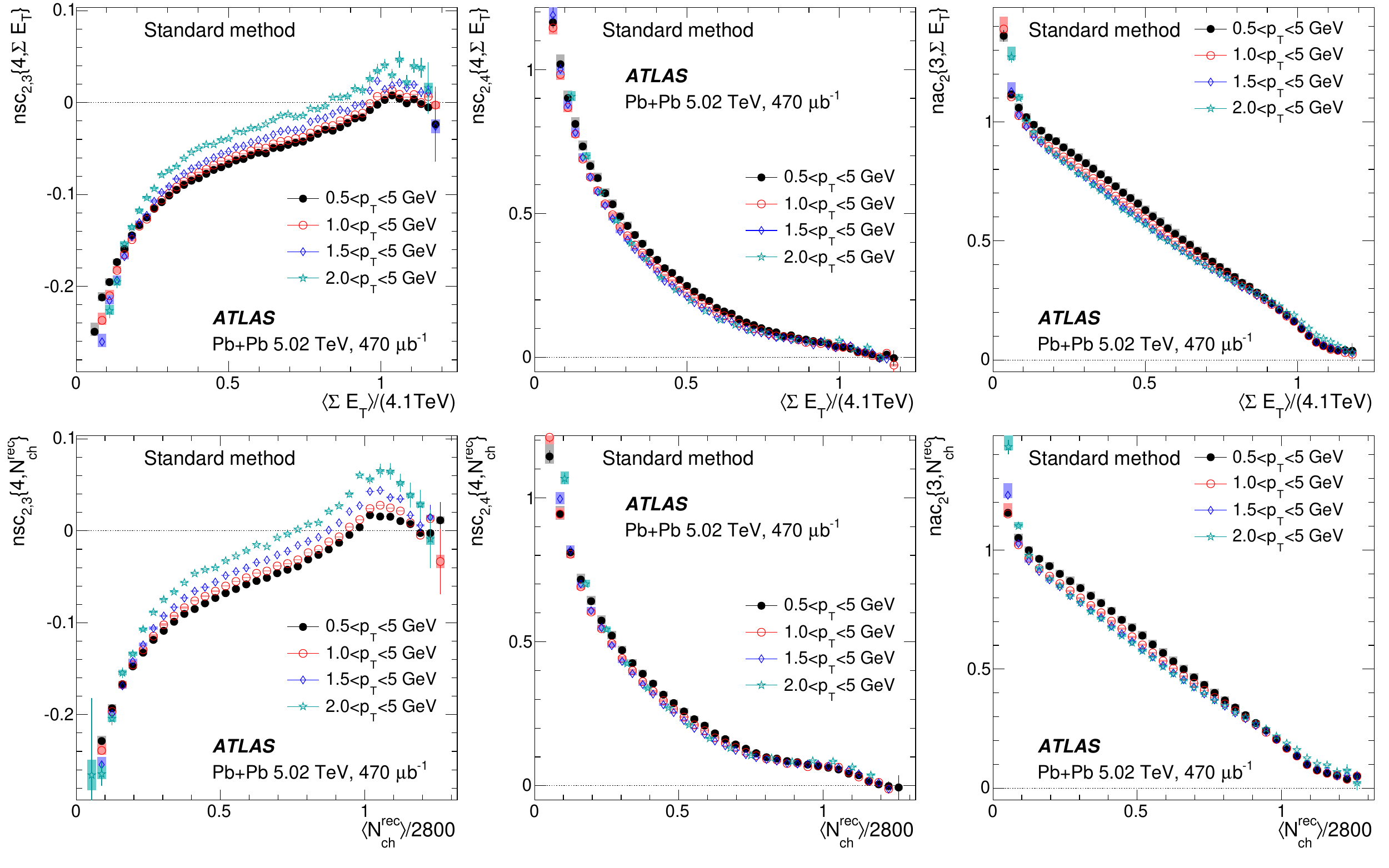}
\end{center}
\caption{\label{fig:c9}  The top row shows the $\lr{\etfcal}$ dependence of normalized cumulants nsc$_{2,3}\{4,\etfcal\}$ (left panel), nsc$_{2,4}\{4,\etfcal\}$ (middle panel) and nac$_{2}\{3,\etfcal\}$ (right panel) for four $\pT$ ranges. The bottom row shows the $\lr{\nchrec}$ dependence of normalized cumulants nsc$_{2,3}\{4,\nchrec\}$ (left panel), nsc$_{2,4}\{4,\nchrec\}$ (middle panel) and nac$_{2}\{3,\nchrec\}$ (right panel) for four $\pT$ ranges. The error bars and shaded boxes represent the statistical and systematic uncertainties, respectively. Zero is indicated by a dotted line.}
\end{figure}
\begin{figure}[h!]
\begin{center}
\includegraphics[width=1\linewidth]{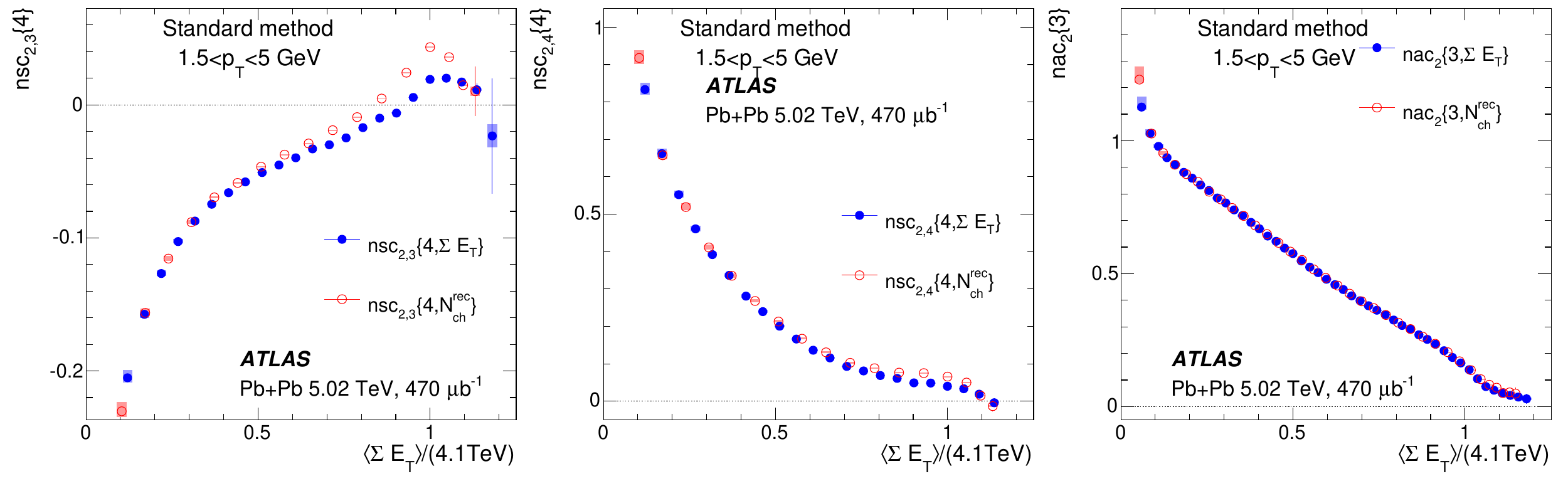}
\end{center}
\caption{\label{fig:c10} Comparison of nsc$_{2,3}\{4,\etfcal\}$ and nsc$_{2,3}\{4,\nchrec\}$ (left panels), nsc$_{2,4}\{4,\etfcal\}$ and nsc$_{2,4}\{4,\nchrec\}$ (middle panels), and nac$_{2}\{3,\etfcal\}$ and nac$_{2}\{3,\nchrec\}$ (right panels) as a function of $\lr{\etfcal}$. The error bars and shaded boxes represent the statistical and systematic uncertainties, respectively. Zero is indicated by a dotted line.}
\end{figure}
The sensitivity to the choice of reference event class is also studied for the symmetric cumulants $\nsca$ and $\nscb$ and the asymmetric cumulant $\naca$. The results obtained with the event class based on $\etfcal$ are shown in the top row of Figure~\ref{fig:c9} as a function of $\lr{\etfcal}$. The $\mathrm{nsc}_{2,3}\{4,\etfcal\}$ values change sign and become positive in ultra-central collisions, and are larger for higher $\pT$. At the largest $\lr{\etfcal}$ values,  $\mathrm{nsc}_{2,4}\{4,\etfcal\}$ reaches zero or even becomes slightly negative while $\mathrm{nac}_{2}\{3,\etfcal\}$ reaches a value of around 0.05. The bottom three panels of Figure~\ref{fig:c9} show the results obtained with the event class based on $\nchrec$ for $\mathrm{nsc}_{2,3}\{4,\nchrec\}$, $\mathrm{nsc}_{2,4}\{4,\nchrec\}$ and $\mathrm{nac}_{2}\{3,\nchrec\}$, respectively. The positive $\mathrm{nsc}_{2,3}\{4,\nchrec\}$ values in the ultra-central region are larger than those for $\mathrm{nsc}_{2,3}\{4,\etfcal\}$. The trends of the other two cumulants are similar to those obtained with the event class based on $\etfcal$.

The direct comparison of $\nsca$, $\nscb$ and $\naca$ obtained with the two reference event classes is shown in Figure~\ref{fig:c10}  as a function of $\lr{\etfcal}$ for particles with $1.5<\pT<5$~$\GeV$. The values of $\mathrm{nsc}_{2,3}\{4,\nchrec\}$ are larger than those for $\mathrm{nsc}_{2,3}\{4,\etfcal\}$ in central and mid-central collisions. However, the values of the other two cumulants are similar between the two reference event classes.

\section{Summary}
\label{sec:sum}
Measurements of multi-particle cumulants for harmonic flow coefficients $v_n$ are presented using 470 $\mu$b$^{-1}$ of Pb+Pb collisions at $\sqrt{s_{\mathrm{NN}}}=5.02$ $\TeV$ with the ATLAS detector at the LHC. The cumulants are designed to provide information about the event-by-event fluctuations of one harmonic, $p(v_n)$, and two different harmonics, $p(v_n,v_m)$. The $p(v_n)$ distribution is studied using $2k$-particle cumulants $c_n\{2k\}$ and normalized cumulants nc$_n\{2k\}$, which provide an estimate of the flow coefficients $v_n\{2k\}$ and cumulant ratios $v_n\{4\}/v_n\{2\}$ and $v_n\{6\}/v_n\{4\}$. The $p(v_n,v_m)$ distribution is studied using the so-called normalized symmetric cumulant $\nscn$ and asymmetric cumulant $\naca$. These normalized cumulants are directly sensitive to  fluctuations of the collision geometry in the initial state. In order to investigate the influence of centrality fluctuations on the flow fluctuations, the cumulants are calculated using events selected with two different reference event-class definitions.

A first observation  of a negative $c_1\{4\}$, and therefore a positive $v_1\{4\}$ is presented, which sheds light on the nature of the dipolar-eccentricity fluctuation in the initial-state geometry. The values of $c_4\{4\}$ are found to be negative in central collisions but change sign around a centrality of 20--25\% and increase quickly for more peripheral collisions. This behaviour is consistent with a non-linear contribution to $v_4$ that is proportional to $v_2^2$. This non-linear contribution increases for more peripheral collisions and makes a positive contribution to $c_4\{4\}$. Over most of the centrality range the $c_2\{4\}$ and $c_3\{4\}$ values are found to be negative but change sign towards the most central collisions, suggesting that the $p(v_2)$ and $p(v_3)$ distributions deviate significantly from a Gaussian shape. The cumulant ratios, $v_2\{4\}/v_2\{2\}$, $v_3\{4\}/v_3\{2\}$, $v_4\{4\}/v_4\{2\}$ and $v_2\{6\}/v_2\{4\}$ exhibit a small but significant $\pT$ dependence, suggesting flow fluctuations may also arise directly in the momentum space through the initial-state correlations or final-state interactions.

This paper also presents a detailed measurement of the four-particle symmetric cumulants $\nsca$ and $\nscb$ and the three-particle asymmetric cumulant $\naca$. The symmetric cumulants probe the correlation between the magnitudes of two flow harmonics, while the asymmetric cumulant is sensitive to correlations involving both the magnitude and phase of flow. Over most of the centrality range, $\nsca$ is found to be negative, reflecting an anti-correlation between $v_2$ and $v_3$. The $\nscb$ and $\naca$ values are found to be positive, and their dependence on centrality is consistent with non-linear mode-mixing effects between $v_2$ and $v_4$. 

In experimental measurements, the flow cumulants are always calculated for events with similar activity. However, for a given activity measure, fluctuations in the particle production process lead to irreducible centrality fluctuations, also known as volume fluctuations. Since $v_n$ changes with centrality, centrality fluctuations lead to an additional fluctuation of $v_n$, and consequently a change in the flow cumulants. In order to study the influence of centrality fluctuations, cumulant observables are calculated for  two reference event classes with different centrality resolution: the total transverse energy in $3.2<|\eta|<4.9$, and number of reconstructed charged particles with $|\eta|<2.5$ and $0.5<\pT<5$~$\GeV$. In ultra-central collisions, the cumulants nc$_2\{4\}$, nc$_3\{4\}$, and $\nsca$ are observed to change sign, indicating a significant influence of centrality fluctuations on the multi-particle cumulants of $p(v_2)$, $p(v_3)$ and $p(v_2,v_3)$. The sign change patterns are more pronounced for the event class based on $\lr{\nchrec}$, consistent with larger centrality fluctuations. The differences between the two event classes are found to extend, with decreasing magnitude, to mid-central collisions, which may suggest that the centrality fluctuations influence the flow fluctuations over a broad centrality range. The sign-change patterns are found to be more pronounced at higher $\pT$, which may indicate that the flow fluctuations have significant $\pT$ dependence. Such $\pT$ dependence cannot be explained by considering only fluctuations in the initial geometry. 

These results provide comprehensive information about the nature of flow fluctuations and the contributions coming from both the initial state and the final state. They also shed light on the influence of centrality fluctuations on flow fluctuations, especially in the ultra-central collisions, which can help to clarify the meaning of centrality and provide insights into the sources of particle production in heavy-ion collisions.
\appendix
\part*{Appendix}
\addcontentsline{toc}{part}{Appendix}

\section{Flow harmonics $v_n\{2k\}$ from $2k$-particle correlations}
\label{sec:app1}
\begin{figure}[h!]
\begin{center}
\includegraphics[width=1\linewidth]{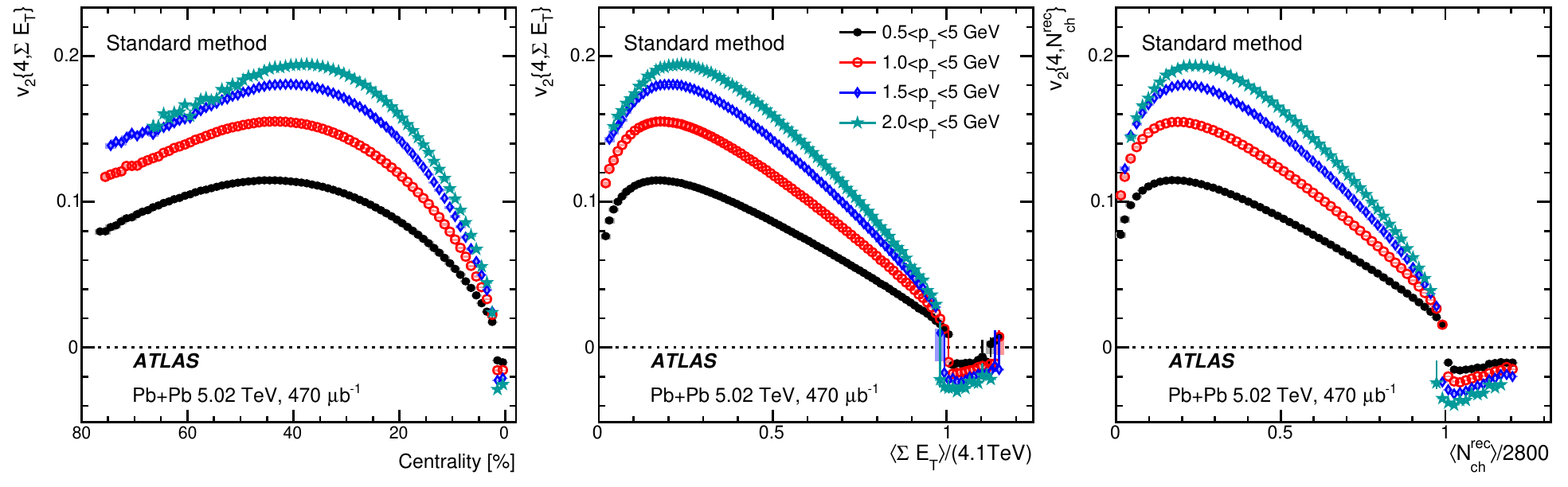}
\end{center}
\caption{\label{fig:app1_1} The $v_2\{4\}$ values calculated for charged particles in four $\pT$ ranges as a function of centrality (left panel), $\etfcal$ (middle panel), and $\nchrec$ (right panel). The error bars and shaded boxes represent the statistical and systematic uncertainties, respectively. Zero is indicated by a dotted line.}
\end{figure}

\begin{figure}[h!]
\begin{center}
\includegraphics[width=1\linewidth]{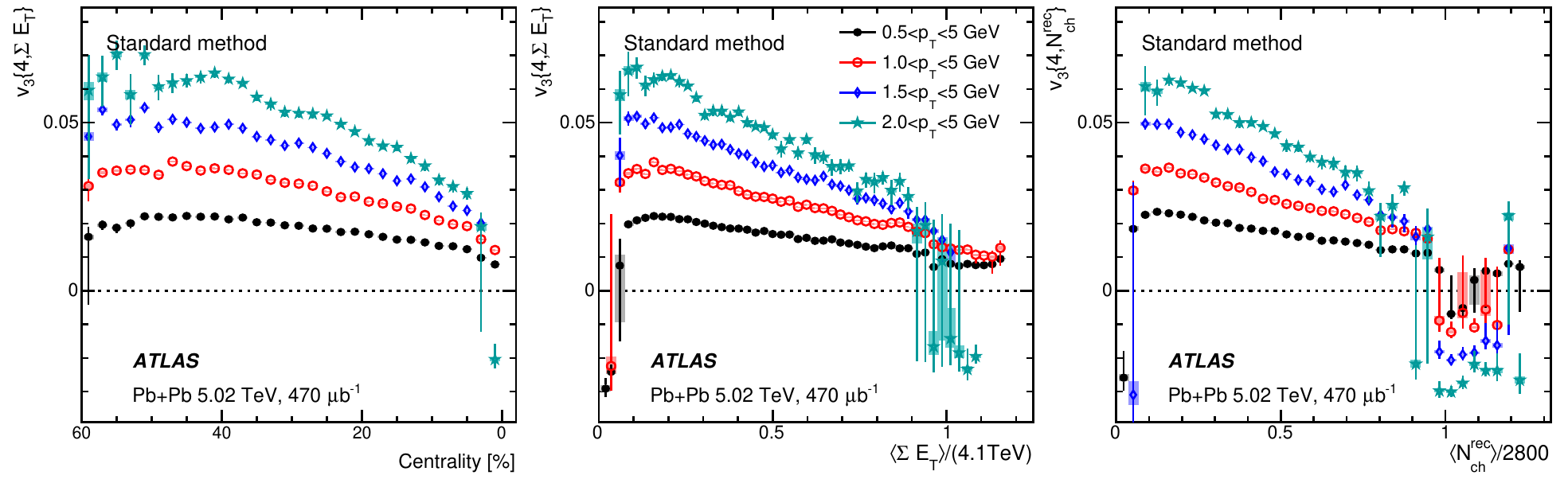}
\end{center}
\caption{\label{fig:app1_2} The $v_3\{4\}$ values calculated for charged particles in four $\pT$ ranges as a function of centrality (left panel), $\etfcal$ (middle panel), and $\nchrec$ (right panel). The error bars and shaded boxes represent the statistical and systematic uncertainties, respectively. Zero is indicated by a dotted line.}
\end{figure}

\begin{figure}[h!]
\begin{center}
\includegraphics[width=1\linewidth]{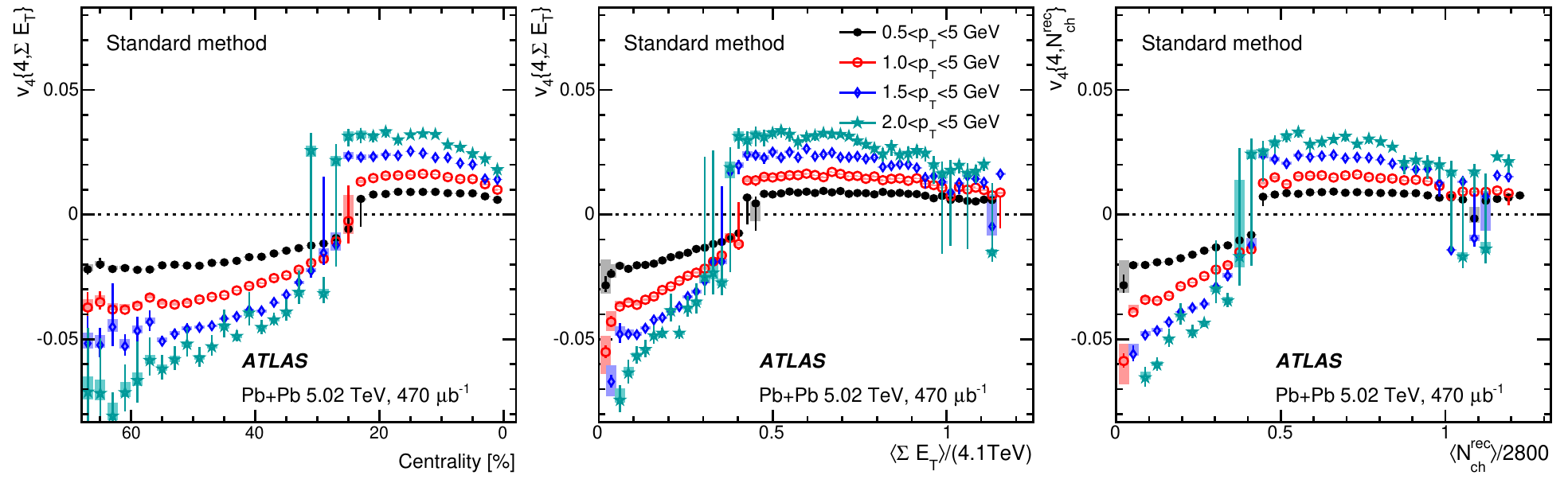}
\end{center}
\caption{\label{fig:app1_3} The $v_4\{4\}$ values calculated for charged particles in four $\pT$ ranges as a function of centrality (left panel), $\etfcal$ (middle panel), and $\nchrec$ (right panel). The error bars and shaded boxes represent the statistical and systematic uncertainties, respectively. Zero is indicated by a dotted line.}
\end{figure}

\begin{figure}[h!]
\begin{center}
\includegraphics[width=1\linewidth]{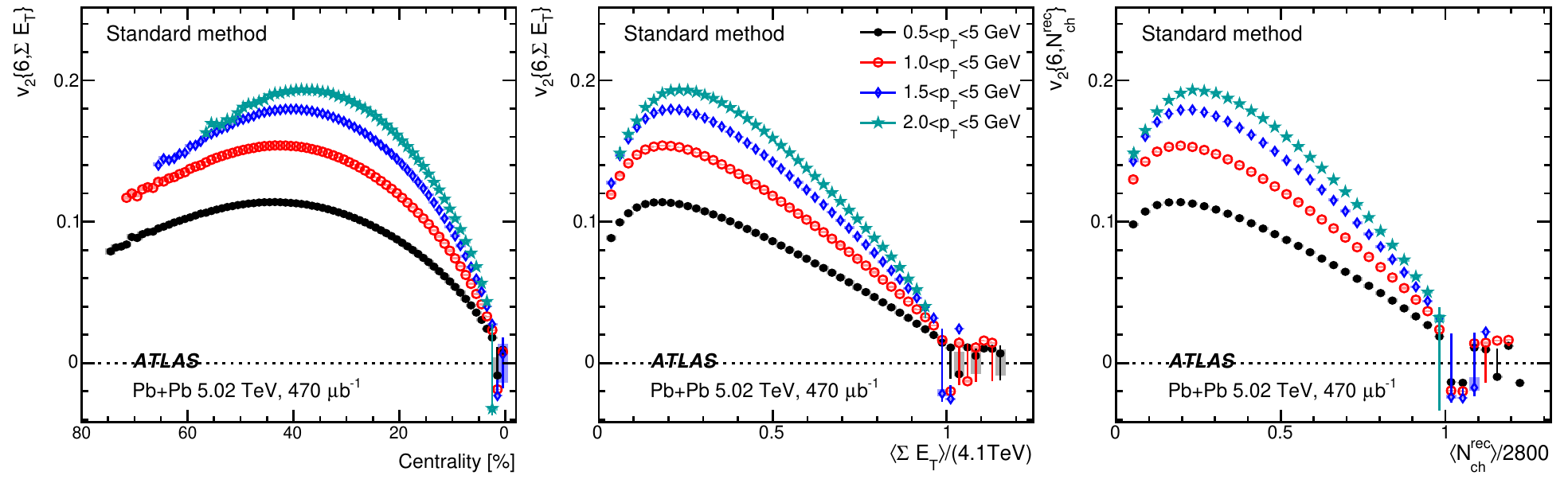}
\end{center}
\caption{\label{fig:app1_4} The $v_2\{6\}$ values calculated for charged particles in four $\pT$ ranges as a function of centrality (left panel), $\etfcal$ (middle panel), and $\nchrec$ (right panel). The error bars and shaded boxes represent the statistical and systematic uncertainties, respectively.  Zero is indicated by a dotted line.}
\end{figure}

Figures~\ref{fig:app1_1}--\ref{fig:app1_3} show the flow coefficients from the four-particle cumulants $v_2\{4\}$, $v_3\{4\}$ and $v_4\{4\}$, respectively.  Figure~\ref{fig:app1_4} shows the elliptic flow coefficient from the six-particle cumulant, $v_2\{6\}$.  They are all obtained from Eq.~\eqref{eq:4} and are shown as a function of centrality, $\etfcal$ and $\nchrec$. The apparent discontinuities correspond to the locations where the corresponding $c_n\{2k\}$ changes sign.

\clearpage
\section{Comparison between standard method and three-subevent method}
\label{sec:app2}
\begin{figure}[h!]
\begin{center}
\includegraphics[width=0.9\linewidth]{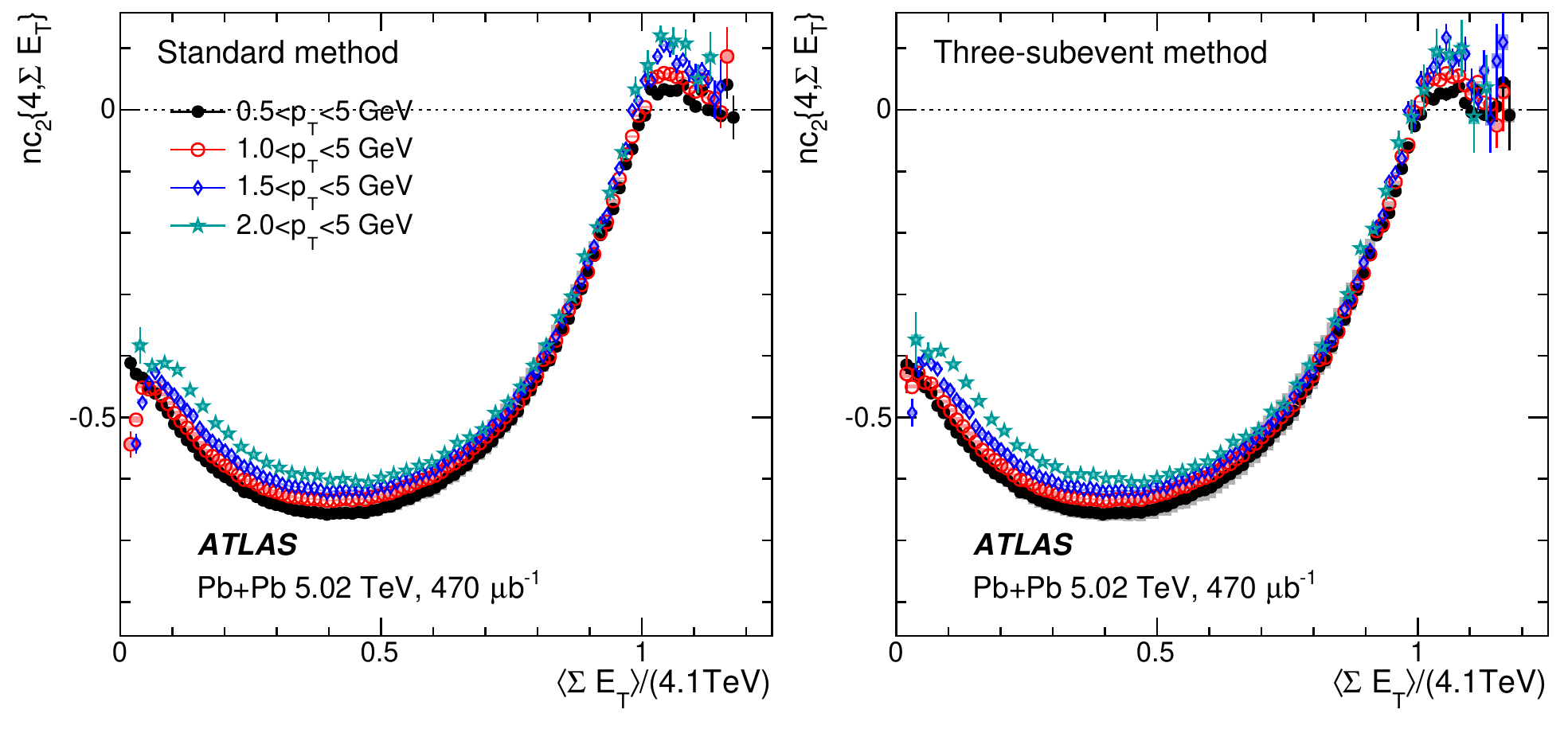}
\end{center}
\caption{\label{fig:app2_1} The nc$_2\{4\}$ values calculated for charged particles in several $\pT$ ranges with the standard cumulant method (left) and three-subevent method (right). The error bars and shaded boxes represent the statistical and systematic uncertainties, respectively. Zero is indicated by a dotted line.}
\end{figure}

\begin{figure}[h!]
\begin{center}
\includegraphics[width=0.9\linewidth]{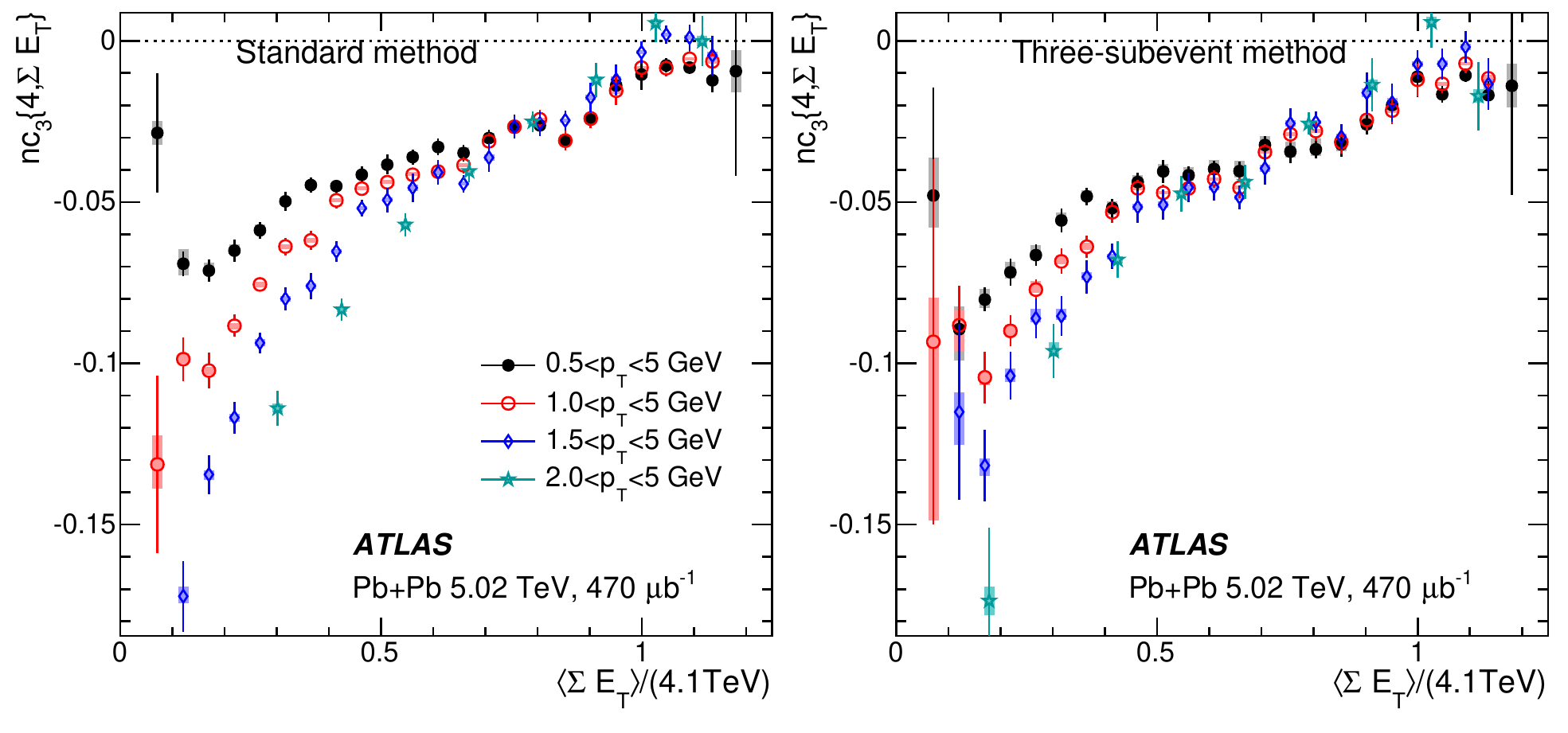}
\end{center}
\caption{\label{fig:app2_2} The nc$_3\{4\}$ values calculated for charged particles in several $\pT$ ranges with the standard cumulant method (left) and three-subevent method (right). The error bars and shaded boxes represent the statistical and systematic uncertainties, respectively. Zero is indicated by a dotted line.}
\end{figure}

\begin{figure}[h!]
\begin{center}
\includegraphics[width=0.9\linewidth]{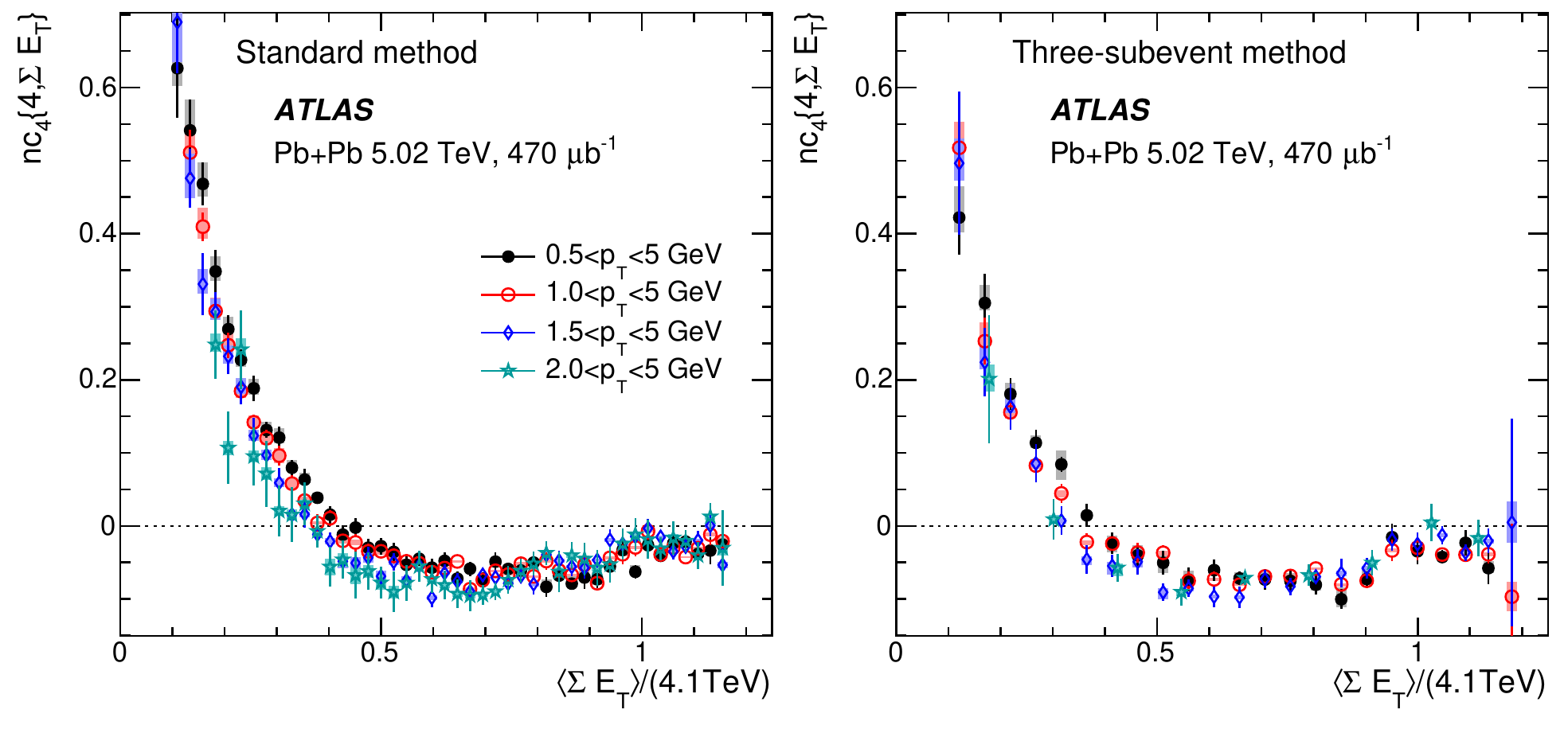}
\end{center}
\caption{\label{fig:app2_3} The nc$_4\{4\}$ values calculated for charged particles in several $\pT$ ranges with the standard cumulant method (left) and three-subevent method (right). The error bars and shaded boxes represent the statistical and systematic uncertainties, respectively.  Zero is indicated by a dotted line.}
\end{figure}

\begin{figure}[h!]
 \begin{center}
\includegraphics[width=0.9\linewidth]{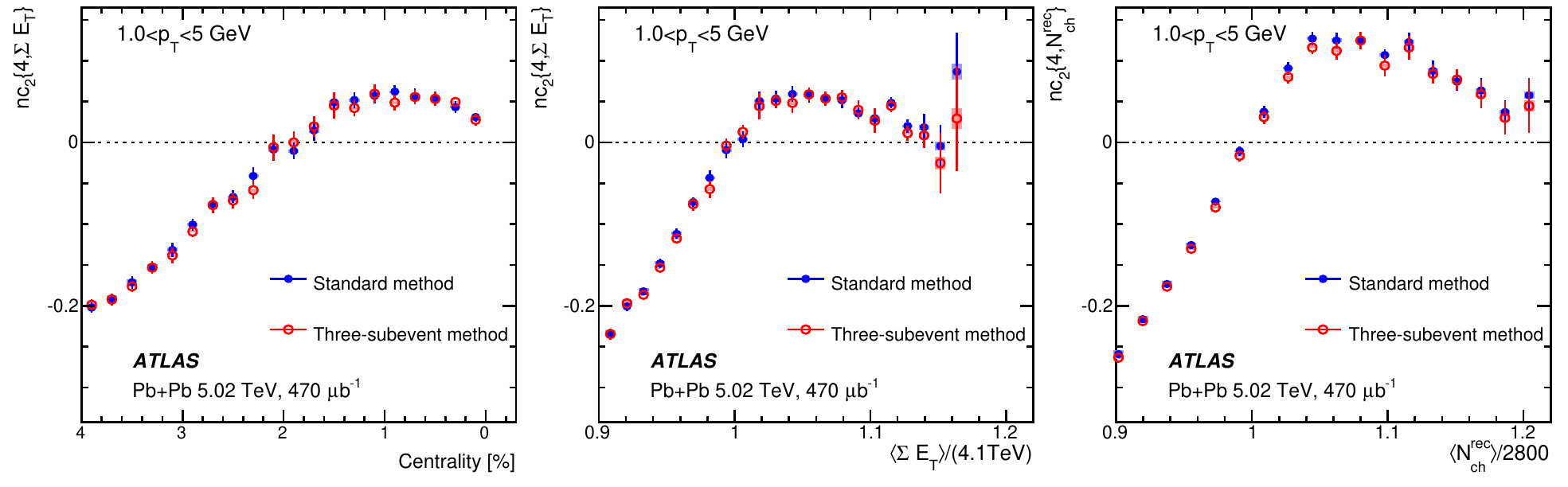}
 \end{center}
 \caption{\label{fig:app2_4} The nc$_2\{4, \etfcal\}$  in ultra-central collisions calculated for charged particles  as a function of centrality (left panel), $\Sigma E_{T}$ (middle panel), and $\nchrec$ (right panel) with the standard cumulant method and the three-subevent method. The error bars and shaded boxes represent the statistical and systematic uncertainties, respectively.}
\end{figure}

\begin{figure}[h!]
\begin{center}
\includegraphics[width=0.9\linewidth]{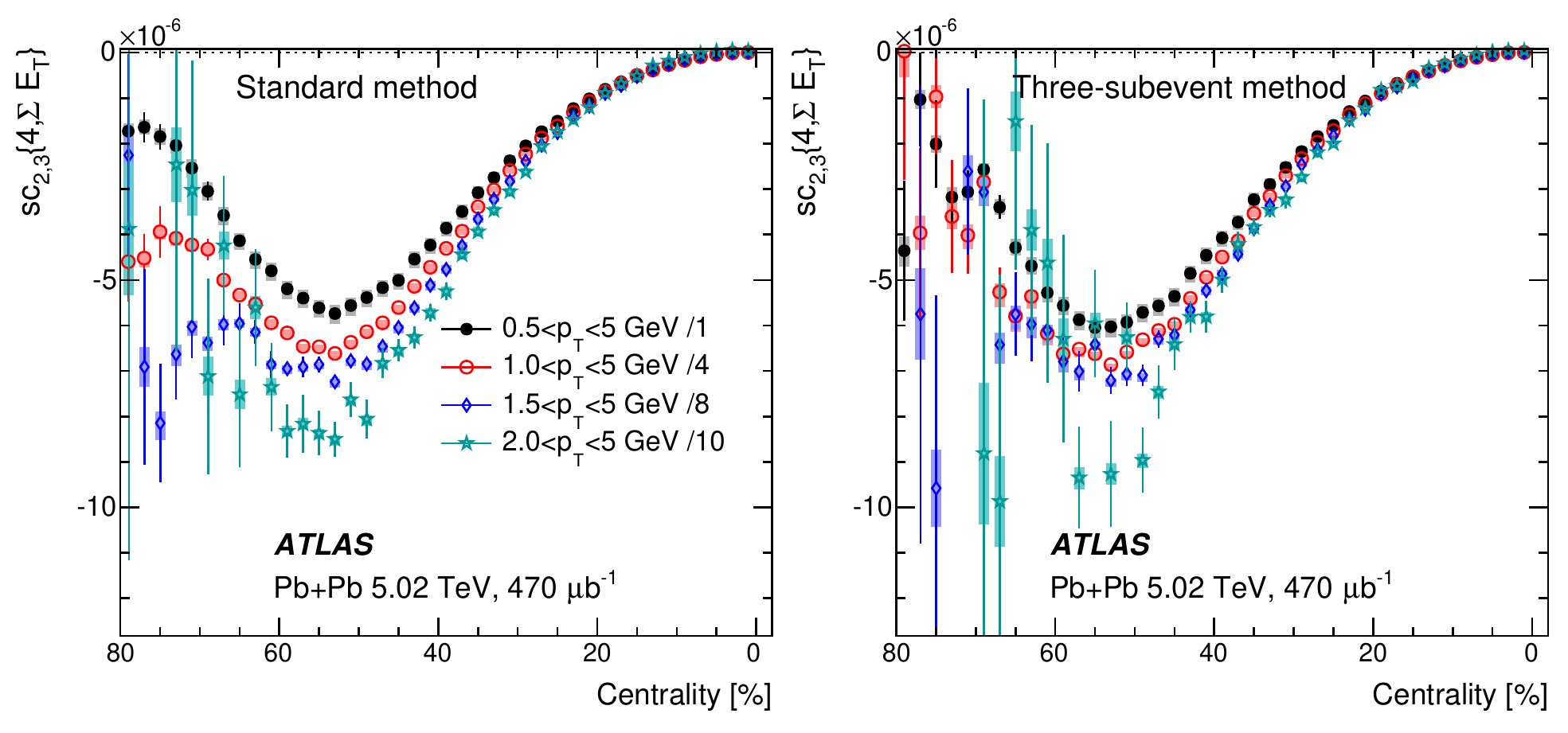}
\end{center}
\caption{\label{fig:app2_5} The $\sca$ values calculated for charged particles in four $\pT$ ranges as a function of centrality for the standard method  (left panel) and three-subevent method (right panel). The error bars and shaded boxes represent the statistical and systematic uncertainties, respectively.}
\end{figure}

\begin{figure}[h!]
\begin{center}
\includegraphics[width=0.9\linewidth]{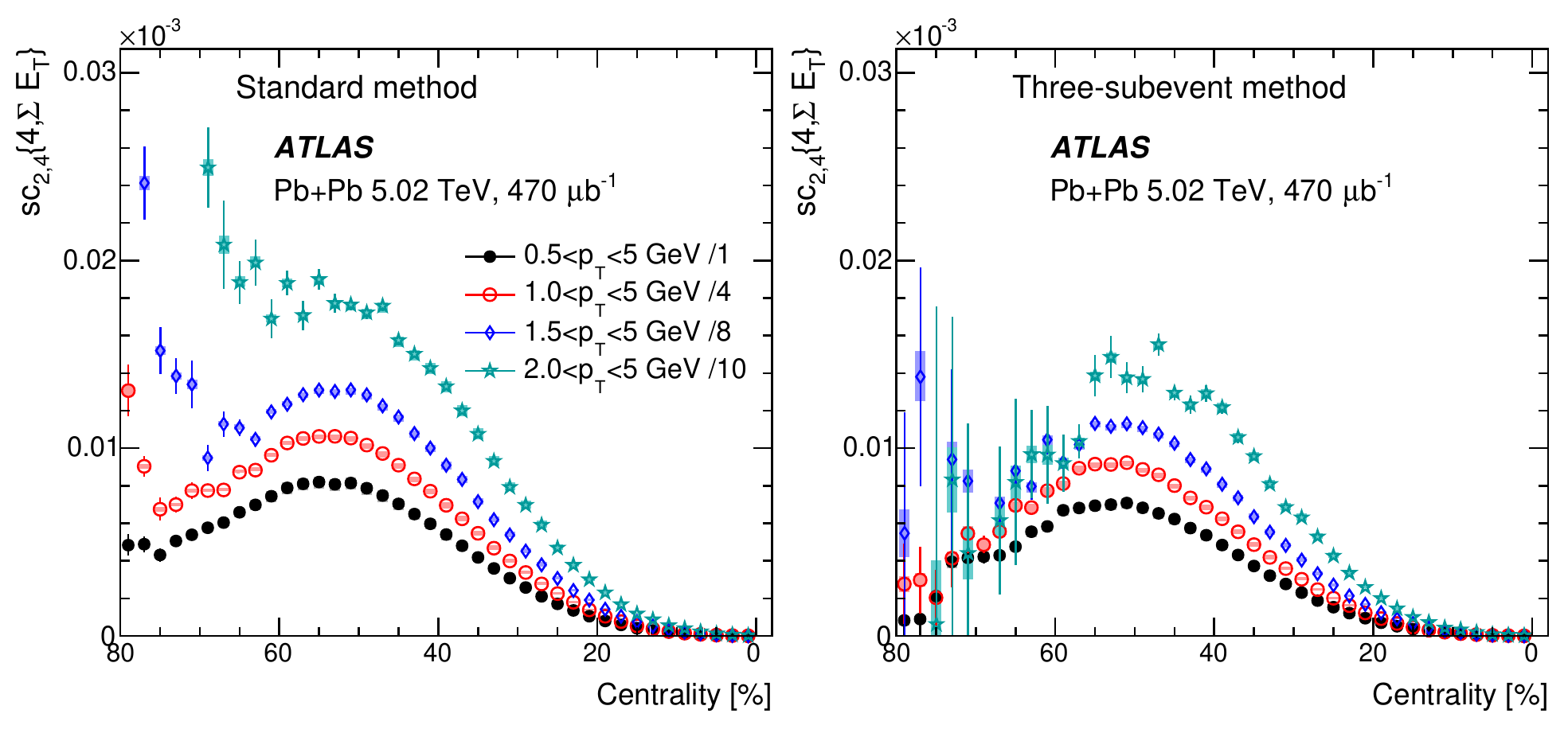}
\end{center}
\caption{\label{fig:app2_6} The $\scb$ values calculated for charged particles in four $\pT$ ranges as a function of centrality for the standard method  (left panel) and three-subevent method (right panel). The error bars and shaded boxes represent the statistical and systematic uncertainties, respectively.}
\end{figure}

\begin{figure}[h!]
\begin{center}
\includegraphics[width=0.9\linewidth]{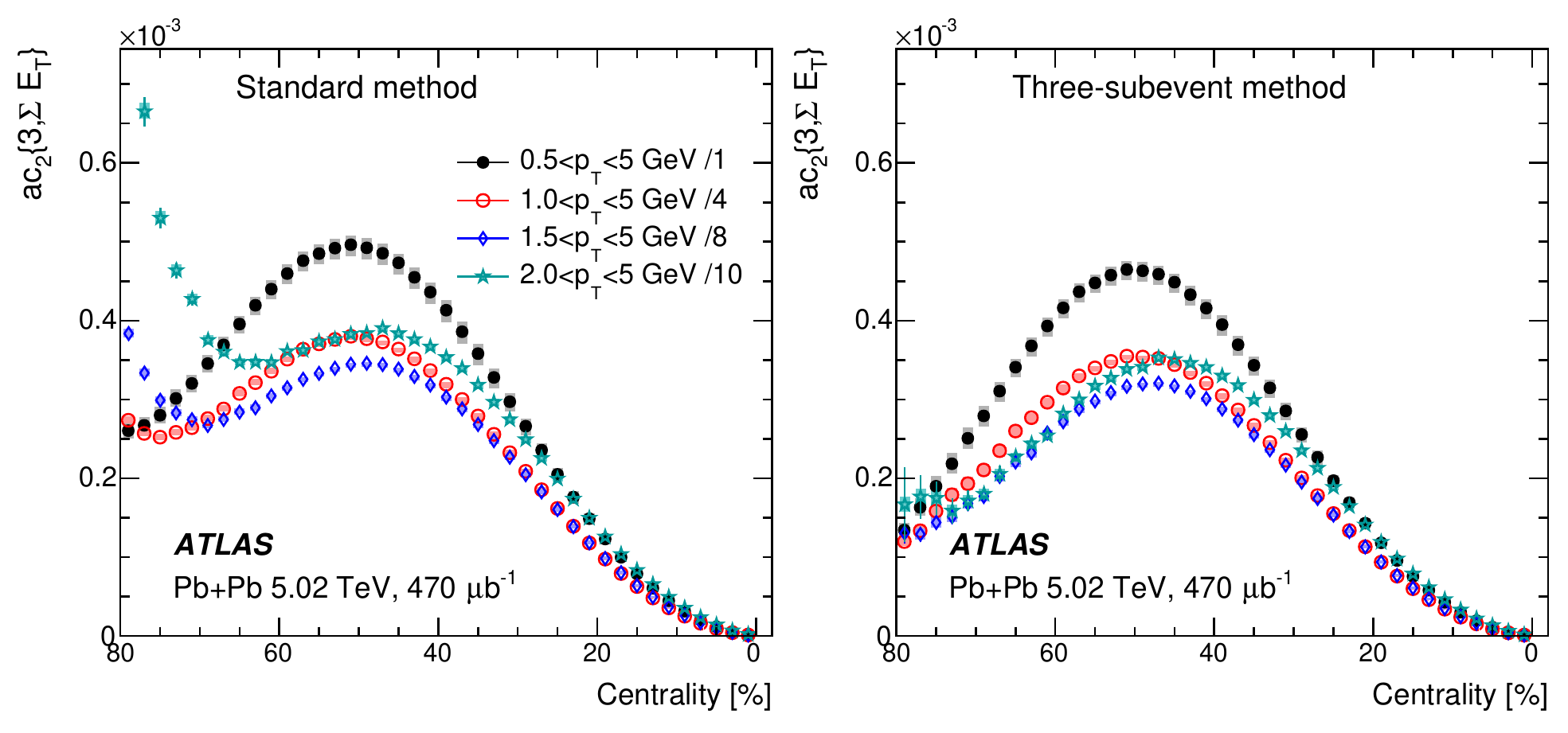}
\end{center}
\caption{\label{fig:app2_7} The $\aca$ values calculated for charged particles in four $\pT$ ranges as a function of centrality for the standard method  (left panel) and three-subevent method (right panel). The error bars and shaded boxes represent the statistical and systematic uncertainties, respectively.}
\end{figure}

\begin{figure}[h!]
\begin{center}
\includegraphics[width=0.7\linewidth]{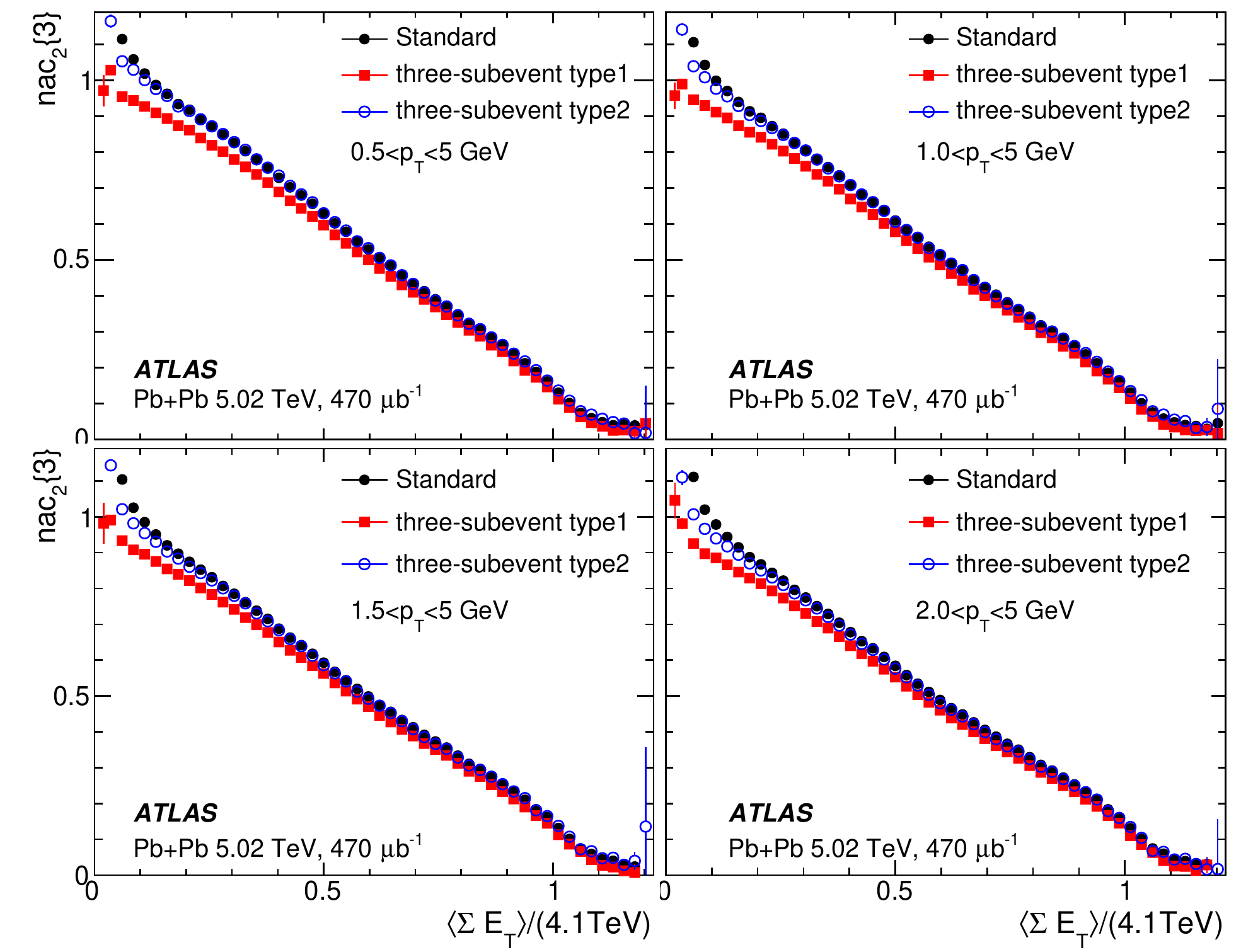}
\end{center}
\caption{\label{fig:app2_8} The nac$_2\{3\}$ from the standard method (solid circles), three-subevent method with ${\bm V}_4$ defined in subevent a or c (solid squares) and three-subevent method with ${\bm V}_4$ defined in subevent b (open circiles). Different panels correspond to different $\pT$ ranges. Only statistical uncertainties are shown.}
\end{figure}

\begin{figure}[h!]
\begin{center}
\includegraphics[width=0.7\linewidth]{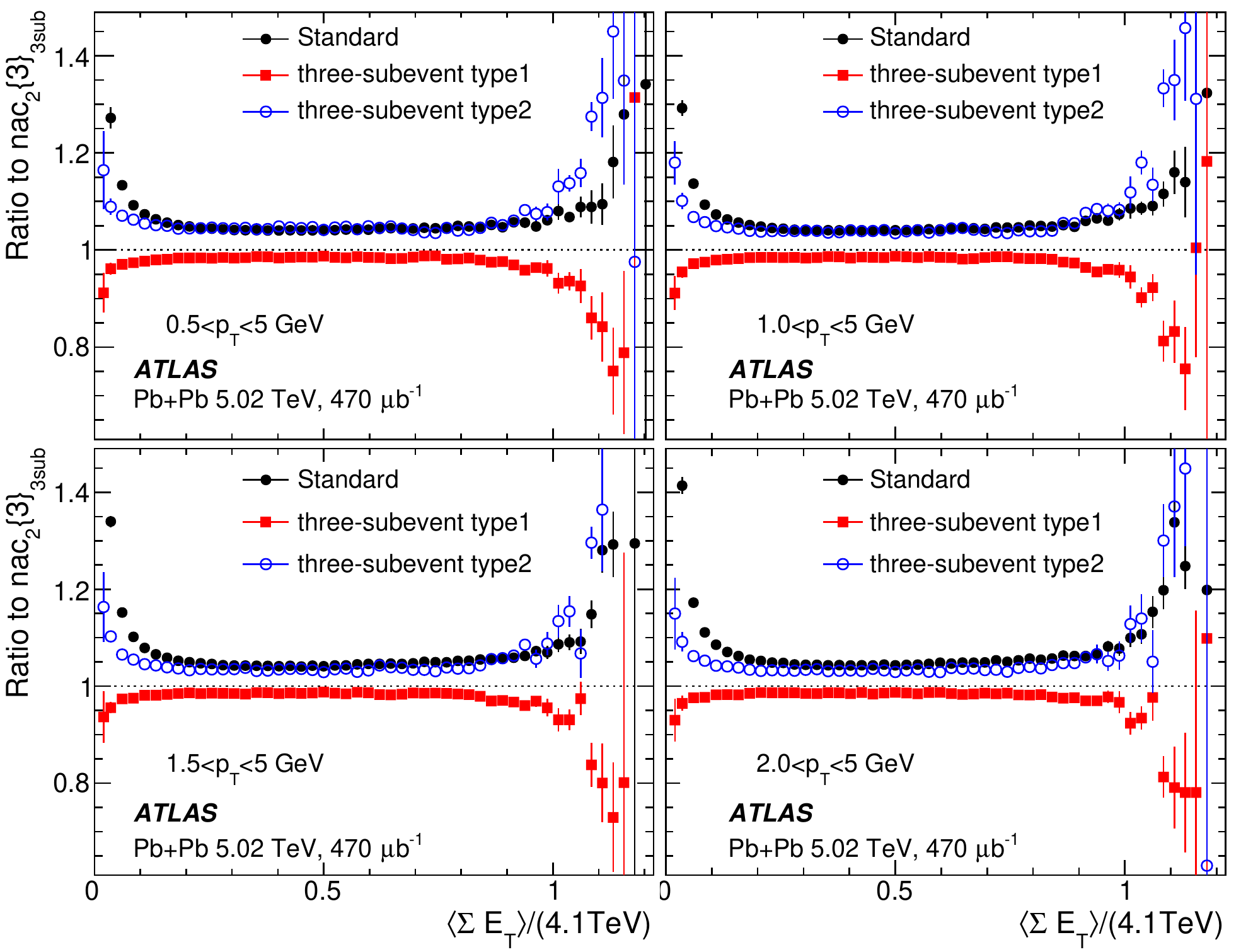}
\end{center}
\caption{\label{fig:app2_9} The ratio of nac$_2\{3\}$ from the standard method (solid circles), three-subevent method with ${\bm V}_4$ defined in subevent a or c (solid squares, type1) and three-subevent method with ${\bm V}_4$ defined in subevent b (open circiles, type2) to nac$_2\{3\}$ from the three-subevent method combined. Different panels correspond to different $\pT$ ranges. Only statistical uncertainties are shown.}
\end{figure}

This appendix shows a comparison between the standard cumulant method and the three-subevent method for various cumulant observables. Figures~\ref{fig:app2_1}--\ref{fig:app2_4} show this comparison for the normalized cumulants nc$_2\{4\}$, nc$_3\{4\}$, and nc$_4\{4\}$ calculated with event class based on $\etfcal$. Figures~\ref{fig:app2_5}--\ref{fig:app2_7} show the comparisons for $\sca$, $\scb$ and $\aca$, respectively. Figures~\ref{fig:app2_8} and \ref{fig:app2_9} compares the standard method and different types of subevent methods. As discussed in~\cite{Aaboud:2018syf}, part of the differences between the standard method and subevent methods can be partially attributed to longitudinal flow decorrelations~\cite{Aaboud:2017tql}. 


\clearpage
\section{Correlation of cumulant ratios}
\label{sec:app3}
\begin{figure}[h!]
\begin{center}
\includegraphics[width=1\linewidth]{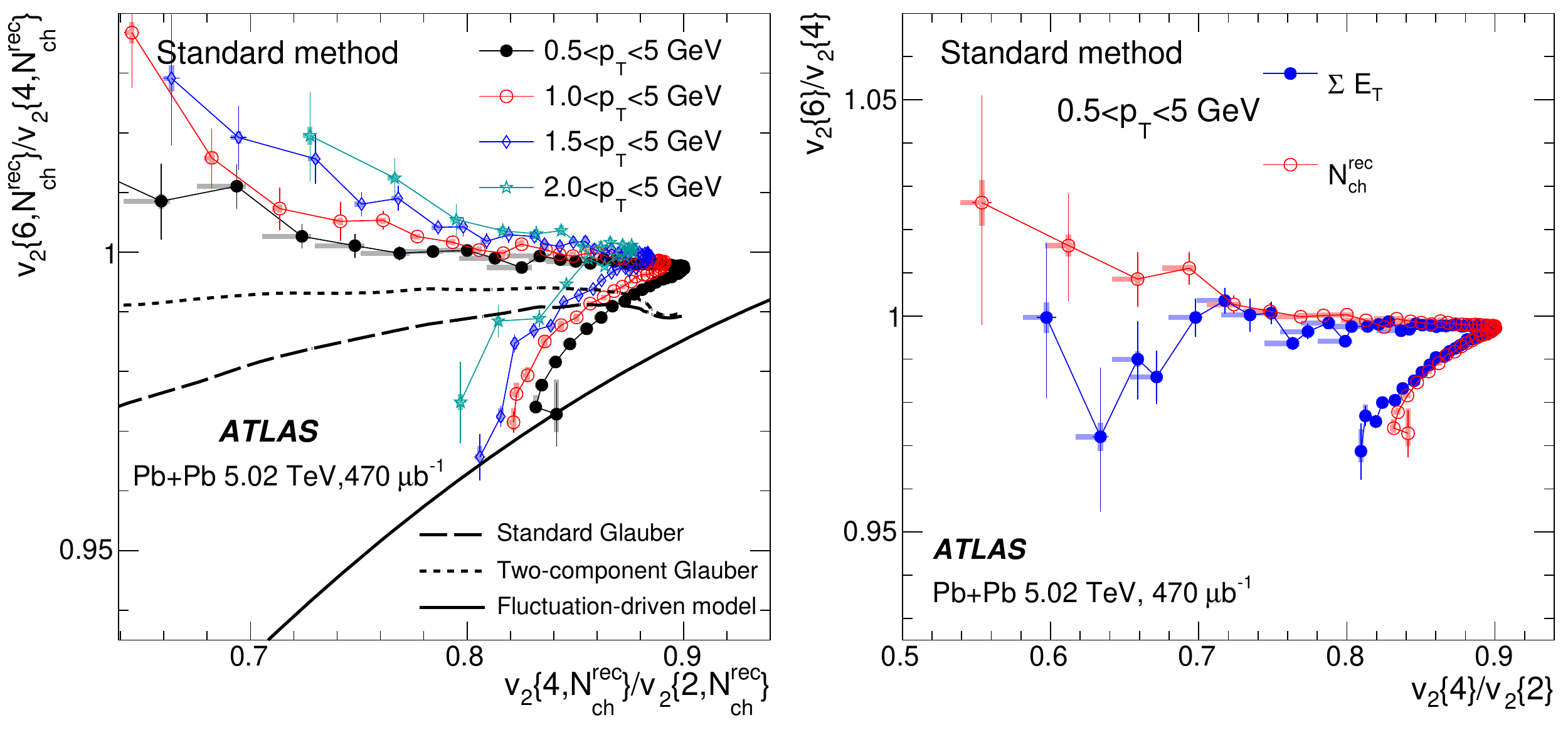}
\end{center}
\vspace*{-0.2cm}
\caption{\label{fig:app3_1} The correlation between $v_2\{6\}/v_2\{4\}$ and $v_2\{4\}/v_2\{2\}$ for four $\pT$ ranges calculated with event class based on $\nchrec$ with models based on initial-state eccentricies (left panel), as well as the correlation compared between the $\etfcal$ event class and $\nchrec$ event class for $0.5<\pT<5$ GeV (right panel). The error bars and shaded boxes represent the statistical and systematic uncertainties, respectively.}
\end{figure}

\begin{figure}[h!]
\begin{center}
\includegraphics[width=1\linewidth]{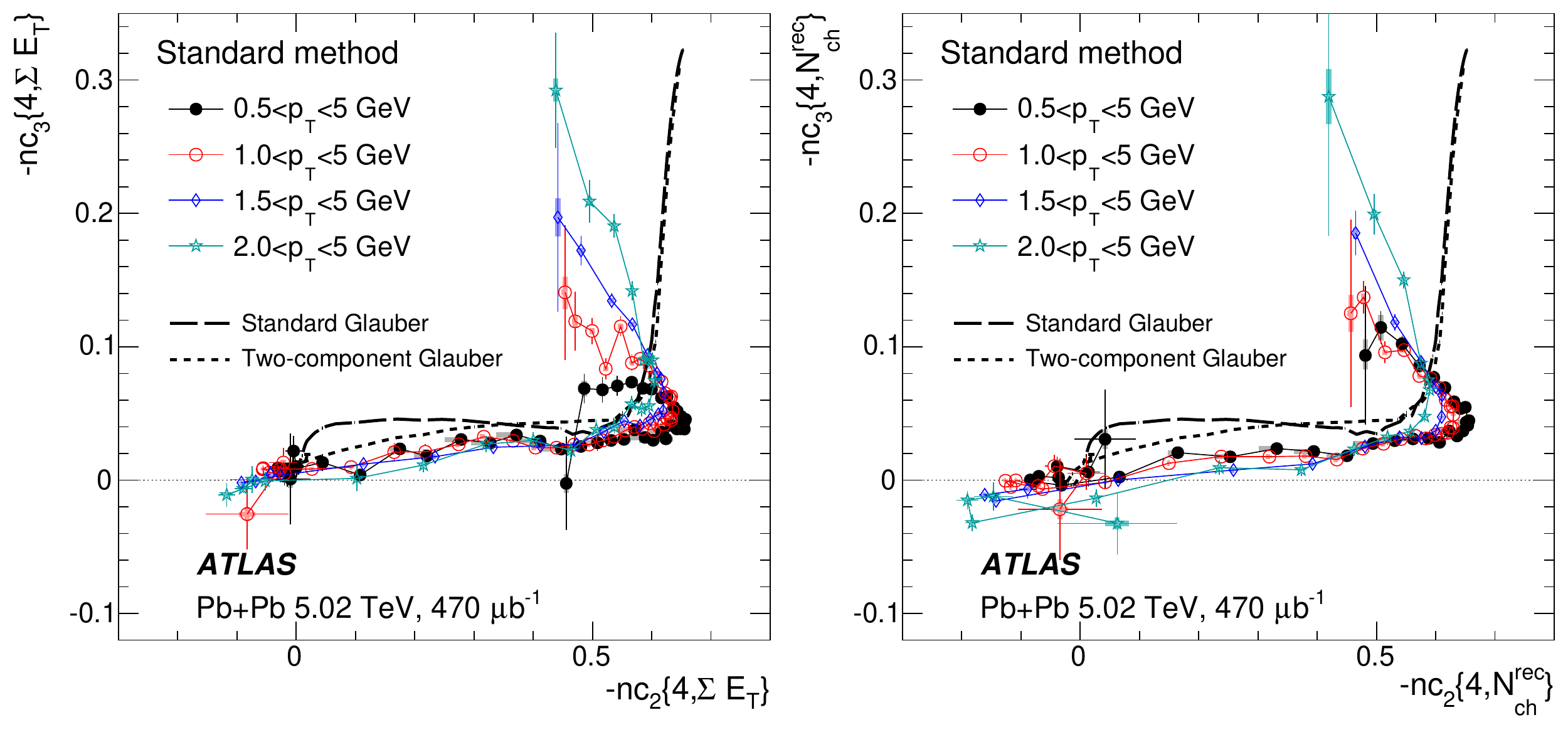}
\end{center}
\vspace*{-0.2cm}
\caption{\label{fig:app3_2} The correlation between $-\mathrm{nc}_3\{4\}$ and $-\mathrm{nc}_2\{4\}$ compared with Glauber model predictions~\cite{Zhou:2018fxx} in four $\pT$ ranges with event class based on $\etfcal$ (left panel) and $\nchrec$ (right panel). The error bars and shaded boxes represent the statistical and systematic uncertainties, respectively.}
\end{figure}

\begin{figure}[h!]
\begin{center}
\includegraphics[width=1\linewidth]{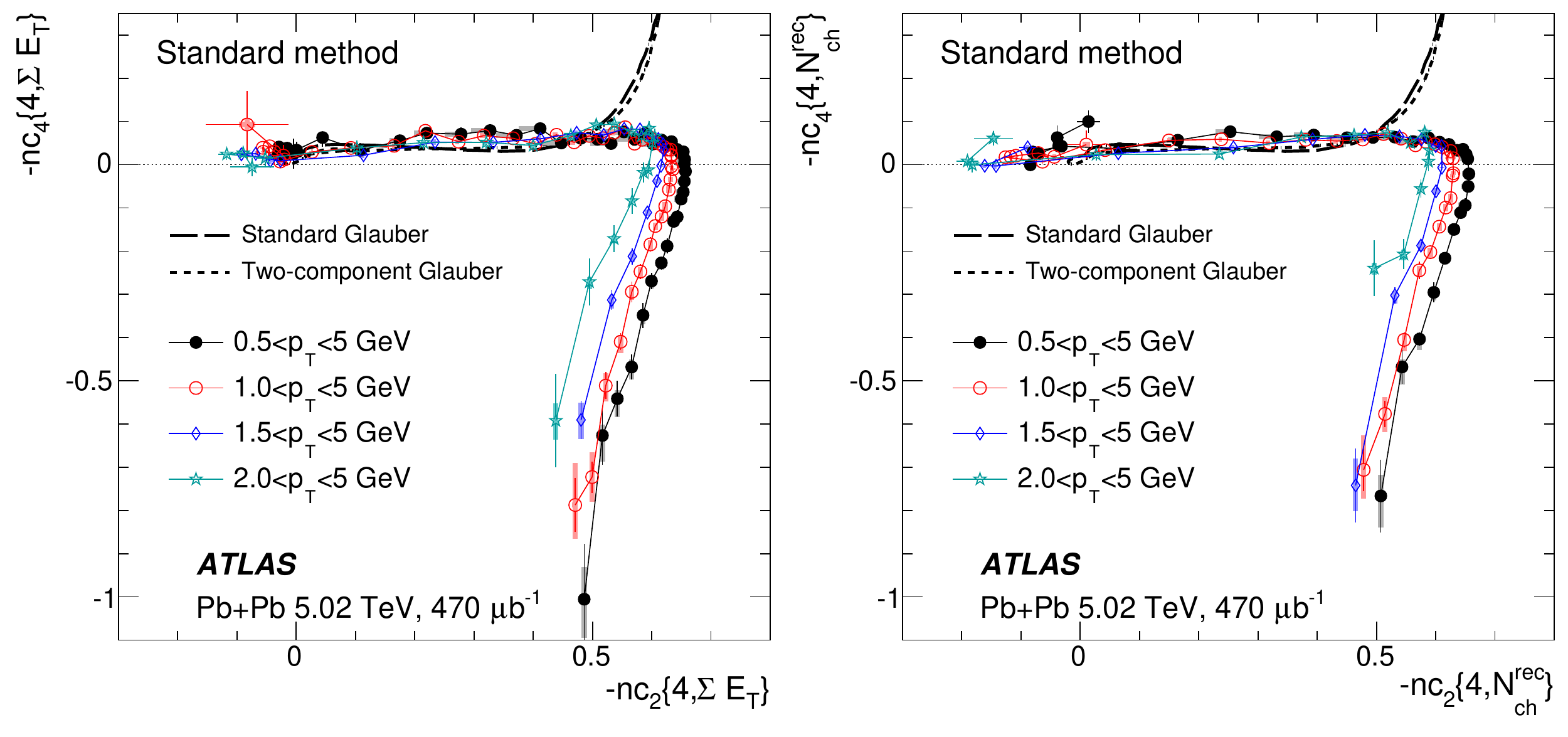}
\end{center}
\vspace*{-0.2cm}
\caption{\label{fig:app3_3} The correlation between $-\mathrm{nc}_4\{4\}$ and $-\mathrm{nc}_2\{4\}$ compared with Glauber model predictions~\cite{Zhou:2018fxx} in four $\pT$ ranges with event class based on $\etfcal$ (left panel) and $\nchrec$ (right panel). The error bars and shaded boxes represent the statistical and systematic uncertainties, respectively.}
\end{figure}

This appendix shows the correlation between different cumulant ratios. Figure~\ref{fig:app3_1} shows the correlation between $v_2\{6\}/v_2\{4\}$ and $v_2\{4\}/v_2\{2\}$ for event class based on $\nchrec$; and this is a complementary plot to the right panel of Figure~\ref{fig:a5}. Figures~\ref{fig:app3_2} and \ref{fig:app3_3} show the correlation between normalized cumulants nc$_n\{4\}$ and nc$_2\{4\}$, these correlations are compared directly with model calculations based on initial-state eccentricities~\cite{Miller:2007ri,Yan:2013laa,Zhou:2018fxx}. 

\vspace*{-0.2cm}

\vspace*{-0.2cm}

\vspace*{-0.2cm}

\clearpage
\section*{Acknowledgements}

We thank CERN for the very successful operation of the LHC, as well as the
support staff from our institutions without whom ATLAS could not be
operated efficiently.

We acknowledge the support of ANPCyT, Argentina; YerPhI, Armenia; ARC, Australia; BMWFW and FWF, Austria; ANAS, Azerbaijan; SSTC, Belarus; CNPq and FAPESP, Brazil; NSERC, NRC and CFI, Canada; CERN; CONICYT, Chile; CAS, MOST and NSFC, China; COLCIENCIAS, Colombia; MSMT CR, MPO CR and VSC CR, Czech Republic; DNRF and DNSRC, Denmark; IN2P3-CNRS, CEA-DRF/IRFU, France; SRNSFG, Georgia; BMBF, HGF, and MPG, Germany; GSRT, Greece; RGC, Hong Kong SAR, China; ISF and Benoziyo Center, Israel; INFN, Italy; MEXT and JSPS, Japan; CNRST, Morocco; NWO, Netherlands; RCN, Norway; MNiSW and NCN, Poland; FCT, Portugal; MNE/IFA, Romania; MES of Russia and NRC KI, Russian Federation; JINR; MESTD, Serbia; MSSR, Slovakia; ARRS and MIZ\v{S}, Slovenia; DST/NRF, South Africa; MINECO, Spain; SRC and Wallenberg Foundation, Sweden; SERI, SNSF and Cantons of Bern and Geneva, Switzerland; MOST, Taiwan; TAEK, Turkey; STFC, United Kingdom; DOE and NSF, United States of America. In addition, individual groups and members have received support from BCKDF, CANARIE, CRC and Compute Canada, Canada; COST, ERC, ERDF, Horizon 2020, and Marie Sk{\l}odowska-Curie Actions, European Union; Investissements d' Avenir Labex and Idex, ANR, France; DFG and AvH Foundation, Germany; Herakleitos, Thales and Aristeia programmes co-financed by EU-ESF and the Greek NSRF, Greece; BSF-NSF and GIF, Israel; CERCA Programme Generalitat de Catalunya, Spain; The Royal Society and Leverhulme Trust, United Kingdom. 

The crucial computing support from all WLCG partners is acknowledged gratefully, in particular from CERN, the ATLAS Tier-1 facilities at TRIUMF (Canada), NDGF (Denmark, Norway, Sweden), CC-IN2P3 (France), KIT/GridKA (Germany), INFN-CNAF (Italy), NL-T1 (Netherlands), PIC (Spain), ASGC (Taiwan), RAL (UK) and BNL (USA), the Tier-2 facilities worldwide and large non-WLCG resource providers. Major contributors of computing resources are listed in Ref.~\cite{ATL-GEN-PUB-2016-002}.

\clearpage
\bibliographystyle{atlasBibStyleWithTitle}
\bibliography{hion-2017-09}

\providecommand{\href}[2]{#2}\begingroup\raggedright\begin{thebibliography}{10}

\bibitem{Gale:2013da}
C.~Gale, S.~Jeon,  and B.~Schenke, {\em {Hydrodynamic Modeling of Heavy-Ion
  Collisions}},
  \href{http://dx.doi.org/10.1142/S0217751X13400113}{Int.~J.~Mod.~Phys. A
  {\bfseries 28} (2013) 1340011},
\href{http://arxiv.org/abs/1301.5893}{{\ttfamily arXiv:1301.5893 [nucl-th]}}.

\bibitem{Heinz:2013th}
U.~Heinz and R.~Snellings, {\em {Collective Flow and Viscosity in Relativistic
  Heavy-ion Collisions}},
  \href{http://dx.doi.org/10.1146/annurev-nucl-102212-170540}{Ann.~Rev.~Nucl.~Part.~Sci.
  {\bfseries 63} (2013) 123},
\href{http://arxiv.org/abs/1301.2826}{{\ttfamily arXiv:1301.2826 [nucl-th]}}.

\bibitem{Busza:2018rrf}
W.~Busza, K.~Rajagopal,  and W.~van~der Schee, {\em {Heavy Ion Collisions: The
  Big Picture, and the Big Questions}},
  \href{http://dx.doi.org/10.1146/annurev-nucl-101917-020852}{Ann. Rev. Nucl.
  Part. Sci. {\bfseries 68} (2018) 339--376},
\href{http://arxiv.org/abs/1802.04801}{{\ttfamily arXiv:1802.04801 [hep-ph]}}.

\bibitem{Gardim:2011xv}
F.~G. Gardim, F.~Grassi, M.~Luzum,  and J.-Y. Ollitrault, {\em {Mapping the
  hydrodynamic response to the initial geometry in heavy-ion collisions}},
  \href{http://dx.doi.org/10.1103/PhysRevC.85.024908}{Phys.~Rev.~C {\bfseries
  85} (2012) 024908}, \href{http://arxiv.org/abs/1111.6538}{{\ttfamily
  arXiv:1111.6538 [nucl-th]}}.

\bibitem{Gale:2012rq}
C.~Gale, S.~Jeon, B.~Schenke, P.~Tribedy,  and R.~Venugopalan, {\em
  {Event-by-Event Anisotropic Flow in Heavy-ion Collisions from Combined
  Yang-Mills and Viscous Fluid Dynamics}},
  \href{http://dx.doi.org/10.1103/PhysRevLett.110.012302}{Phys.~Rev.~Lett.
  {\bfseries 110} (2013) 012302},
  \href{http://arxiv.org/abs/1209.6330}{{\ttfamily arXiv:1209.6330 [nucl-th]}}.

\bibitem{Teaney:2010vd}
D.~Teaney and L.~Yan, {\em {Triangularity and dipole asymmetry in relativistic
  heavy ion collisions}},
  \href{http://dx.doi.org/10.1103/PhysRevC.83.064904}{Phys.~Rev.~C {\bfseries
  83} (2011) 064904}, \href{http://arxiv.org/abs/1010.1876}{{\ttfamily
  arXiv:1010.1876 [nucl-th]}}.

\bibitem{Niemi:2012aj}
H.~Niemi, G.~S. Denicol, H.~Holopainen,  and P.~Huovinen, {\em {Event-by-event
  distributions of azimuthal asymmetries in ultrarelativistic heavy-ion
  collisions}},
  \href{http://dx.doi.org/10.1103/PhysRevC.87.054901}{Phys.~Rev.~C {\bfseries
  87} (2013) 054901}, \href{http://arxiv.org/abs/1212.1008}{{\ttfamily
  arXiv:1212.1008 [nucl-th]}}.

\bibitem{Adare:2011tg}
{PHENIX Collaboration}, {\em {Measurements of Higher-Order Flow Harmonics in
  Au+Au Collisions at $\sqrt{s_{NN}} = 200$ GeV}},
  \href{http://dx.doi.org/10.1103/PhysRevLett.107.252301}{Phys.~Rev.~Lett.
  {\bfseries 107} (2011) 252301},
  \href{http://arxiv.org/abs/1105.3928}{{\ttfamily arXiv:1105.3928 [nucl-ex]}}.

\bibitem{ALICE:2011ab}
{ALICE Collaboration}, {\em {Higher Harmonic Anisotropic Flow Measurements of
  Charged Particles in Pb-Pb Collisions at $\sqrt{s_{NN}}=2.76$ TeV}},
  \href{http://dx.doi.org/10.1103/PhysRevLett.107.032301}{Phys.~Rev.~Lett.
  {\bfseries 107} (2011) 032301},
  \href{http://arxiv.org/abs/1105.3865}{{\ttfamily arXiv:1105.3865 [nucl-ex]}}.

\bibitem{Aad:2012bu}
{ATLAS Collaboration}, {\em {Measurement of the azimuthal anisotropy for
  charged particle production in $\sqrt{s_{NN}}=2.76$ TeV lead-lead collisions
  with the ATLAS detector}},
  \href{http://dx.doi.org/10.1103/PhysRevC.86.014907}{Phys.~Rev.~C {\bfseries
  86} (2012) 014907}, \href{http://arxiv.org/abs/1203.3087}{{\ttfamily
  arXiv:1203.3087 [hep-ex]}}.

\bibitem{Chatrchyan:2013kba}
{CMS Collaboration}, {\em {Measurement of higher-order harmonic azimuthal
  anisotropy in PbPb collisions at $\sqrt{s_{NN}}$ = 2.76 TeV}},
  \href{http://dx.doi.org/10.1103/PhysRevC.89.044906}{Phys.~Rev.~C {\bfseries
  89} (2014) 044906}, \href{http://arxiv.org/abs/1310.8651}{{\ttfamily
  arXiv:1310.8651 [nucl-ex]}}.

\bibitem{Aad:2013xma}
{ATLAS Collaboration}, {\em {Measurement of the distributions of event-by-event
  flow harmonics in lead-lead collisions at $\sqrt{s_{NN}}$= 2.76 TeV with the
  ATLAS detector at the LHC}},
  \href{http://dx.doi.org/10.1007/JHEP11(2013)183}{JHEP {\bfseries 11} (2013)
  183}, \href{http://arxiv.org/abs/1305.2942}{{\ttfamily arXiv:1305.2942
  [hep-ex]}}.

\bibitem{Aad:2014fla}
{ATLAS Collaboration}, {\em {Measurement of event-plane correlations in
  $\sqrt{s_{NN}}=2.76$ TeV lead-lead collisions with the ATLAS detector}},
  \href{http://dx.doi.org/10.1103/PhysRevC.90.024905}{Phys.~Rev.~C {\bfseries
  90} (2014) 024905}, \href{http://arxiv.org/abs/1403.0489}{{\ttfamily
  arXiv:1403.0489 [hep-ex]}}.

\bibitem{Aad:2015lwa}
{ATLAS Collaboration}, {\em {Measurement of the correlation between flow
  harmonics of different order in lead-lead collisions at $\sqrt{s_{NN}}$=2.76
  TeV with the ATLAS detector}},
  \href{http://dx.doi.org/10.1103/PhysRevC.92.034903}{Phys. Rev. C {\bfseries
  92} (2015) 034903},
\href{http://arxiv.org/abs/1504.01289}{{\ttfamily arXiv:1504.01289 [hep-ex]}}.

\bibitem{ALICE:2016kpq}
{ALICE Collaboration}, {\em {Correlated Event-by-Event Fluctuations of Flow
  Harmonics in Pb-Pb Collisions at $\sqrt{s_{\mathrm{NN}}}=2.76$ TeV}},
  \href{http://dx.doi.org/10.1103/PhysRevLett.117.182301}{Phys. Rev. Lett.
  {\bfseries 117} (2016) 182301},
\href{http://arxiv.org/abs/1604.07663}{{\ttfamily arXiv:1604.07663 [nucl-ex]}}.

\bibitem{Luzum:2012wu}
M.~Luzum and J.-Y. Ollitrault, {\em {Extracting the shear viscosity of the
  quark-gluon plasma from flow in ultra-central heavy-ion collisions}},
  \href{http://dx.doi.org/10.1016/j.nuclphysa.2013.02.028}{Nucl.~Phys.~A
  {\bfseries 904} (2013) 377c},
  \href{http://arxiv.org/abs/1210.6010}{{\ttfamily arXiv:1210.6010 [nucl-th]}}.

\bibitem{Qiu:2012uy}
Z.~Qiu and U.~Heinz, {\em {Hydrodynamic event-plane correlations in Pb+Pb
  collisions at $\sqrt{s}=2.76$A~TeV}},
  \href{http://dx.doi.org/10.1016/j.physletb.2012.09.030}{Phys.~Lett.~B
  {\bfseries 717} (2012) 261}, \href{http://arxiv.org/abs/1208.1200}{{\ttfamily
  arXiv:1208.1200 [nucl-th]}}.

\bibitem{Teaney:2013dta}
D.~Teaney and L.~Yan, {\em {Event-plane correlations and hydrodynamic
  simulations of heavy ion collisions}},
  \href{http://dx.doi.org/10.1103/PhysRevC.90.024902}{Phys.~Rev.~C {\bfseries
  90} (2014) 024902}, \href{http://arxiv.org/abs/1312.3689}{{\ttfamily
  arXiv:1312.3689 [nucl-th]}}.

\bibitem{Borghini:2000sa}
N.~Borghini, P.~M. Dinh,  and J.-Y. Ollitrault, {\em {New method for measuring
  azimuthal distributions in nucleus-nucleus collisions}},
  \href{http://dx.doi.org/10.1103/PhysRevC.63.054906}{Phys.~Rev.~C {\bfseries
  63} (2001) 054906}, \href{http://arxiv.org/abs/nucl-th/0007063}{{\ttfamily
  arXiv:nucl-th/0007063}}.

\bibitem{Borghini:2001vi}
N.~Borghini, P.~M. Dinh,  and J.-Y. Ollitrault, {\em {Flow analysis from
  multiparticle azimuthal correlations}},
  \href{http://dx.doi.org/10.1103/PhysRevC.64.054901}{Phys. Rev. C {\bfseries
  64} (2001) 054901},
\href{http://arxiv.org/abs/nucl-th/0105040}{{\ttfamily arXiv:nucl-th/0105040
  [nucl-th]}}.

\bibitem{Bilandzic:2010jr}
A.~Bilandzic, R.~Snellings,  and S.~Voloshin, {\em {Flow analysis with
  cumulants: Direct calculations}},
  \href{http://dx.doi.org/10.1103/PhysRevC.83.044913}{Phys.~Rev.~C {\bfseries
  83} (2011) 044913}, \href{http://arxiv.org/abs/1010.0233}{{\ttfamily
  arXiv:1010.0233 [nucl-ex]}}.

\bibitem{Bilandzic:2013kga}
A.~Bilandzic, C.~H. Christensen, K.~Gulbrandsen, A.~Hansen,  and Y.~Zhou, {\em
  {Generic framework for anisotropic flow analyses with multiparticle azimuthal
  correlations}}, \href{http://dx.doi.org/10.1103/PhysRevC.89.064904}{Phys.
  Rev. C {\bfseries 89} (2014) 064904},
\href{http://arxiv.org/abs/1312.3572}{{\ttfamily arXiv:1312.3572 [nucl-ex]}}.

\bibitem{Jia:2014jca}
J.~Jia, {\em {Event-shape fluctuations and flow correlations in
  ultra-relativistic heavy-ion collisions}},
  \href{http://dx.doi.org/10.1088/0954-3899/41/12/124003}{J.~Phys.~G {\bfseries
  41} (2014) 124003}, \href{http://arxiv.org/abs/1407.6057}{{\ttfamily
  arXiv:1407.6057 [nucl-ex]}}.

\bibitem{Aad:2014vba}
{ATLAS Collaboration}, {\em {Measurement of flow harmonics with multi-particle
  cumulants in Pb+Pb collisions at $\sqrt{s_{\mathrm{NN}}}=2.76$ TeV with the
  ATLAS detector}},
  \href{http://dx.doi.org/10.1140/epjc/s10052-014-3157-z}{Eur. Phys. J. C
  {\bfseries 74} (2014) 3157},
\href{http://arxiv.org/abs/1408.4342}{{\ttfamily arXiv:1408.4342 [hep-ex]}}.

\bibitem{Abelev:2014mda}
{ALICE Collaboration}, {\em {Multiparticle azimuthal correlations in p-Pb and
  Pb-Pb collisions at the CERN Large Hadron Collider}},
  \href{http://dx.doi.org/10.1103/PhysRevC.90.054901}{Phys. Rev. C {\bfseries
  90} (2014) 054901},
\href{http://arxiv.org/abs/1406.2474}{{\ttfamily arXiv:1406.2474 [nucl-ex]}}.

\bibitem{Chatrchyan:2013nka}
{CMS Collaboration}, {\em {Multiplicity and transverse momentum dependence of
  two- and four-particle correlations in pPb and PbPb collisions}},
  \href{http://dx.doi.org/10.1016/j.physletb.2013.06.028}{Phys. Lett. B
  {\bfseries 724} (2013) 213},
\href{http://arxiv.org/abs/1305.0609}{{\ttfamily arXiv:1305.0609 [nucl-ex]}}.

\bibitem{Adamczyk:2015obl}
{STAR Collaboration}, {\em {Azimuthal Anisotropy in U$+$U and Au$+$Au
  Collisions at RHIC}},
  \href{http://dx.doi.org/10.1103/PhysRevLett.115.222301}{Phys. Rev. Lett.
  {\bfseries 115} (2015) 222301},
\href{http://arxiv.org/abs/1505.07812}{{\ttfamily arXiv:1505.07812 [nucl-ex]}}.

\bibitem{Sirunyan:2017pan}
{CMS Collaboration}, {\em {Azimuthal anisotropy of charged particles with
  transverse momentum up to 100 GeV/ c in PbPb collisions at
  $\sqrt{s_{NN}}$=5.02 TeV}},
  \href{http://dx.doi.org/10.1016/j.physletb.2017.11.041}{Phys. Lett. B
  {\bfseries 776} (2018) 195},
\href{http://arxiv.org/abs/1702.00630}{{\ttfamily arXiv:1702.00630 [hep-ex]}}.

\bibitem{Aaboud:2017blb}
{ATLAS Collaboration}, {\em {Measurement of long-range multiparticle azimuthal
  correlations with the subevent cumulant method in $pp$ and $p + Pb$
  collisions with the ATLAS detector at the CERN Large Hadron Collider}},
  \href{http://dx.doi.org/10.1103/PhysRevC.97.024904}{Phys. Rev. C {\bfseries
  97} (2018) 024904},
\href{http://arxiv.org/abs/1708.03559}{{\ttfamily arXiv:1708.03559 [hep-ex]}}.

\bibitem{Sirunyan:2017fts}
{CMS Collaboration}, {\em {Non-Gaussian elliptic-flow fluctuations in PbPb
  collisions at $\sqrt{\smash[b]{s_{_\text{NN}}}} = 5.02$ TeV}},
  \href{http://dx.doi.org/10.1016/j.physletb.2018.11.063}{Phys. Lett. B
  {\bfseries 789} (2019) 643},
\href{http://arxiv.org/abs/1711.05594}{{\ttfamily arXiv:1711.05594 [nucl-ex]}}.

\bibitem{Acharya:2019vdf}
{ALICE Collaboration}, {\em {Investigations of Anisotropic Flow Using
  Multiparticle Azimuthal Correlations in pp, p-Pb, Xe-Xe, and Pb-Pb Collisions
  at the LHC}}, \href{http://dx.doi.org/10.1103/PhysRevLett.123.142301}{Phys.
  Rev. Lett. {\bfseries 123} (2019) 142301},
\href{http://arxiv.org/abs/1903.01790}{{\ttfamily arXiv:1903.01790 [nucl-ex]}}.

\bibitem{Yan:2013laa}
L.~Yan and J.-Y. Ollitrault, {\em {Universal Fluctuation-Driven Eccentricities
  in Proton-Proton, Proton-Nucleus and Nucleus-Nucleus Collisions}},
  \href{http://dx.doi.org/10.1103/PhysRevLett.112.082301}{Phys. Rev. Lett.
  {\bfseries 112} (2014) 082301},
\href{http://arxiv.org/abs/1312.6555}{{\ttfamily arXiv:1312.6555 [nucl-th]}}.

\bibitem{Yan:2014nsa}
L.~Yan, J.-Y. Ollitrault,  and A.~M. Poskanzer, {\em {Azimuthal anisotropy
  distributions in high-energy collisions}},
  \href{http://dx.doi.org/10.1016/j.physletb.2015.01.039}{Phys. Lett. B
  {\bfseries 742} (2015) 290--295},
\href{http://arxiv.org/abs/1408.0921}{{\ttfamily arXiv:1408.0921 [nucl-th]}}.

\bibitem{Acharya:2018lmh}
{ALICE Collaboration}, {\em {Energy dependence and fluctuations of anisotropic
  flow in Pb-Pb collisions at $ \sqrt{s_{\mathrm{NN}}}=5.02$ and 2.76 TeV}},
  \href{http://dx.doi.org/10.1007/JHEP07(2018)103}{JHEP {\bfseries 07} (2018)
  103},
\href{http://arxiv.org/abs/1804.02944}{{\ttfamily arXiv:1804.02944 [nucl-ex]}}.

\bibitem{Jia:2017hbm}
J.~Jia, M.~Zhou,  and A.~Trzupek, {\em {Revealing long-range multiparticle
  collectivity in small collision systems via subevent cumulants}},
  \href{http://dx.doi.org/10.1103/PhysRevC.96.034906}{Phys. Rev. C {\bfseries
  96} (2017) 034906},
\href{http://arxiv.org/abs/1701.03830}{{\ttfamily arXiv:1701.03830 [nucl-th]}}.

\bibitem{Acharya:2017gsw}
{ALICE} Collaboration, S.~Acharya {et~al.}, {\em {Systematic studies of
  correlations between different order flow harmonics in Pb-Pb collisions at
  $\sqrt{s_{\textrm{ NN}}}$ = 2.76 TeV}},
  \href{http://dx.doi.org/10.1103/PhysRevC.97.024906}{Phys. Rev. {\bfseries
  C97} (2018) 024906},
\href{http://arxiv.org/abs/1709.01127}{{\ttfamily arXiv:1709.01127 [nucl-ex]}}.

\bibitem{STAR:2018fpo}
{STAR Collaboration}, {\em {Correlation measurements between flow harmonics in
  Au+Au collisions at RHIC}},
  \href{http://dx.doi.org/10.1016/j.physletb.2018.05.076}{Phys. Lett. B
  {\bfseries 783} (2018) 459},
\href{http://arxiv.org/abs/1803.03876}{{\ttfamily arXiv:1803.03876 [nucl-ex]}}.

\bibitem{Giacalone:2017uqx}
G.~Giacalone, J.~Noronha-Hostler,  and J.-Y. Ollitrault, {\em {Relative flow
  fluctuations as a probe of initial state fluctuations}},
  \href{http://dx.doi.org/10.1103/PhysRevC.95.054910}{Phys. Rev. C {\bfseries
  95} (2017) 054910},
\href{http://arxiv.org/abs/1702.01730}{{\ttfamily arXiv:1702.01730 [nucl-th]}}.

\bibitem{Gardim:2012im}
F.~G. Gardim, F.~Grassi, M.~Luzum,  and J.-Y. Ollitrault, {\em {Breaking of
  factorization of two-particle correlations in hydrodynamics}},
  \href{http://dx.doi.org/10.1103/PhysRevC.87.031901}{Phys.~Rev.~C {\bfseries
  87} (2013) 031901},
\href{http://arxiv.org/abs/1211.0989}{{\ttfamily arXiv:1211.0989 [nucl-th]}}.

\bibitem{Heinz:2013bua}
U.~Heinz, Z.~Qiu,  and C.~Shen, {\em {Fluctuating flow angles and anisotropic
  flow measurements}},
  \href{http://dx.doi.org/10.1103/PhysRevC.87.034913}{Phys. Rev. C {\bfseries
  87} (2013) 034913},
\href{http://arxiv.org/abs/1302.3535}{{\ttfamily arXiv:1302.3535 [nucl-th]}}.

\bibitem{Zhou:2018fxx}
M.~Zhou and J.~Jia, {\em {Centrality fluctuations in heavy-ion collisions}},
  \href{http://dx.doi.org/10.1103/PhysRevC.98.044903}{Phys. Rev. C {\bfseries
  98} (2018) 044903},
\href{http://arxiv.org/abs/1803.01812}{{\ttfamily arXiv:1803.01812 [nucl-th]}}.

\bibitem{Skokov:2012ds}
V.~Skokov, B.~Friman,  and K.~Redlich, {\em {Volume fluctuations and higher
  order cumulants of the net baryon number}},
  \href{http://dx.doi.org/10.1103/PhysRevC.88.034911}{Phys. Rev. C {\bfseries
  88} (2013) 034911},
\href{http://arxiv.org/abs/1205.4756}{{\ttfamily arXiv:1205.4756 [hep-ph]}}.

\bibitem{Luo:2013bmi}
X.~Luo, J.~Xu, B.~Mohanty,  and N.~Xu, {\em {Volume fluctuation and
  auto-correlation effects in the moment analysis of net-proton multiplicity
  distributions in heavy-ion collisions}},
  \href{http://dx.doi.org/10.1088/0954-3899/40/10/105104}{J. Phys. G {\bfseries
  40} (2013) 105104},
\href{http://arxiv.org/abs/1302.2332}{{\ttfamily arXiv:1302.2332 [nucl-ex]}}.

\bibitem{Xu:2016qzd}
H.-j. Xu, {\em {Cumulants of multiplicity distributions in most-central
  heavy-ion collisions}},
  \href{http://dx.doi.org/10.1103/PhysRevC.94.054903}{Phys. Rev. C {\bfseries
  94} (2016) 054903},
\href{http://arxiv.org/abs/1602.07089}{{\ttfamily arXiv:1602.07089 [nucl-th]}}.

\bibitem{Aad:2008zzm}
{ATLAS Collaboration}, {\em {The ATLAS Experiment at the CERN Large Hadron
  Collider}},
\href{http://dx.doi.org/10.1088/1748-0221/3/08/S08003}{JINST {\bfseries 3}
  (2008) S08003}.

\bibitem{Aad:ID}
{ATLAS Collaboration}, {\em {The ATLAS Inner Detector commissioning and
  calibration}},
  \href{http://dx.doi.org/10.1140/epjc/s10052-010-1366-7}{Eur.~Phys.~J.~C
  {\bfseries 70} (2010) 787}, \href{http://arxiv.org/abs/1004.5293}{{\ttfamily
  arXiv:1004.5293 [physics.ins-det]}}.

\bibitem{ATLAS-TDR-19}
{ATLAS Collaboration}, {\em ATLAS Insertable B-Layer Technical Design Report},
  Atlas-tdr-19, 2010, \url{https://cds.cern.ch/record/1291633}, {\textit{ATLAS
  Insertable B-Layer Technical Design Report Addendum}}, ATLAS-TDR-19-ADD-1,
  2012, \url{https://cds.cern.ch/record/1451888}.

\bibitem{Abbott:2018ikt}
B.~Abbott {et~al.}, {\em {Production and Integration of the ATLAS Insertable
  B-Layer}}, \href{http://dx.doi.org/10.1088/1748-0221/13/05/T05008}{JINST
  {\bfseries 13} (2018) T05008},
\href{http://arxiv.org/abs/1803.00844}{{\ttfamily arXiv:1803.00844
  [physics.ins-det]}}.

\bibitem{Aaboud:2016leb}
{ATLAS Collaboration}, {\em {Performance of the ATLAS Trigger System in 2015}},
  \href{http://dx.doi.org/10.1140/epjc/s10052-017-4852-3}{Eur. Phys. J. C
  {\bfseries 77} (2017) 317},
\href{http://arxiv.org/abs/1611.09661}{{\ttfamily arXiv:1611.09661 [hep-ex]}}.

\bibitem{Aad:2011yr}
{ATLAS Collaboration}, {\em {Measurement of the centrality dependence of the
  charged particle pseudorapidity distribution in lead-lead collisions at
  $\sqrt{s_{NN}}=2.76$ TeV with the ATLAS detector}},
  \href{http://dx.doi.org/10.1016/j.physletb.2012.02.045}{Phys.~Lett.~B
  {\bfseries 710} (2012) 363}, \href{http://arxiv.org/abs/1108.6027}{{\ttfamily
  arXiv:1108.6027 [hep-ex]}}.

\bibitem{Miller:2007ri}
M.~L. Miller, K.~Reygers, S.~J. Sanders,  and P.~Steinberg, {\em {Glauber
  Modeling in High-Energy Nuclear Collisions}},
  \href{http://dx.doi.org/10.1146/annurev.nucl.57.090506.123020}{Ann.~Rev.~Nucl.~Part.~Sci.
  {\bfseries 57} (2007) 205},
\href{http://arxiv.org/abs/nucl-ex/0701025}{{\ttfamily arXiv:nucl-ex/0701025}}.

\bibitem{Aad:2015wga}
{ATLAS Collabration}, {\em {Measurement of charged-particle spectra in Pb+Pb
  collisions at $\sqrt{s_{NN}} = 2.76$ TeV with the ATLAS detector at the
  LHC}}, \href{http://dx.doi.org/10.1007/JHEP09(2015)050}{JHEP {\bfseries 09}
  (2015) 050},
\href{http://arxiv.org/abs/1504.04337}{{\ttfamily arXiv:1504.04337 [hep-ex]}}.

\bibitem{Aad:2011yk}
{ATLAS Collaboration}, {\em {Measurement of the pseudorapidity and transverse
  momentum dependence of the elliptic flow of charged particles in lead-lead
  collisions at $\sqrt{s_{NN}}=2.76$ TeV with the ATLAS detector}},
  \href{http://dx.doi.org/10.1016/j.physletb.2011.12.056}{Phys.~Lett.~B
  {\bfseries 707} (2012) 330}, \href{http://arxiv.org/abs/1108.6018}{{\ttfamily
  arXiv:1108.6018 [hep-ex]}}.

\bibitem{Gyulassy:1994ew}
M.~Gyulassy and X.-N. Wang, {\em {HIJING 1.0: A Monte Carlo program for parton
  and particle production in high-energy hadronic and nuclear collisions}},
  \href{http://dx.doi.org/10.1016/0010-4655(94)90057-4}{Comput.~Phys.~Commun.
  {\bfseries 83} (1994) 307},
  \href{http://arxiv.org/abs/nucl-th/9502021}{{\ttfamily
  arXiv:nucl-th/9502021}}.

\bibitem{Masera:2009zz}
M.~Masera, G.~Ortona, M.~Poghosyan,  and F.~Prino, {\em {Anisotropic transverse
  flow introduction in Monte Carlo generators for heavy ion collisions}},
  \href{http://dx.doi.org/10.1103/PhysRevC.79.064909}{Phys.~Rev.~C {\bfseries
  79} (2009) 064909}.

\bibitem{Agostinelli:2002hh}
{GEANT4} Collaboration, S.~Agostinelli {et~al.}, {\em {GEANT4: A Simulation
  toolkit}},
\href{http://dx.doi.org/10.1016/S0168-9002(03)01368-8}{Nucl.~Instrum.~Meth.~A
  {\bfseries 506} (2003) 250}.

\bibitem{Aad:2010ah}
{ATLAS Collaboration}, {\em {The ATLAS Simulation Infrastructure}},
  \href{http://dx.doi.org/10.1140/epjc/s10052-010-1429-9}{Eur.~Phys.~J.~C
  {\bfseries 70} (2010) 823}, \href{http://arxiv.org/abs/1005.4568}{{\ttfamily
  arXiv:1005.4568 [physics.ins-det]}}.

\bibitem{Aaboud:2018syf}
{ATLAS Collaboration}, {\em {Correlated long-range mixed-harmonic fluctuations
  measured in $pp$, $p$+Pb and low-multiplicity Pb+Pb collisions with the ATLAS
  detector}}, \href{http://dx.doi.org/10.1016/j.physletb.2018.11.065}{Phys.
  Lett. B {\bfseries 789} (2019) 444},
\href{http://arxiv.org/abs/1807.02012}{{\ttfamily arXiv:1807.02012 [nucl-ex]}}.

\bibitem{DiFrancesco:2016srj}
P.~Di~Francesco, M.~Guilbaud, M.~Luzum,  and J.-Y. Ollitrault, {\em {Systematic
  procedure for analyzing cumulants at any order}},
  \href{http://dx.doi.org/10.1103/PhysRevC.95.044911}{Phys. Rev. {\bfseries
  C95} (2017) 044911},
\href{http://arxiv.org/abs/1612.05634}{{\ttfamily arXiv:1612.05634 [nucl-th]}}.

\bibitem{Voloshin:2008dg}
S.~A. Voloshin, A.~M. Poskanzer,  and R.~Snellings, {\em {Collective Phenomena
  in Non-Central Nuclear Collisions}},
  \href{http://dx.doi.org/10.1007/978-3-642-01539-7_10}{Landolt-Bornstein,
  Elementary Particles, Nuclei and Atoms {\bfseries 23} (2010) 293},
\href{http://arxiv.org/abs/0809.2949}{{\ttfamily arXiv:0809.2949 [nucl-ex]}}.

\bibitem{Jia:2014pza}
J.~Jia and S.~Radhakrishnan, {\em {Limitation of multiparticle correlations for
  studying the event-by-event distribution of harmonic flow in heavy-ion
  collisions}}, \href{http://dx.doi.org/10.1103/PhysRevC.92.024911}{Phys. Rev.
  C {\bfseries 92} (2015) 024911},
\href{http://arxiv.org/abs/1412.4759}{{\ttfamily arXiv:1412.4759 [nucl-ex]}}.

\bibitem{Aaboud:2017acw}
{ATLAS Collaboration}, {\em {Measurement of multi-particle azimuthal
  correlations in $pp$, $p+$Pb and low-multiplicity Pb$+$Pb collisions with the
  ATLAS detector}},
  \href{http://dx.doi.org/10.1140/epjc/s10052-017-4988-1}{Eur. Phys. J. C
  {\bfseries 77} (2017) 428},
\href{http://arxiv.org/abs/1705.04176}{{\ttfamily arXiv:1705.04176 [hep-ex]}}.

\bibitem{Huo:2017nms}
P.~Huo, K.~Gajdošová, J.~Jia,  and Y.~Zhou, {\em {Importance of non-flow in
  mixed-harmonic multi-particle correlations in small collision systems}},
  \href{http://dx.doi.org/10.1016/j.physletb.2017.12.035}{Phys. Lett. B
  {\bfseries 777} (2018) 201--206},
\href{http://arxiv.org/abs/1710.07567}{{\ttfamily arXiv:1710.07567 [nucl-ex]}}.

\bibitem{Zhang:2018lls}
C.~Zhang, J.~Jia,  and J.~Xu, {\em {Non-flow effects in three-particle
  mixed-harmonic azimuthal correlations in small collision systems}},
  \href{http://dx.doi.org/10.1016/j.physletb.2019.03.035}{Phys. Lett. B
  {\bfseries 792} (2019) 138--141},
\href{http://arxiv.org/abs/1812.03536}{{\ttfamily arXiv:1812.03536 [nucl-th]}}.

\bibitem{Giacalone:2016afq}
G.~Giacalone, L.~Yan, J.~Noronha-Hostler,  and J.-Y. Ollitrault, {\em
  {Symmetric cumulants and event-plane correlations in Pb + Pb collisions}},
  \href{http://dx.doi.org/10.1103/PhysRevC.94.014906}{Phys. Rev. C {\bfseries
  94} (2016) 014906},
\href{http://arxiv.org/abs/1605.08303}{{\ttfamily arXiv:1605.08303 [nucl-th]}}.

\bibitem{Das:2017ned}
S.~J. Das, G.~Giacalone, P.-A. Monard,  and J.-Y. Ollitrault, {\em {Relating
  centrality to impact parameter in nucleus-nucleus collisions}},
  \href{http://dx.doi.org/10.1103/PhysRevC.97.014905}{Phys. Rev. C {\bfseries
  97} (2018) 014905},
\href{http://arxiv.org/abs/1708.00081}{{\ttfamily arXiv:1708.00081 [nucl-th]}}.

\bibitem{Akamatsu:2016llw}
Y.~Akamatsu, A.~Mazeliauskas,  and D.~Teaney, {\em {Kinetic regime of
  hydrodynamic fluctuations and long time tails for a Bjorken expansion}},
  \href{http://dx.doi.org/10.1103/PhysRevC.95.014909}{Phys. Rev. C {\bfseries
  95} (2017) 014909},
\href{http://arxiv.org/abs/1606.07742}{{\ttfamily arXiv:1606.07742 [nucl-th]}}.

\bibitem{Teaney:2012ke}
D.~Teaney and L.~Yan, {\em {Nonlinearities in the harmonic spectrum of heavy
  ion collisions with ideal and viscous hydrodynamics}},
  \href{http://dx.doi.org/10.1103/PhysRevC.86.044908}{Phys.~Rev. C {\bfseries
  86} (2012) 044908}, \href{http://arxiv.org/abs/1206.1905}{{\ttfamily
  arXiv:1206.1905 [nucl-th]}}.

\bibitem{Adler:2002pu}
{STAR Collaboration}, {\em {Elliptic flow from two and four particle
  correlations in Au+Au collisions at 130-GeV}},
  \href{http://dx.doi.org/10.1103/PhysRevC.66.034904}{Phys. Rev. C {\bfseries
  66} (2002) 034904},
\href{http://arxiv.org/abs/nucl-ex/0206001}{{\ttfamily arXiv:nucl-ex/0206001
  [nucl-ex]}}.

\bibitem{Alver:2008zza}
B.~Alver {et~al.}, {\em {Importance of correlations and fluctuations on the
  initial source eccentricity in high-energy nucleus-nucleus collisions}},
  \href{http://dx.doi.org/10.1103/PhysRevC.77.014906}{Phys. Rev. C {\bfseries
  77} (2008) 014906},
\href{http://arxiv.org/abs/0711.3724}{{\ttfamily arXiv:0711.3724 [nucl-ex]}}.

\bibitem{Giacalone:2016mdr}
G.~Giacalone, L.~Yan, J.~Noronha-Hostler,  and J.-Y. Ollitrault, {\em {The
  fluctuations of quadrangular flow}},
  \href{http://dx.doi.org/10.1088/1742-6596/779/1/012064}{J. Phys. Conf. Ser.
  {\bfseries 779} (2017) 012064},
\href{http://arxiv.org/abs/1608.06022}{{\ttfamily arXiv:1608.06022 [nucl-th]}}.

\bibitem{Borghini:2002mv}
N.~Borghini, P.~M. Dinh, J.-Y. Ollitrault, A.~M. Poskanzer,  and S.~A.
  Voloshin, {\em {Effects of momentum conservation on the analysis of
  anisotropic flow}}, \href{http://dx.doi.org/10.1103/PhysRevC.66.014901}{Phys.
  Rev. C {\bfseries 66} (2002) 014901},
  \href{http://arxiv.org/abs/nucl-th/0202013}{{\ttfamily
  arXiv:nucl-th/0202013}}.

\bibitem{Zhu:2016puf}
X.~Zhu, Y.~Zhou, H.~Xu,  and H.~Song, {\em {Correlations of flow harmonics in
  2.76A TeV Pb--Pb collisions}},
  \href{http://dx.doi.org/10.1103/PhysRevC.95.044902}{Phys. Rev. C {\bfseries
  95} (2017) 044902},
\href{http://arxiv.org/abs/1608.05305}{{\ttfamily arXiv:1608.05305 [nucl-th]}}.

\bibitem{Agakishiev:2011eq}
{STAR Collaboration}, {\em {Energy and system-size dependence of two- and
  four-particle $v_2$ measurements in heavy-ion collisions at
  $\sqrt{s_{\mathrm{NN}}}=62.4$ and 200 GeV and their implications on flow
  fluctuations and nonflow}},
  \href{http://dx.doi.org/10.1103/PhysRevC.86.014904}{Phys. Rev. C {\bfseries
  86} (2012) 014904},
\href{http://arxiv.org/abs/1111.5637}{{\ttfamily arXiv:1111.5637 [nucl-ex]}}.

\bibitem{Aaboud:2017tql}
{ATLAS Collaboration}, {\em {Measurement of longitudinal flow decorrelations in
  Pb+Pb collisions at $\sqrt{s_{\text {NN}}}=2.76$ and 5.02 TeV with the ATLAS
  detector}}, \href{http://dx.doi.org/10.1140/epjc/s10052-018-5605-7}{Eur.
  Phys. J. C {\bfseries 78} (2018) 142},
\href{http://arxiv.org/abs/1709.02301}{{\ttfamily arXiv:1709.02301 [nucl-ex]}}.

\bibitem{ATL-GEN-PUB-2016-002}
{ATLAS Collaboration}, {\em {ATLAS Computing Acknowledgements 2016--2017}},
  {ATL-GEN-PUB-2016-002}, \url{https://cds.cern.ch/record/2202407}.

\end{thebibliography}\endgroup
\clearpage 
 
\begin{flushleft}
{\Large The ATLAS Collaboration}

\bigskip

M.~Aaboud$^\textrm{\scriptsize 35d}$,    
G.~Aad$^\textrm{\scriptsize 101}$,    
B.~Abbott$^\textrm{\scriptsize 128}$,    
D.C.~Abbott$^\textrm{\scriptsize 102}$,    
O.~Abdinov$^\textrm{\scriptsize 13,*}$,    
A.~Abed~Abud$^\textrm{\scriptsize 70a,70b}$,    
D.K.~Abhayasinghe$^\textrm{\scriptsize 93}$,    
S.H.~Abidi$^\textrm{\scriptsize 167}$,    
O.S.~AbouZeid$^\textrm{\scriptsize 40}$,    
N.L.~Abraham$^\textrm{\scriptsize 156}$
% ...
\bigskip
\\

% ...
$^{13}$Institute of Physics, Azerbaijan Academy of Sciences, Baku; Azerbaijan.\\
% ...
$^{35}$$^{(a)}$Facult\'e des Sciences Ain Chock, R\'eseau Universitaire de Physique des Hautes Energies - Universit\'e Hassan II, Casablanca;$^{(b)}$Facult\'{e} des Sciences, Universit\'{e} Ibn-Tofail, K\'{e}nitra;$^{(c)}$Facult\'e des Sciences Semlalia, Universit\'e Cadi Ayyad, LPHEA-Marrakech;$^{(d)}$Facult\'e des Sciences, Universit\'e Mohamed Premier and LPTPM, Oujda;$^{(e)}$Facult\'e des sciences, Universit\'e Mohammed V, Rabat; Morocco.\\
% ...
$^{40}$Niels Bohr Institute, University of Copenhagen, Copenhagen; Denmark.\\
% ...
$^{70}$$^{(a)}$INFN Sezione di Pavia;$^{(b)}$Dipartimento di Fisica, Universit\`a di Pavia, Pavia; Italy.\\
% ...
$^{93}$Department of Physics, Royal Holloway University of London, Egham; United Kingdom.\\
% ...
$^{101}$CPPM, Aix-Marseille Universit\'e, CNRS/IN2P3, Marseille; France.\\
$^{102}$Department of Physics, University of Massachusetts, Amherst MA; United States of America.\\
% ...
$^{128}$Homer L. Dodge Department of Physics and Astronomy, University of Oklahoma, Norman OK; United States of America.\\
% ...
$^{156}$Department of Physics and Astronomy, University of Sussex, Brighton; United Kingdom.\\
% ...
$^{167}$Department of Physics, University of Toronto, Toronto ON; Canada.\\
% ...
$^{a}$ Also at Borough of Manhattan Community College, City University of New York, New York NY; United States of America.\\
$^{b}$ Also at Centre for High Performance Computing, CSIR Campus, Rosebank, Cape Town; South Africa.\\
$^{c}$ Also at CERN, Geneva; Switzerland.\\
$^{d}$ Also at CPPM, Aix-Marseille Universit\'e, CNRS/IN2P3, Marseille; France.\\
$^{e}$ Also at D\'epartement de Physique Nucl\'eaire et Corpusculaire, Universit\'e de Gen\`eve, Gen\`eve; Switzerland.\\
% ...
$^{*}$ Deceased

\end{flushleft}


\end{document}